# A survey on software testability

Vahid Garousi
Information Technology Group
Wageningen University, Netherlands
vahid.garousi@wur.nl

Michael Felderer
Blekinge Institute of Technology, Sweden &
University of Innsbruck, Austria
michael.felderer@uibk.ac.at

Feyza Nur Kılıçaslan
Department of Computer Engineering
Hacettepe University, Ankara, Turkey
feyzanur@cs.hacettepe.edu.tr

**Abstract**:

*Context*: Software testability is the degree to which a software system or a unit under test supports its own testing. To predict and improve software testability, a large number of techniques and metrics have been proposed by both practitioners and researchers in the last several decades. Reviewing and getting an overview of the entire state-of-the-art and state-of-the-practice in this area is often challenging for a practitioner or a new researcher.

*Objective*: Our objective is to summarize the body of knowledge in this area and to benefit the readers (both practitioners and researchers) in preparing, measuring and improving software testability.

*Method*: To address the above need, the authors conducted a survey in the form of a systematic literature mapping (classification) to find out what we as a community know about this topic. After compiling an initial pool of 303 papers, and applying a set of inclusion/exclusion criteria, our final pool included 208 papers.

*Results*: The area of software testability has been comprehensively studied by researchers and practitioners. Approaches for measurement of testability and improvement of testability are the most-frequently addressed in the papers. The two most often mentioned factors affecting testability are observability and controllability. Common ways to improve testability are testability transformation, improving observability, adding assertions, and improving controllability.

*Conclusion*: This paper serves for both researchers and practitioners as an "index" to the vast body of knowledge in the area of testability. The results could help practitioners measure and improve software testability in their projects. To assess potential benefits of this review paper, we shared its draft version with two of our industrial collaborators. They stated that they found the review useful and beneficial in their testing activities. Our results can also benefit researchers in observing the trends in this area and identify the topics that require further investigation.

## TABLE OF CONTENTS

## 1 INTRODUCTION

Software testing is a fundamental and the most widespread activity to ensure quality of software systems. However, not all software systems are easily testable. Some are easier to be tested than others. Software testability is the degree to which a software artifact (i.e. a software system, software module, requirements or design document) supports testing in a given test context [1]. If the testability of the software artifact is high, then finding faults in the system (if it has any) by means of testing is easier. A lower degree of testability results in increased test effort, and thus in less testing performed in a given fixed amount of time, and thus less chances for findings software defects [1]. As one test practitioner puts it [2]: "*Anything that makes testing harder or slower gives bugs more time or more opportunities to hide*". If software engineers can increase testability of a given software, they will be able to decrease cost, increase quality of test activities and, as a result, produce higher quality software.

Broadly conceived, software testability is a result of six factors [3]: (1) characteristics of the representation (e.g., requirements), (2) characteristics of the implementation; (3) built-in test capabilities; (4) the test suite (test cases); (5) the test support environment; and (6) the software process in which testing is conducted.

To predict and improve software testability, a large number of techniques, metrics and frameworks have been proposed over the last few decades both by researchers and practitioners. This makes it challenging to review and get an overview of the entire state-of-the-art and state-of-the-practice in the area of testability. Furthermore, we observed that there is no clear guidance on how to measure testability and how to handle issues with it. Especially also in our industrial software testing projects and interactions with the industry in the last 15 years (e.g., [4-8]), we often observed that many companies struggle with predicting and improving software testability in their context, due to not having an adequate overview of what already exists in this area and due to missing systematic support in general.

Knowing that they can adapt/customize an existing technique to predict and improve software testability in their own context can potentially save companies and test engineers a lot of time and money. Furthermore, there are also a lot of open research challenges on testability that require a summary and a classification of the entire field. Furthermore, a recent insightful paper in IEEE Software [9] highlighted "*the practical value of historical data [and approaches published in the past]*" and a "*vicious cycle of inflation of software engineering terms and knowledge*" (due to many papers not adequately reviewing the state of art). We believe survey papers like the current one aim at addressing the above problem, and also the following challenge as mentioned in that paper: "*Previous work is hard to find*".

Although there have been review and survey papers on specific aspects of software testability (see Table 2), no paper has so far studied the entire body of knowledge in a holistic manner, which is essential for the field of software testability that is equally driven by academia and industry.

To address the above need and to find out what we, as a community, know about software testability, we recently conducted a systematic mapping on the technical papers written by practitioners and researchers and we present a summary of its results in this article. Our review pool included 208 technical papers published in conferences and journals, and the earliest paper [P109][1] was published in 1982, entitled "*On testing non-testable programs*" which highlights the significance of testability more than three decades ago. Previous "review" (survey) papers like this article have appeared on other topics, e.g., Agile development [10], developer motivation [11], or testing embedded software [12], and have shown to be useful in providing concise overviews on a given area.

The remainder of this article is structured as follows. Section 2 provides background and related work on testability. Section 3 describes the research method and the review planning. Section 4 presents the search phase and the selection of pool of sources to be reviewed. Section 5 discusses the development of the systematic map and data-extraction plan. Section 6 presents the results of the literature review. Section 7 summarizes the findings and discusses the lessons learned. Finally, in Section 8, we draw conclusions, and suggest areas for further research.

## 2 Background and related work

In this section, we first provide a brief overview of the concept of software testability. We then briefly review the related work, i.e., survey (review) papers on software testability in the literature.

### 2.1 A brief overview of the concept of software testability

A large number of definitions for software testability has been offered in the literature. As expected, there were similarities among different definitions. Based on our pool of studies (Section 4), we summarize a non-exhaustive list of those definitions in Table 1.

The first three rows in the table are the definitions by the following three standards: IEEE standard 610.12-1990 (glossary of software engineering terminology) [13], ISO standard 9126-1: 2001 (Software engineering - product quality) [14], and military standard MIL-STD-2165 (Testability program for electronic systems and equipment) [15]. The other definitions are those provided by the papers in the pool.

We have furthermore classified the definitions of software testability regarding their focus into three groups: (1) facilitation (ease) of testing; (2) facilitation (ease) of revealing faults; and (3) other focus areas. In the first group of definitions, testability is interpreted as the factor related to facilitation (ease) of testing which is often viewed as related to test efficiency. The second group of definitions interprets testability as facilitation (ease) of revealing faults which is related to test effectiveness.

As one can see in Table 1., the definitions are quite equally partitioned between the first two groups. Some sources have also defined testability considering both aspects, e.g., MIL-STD-2165 [15], [P141, P200]]. We also personally believe that a definition considering both aspects is more appropriate, i.e., a definition such as the following: *the degree to which a software system or component facilitates the establishment of test criteria, generation and execution of test cases, and also the probability that a piece of software will fail during testing if it includes a fault*.

[1] Citations in the form of [Pn], such as [P98], refer to the IDs of the primary studies (papers) reviewed in our study. They are available in the online dataset of this study: www.goo.gl/boNuFD

In addition, one can also see from Table 1 that some sources have extended (specialized) the definition of software testability, e.g., "runtime testability" [P103] which is the "*degree to which a system can be runtime tested*", "design for testability" which "*is designing a software so that the testing of its implementation becomes more economical and manageable*" [P188], and evolutionary (search-based) testability which is defined as "*the degree of difficulty to successfully apply evolutionary testing to a particular software*" [P97].

**Table 1: A subset of definitions for software testability in the literature**

| Reference | Definition | Classification of definitions | | |
|---|---|---|---|---|
| | | Facilitation of testing (test efficiency) | Facilitation of revealing faults (test effectiveness) | Other focus areas |
| IEEE standard 610.12-1990 [13] | *"the degree to which a system or component facilitates the establishment of test criteria and the performance of tests to determine whether those criteria have been met; the degree to which a requirement is stated in terms that permit the establishment of test criteria and performance of tests to determine whether those criteria have been met"* | x | | Requirements in support of testing, requirements testability |
| ISO standard 9126-1: 2001 [14] | *"attributes of software that bear on the effort needed to validate the software product"* | x | | |
| ISO standard 25010:2011 [16] | *"degree of effectiveness and efficiency with which test criteria can be established for a system, product or component and tests can be performed to determine whether those criteria have been met"* | x | x | |
| ISTQB standard glossary [17] | *"capability of the software product to enable modified software to be tested"* | x | | |
| ISO standard 12207:2008 [18] | *"extent to which an objective and feasible test can be designed to determine whether a requirement is met"* | x | | |
| ISO/IEC/IEEE 29148:2011 [19] | *"degree to which a requirement is stated in terms that permit establishment of test criteria and performance of tests to determine whether those criteria have been met"* | x | | |
| ISO/IEEE standard 24765:2010 [13] | *"degree to which a system can be unit tested and system tested!* <br> *"effort required to test software"* <br> *"degree to which a system or component facilitates the establishment of test criteria and the performance of tests to determine whether those criteria have been met"* | x | | |
| Military standard MIL-STD-2165 [15] | *"a design characteristic which allows the status (operable, inoperable, or degraded) to be determined and the isolation of faults within the terms to be performed in a timely and efficient manner"* | x | x | |
| [P6] in the pool of studies reviewed in this survey | *"…by testability, we will understand how difficult it is to test systems, measured in terms of the size required by test suites to be complete."* | x | | |
| [P17] | *"Domain testability refers to the ease of modifying a program so that it is observable and controllable"* | | | Relies on observability and controllability |
| [P21] | *"degree to which a component or system can be tested in isolation"* | x | | |
| [P41] | *"ease of testing a piece of software design using structural testing strategies"* | x | | |
| [P43] | *"tendency for software to reveal its faults during testing"* | | x | |
| [P65] | *"relative ease and expense of revealing software faults"* | | x | |
| [P68] | *"property of an object that facilitates the testing process"* | x | | |
| [P79] | *"prediction of the tendency for failures to be observed during random black-box testing when faults are present"* | | x | |
| [P81] | *"how easy it is to test by a particular tester and test process, in a given context"* | x | | |
| [P94] | *"the minimum number of test cases to provide total test coverage, assuming that such coverage is possible"* | x | | |
| [P97] | *"Evolutionary testability can be defined as the degree of difficulty to successfully apply evolutionary testing to a particular software"* | x | | |
| [P103] | *"Runtime testability is the degree to which a system can be runtime tested"* | x | | |
| [P106] | *"Testability is an measure of how easily software exposes faults when tested"* | | x | |
| [P108] | *"how well a component is structured to facilitate software testing"* | x | | |
| [P112] | *"Design For Testability (DFI') is understood as the process of introducing some features into a protocol entity that facilitate the testing process of protocol implementations."* | x | | |
| [P117] | *"The testability of a program is the probability that a test of the program on an input diagram from a specified probability distribution of the inputs is rejected, given a specified oracle and given that the program is faulty."* | | x | |
| [P122] | *"probability that existing faults will be revealed by existing test cases"* | | x | |
| [P127] | *"tendency of code to reveal existing faults during random testing"* | | x | |
| [P141] | *"Testability is a property of both the software and the process and refers to the easiness for applying all the [testing]steps and on the inherent of the software to reveal faults during testing"* | x | x | |
| [P150] | *"a program's property that is introduced with the intention of predicting efforts required for testing the program"* | x | | |
| [P151] | *"the probability that a piece of software will fail on its next execution during testing … if the software includes a fault"* | | x | |
| [P168] | *"a module is said to be testable if it is observable and controllable"* | | | Observability and controllability |
| [P188] | *"software design for testability is designing a software so that the testing of its implementation becomes more economical and manageable"* | x | | |
| [P193] | *"We define testability with an intuitive static approximation to the Voas' execution probability. Testability of each node in the graph is the proportion of paths passing through that node. A testability of 1 means the node has perfect testability. A testability of 0 means the node is unreachable."* | | | Ratio of paths covering a node (statement) |
| [P200] | *"Software testability is a characteristic that either suggests how easy software is to test, how well the tests are able to interact with the code to detect defects or a combination of the two"* | x | x | |

## 2.2 Related works: other survey (review) papers on software testability

Several survey/review papers (secondary studies) have been reported in the scope of software testability. We were able to find 13 such studies [20-32] and provide their list in Table 2. For each study, we include its publication year, its type (regular survey or systematic mapping/review), number of artifacts reviewed (papers, tools, etc.) by the study, and some explanatory notes. As one can see in Table 2, the earliest survey paper in this area was published in 1991 [20], which was a survey of tools (not papers) for testability, reliability and maintainability assessment, published by the US Department of Defense (DoD). Most of the other studies are published in recent years (since 2010). Most papers reviewed papers as their artifacts under review, while [20] reviewed tools for testability, and [23] reviewed testability metrics.

While most of these studies are surveys on general scope of testability, some focused on specific aspects in relation to testability, e.g., [28] was a conventional survey on design testability, [30] was a systematic literature review (SLR) on the relationship of software robustness and testability, and [32] was a systematic mapping (SM) study on relationship of software robustness and software performance.

In term of the number of artifacts reviewed (papers, tools, etc.) by each study, the numbers range between 10 and 42 papers. However, our survey is the most up-to-date and comprehensive systematic review in the area by considering all 208 papers published in this area. Note that our survey does not do any "comparison" or "contrasting" with any of the existing surveys, but instead the pool of papers reviewed in this work is systematically collected and substantially larger than existing surveys in this area.

We should also mention that this work is a major extension of an initial systematic literature mapping reported in a regional conference in Turkey in 2016, which had reviewed only a pool of 29 papers [33]. Compared to that initial work, this substantially-extended mapping study has more RQs and also a much larger paper pool (208 papers).

**Table 2: A list of other survey (review) papers on software testability (sorted by publication years)**

| Paper title | Year | Reference | Type of survey | | Num. of artifacts reviewed (papers, tools, etc.) | Notes |
|---|---|---|---|---|---|---|
| | | | Regular survey | Systematic mapping/review | | |
| A survey of reliability, maintainability, supportability, and testability software tools | 1991 | [20] | x | | 29 testability tools | Section 1.3 is on testability tools |
| An evaluation for model testability approaches | 2010 | [21] | x | | 10 primary studies (papers) | Focused on model testability |
| Present and future of software testability analysis | 2010 | [22] | x | | 42 papers | |
| Survey of source code metrics for evaluating testability of object-oriented (OO) systems | 2010 | [23] | x | | 12 metrics (see Table 1) | Focused on source code metrics for evaluating testability |
| Adaptation of software testability concept for test suite generation: a systematic review | 2012 | [24] | x | | 36 papers | |
| An empirical study into model testability | 2013 | [25] | x | | 10 papers | Focused on model testability. Same authors and similar contents as [21] |
| OO software testability survey at designing and implementation phase | 2013 | [26] | x | | 14 papers | Focused on testability at design and implementation phase |
| Measuring testability of OO design: a systematic review | 2014 | [27] | | x | 17 papers | Focused on design testability |
| OO design for testability: a systematic review | 2014 | [28] | | x | 13 papers | Focused on design testability |
| Testability estimation of object oriented software: a systematic review | 2014 | [29] | | x | 13 papers | |
| Testability and software robustness: a systematic literature review | 2015 | [30] | | x | 27 papers | Focused on relationship of software robustness and testability |
| OO software testability (OOST) metrics analysis | 2015 | [31] | x | | 15 papers | A high-level survey |
| Testability and software performance: a systematic mapping study | 2016 | [32] | | x | 26 papers | Focused on relationship of software robustness and software performance |
| Software testability: a systematic literature mapping | 2016 | [33] | | x | 29 papers | Our initial work, on which this review has been built upon |
| (This study) | 2018 | | | x | 208 papers | The most up-to-date and comprehensive systematic review in the area |

## 3 Research method and SLM planning

In this section an overview of our research method (Section 3.1) as well as then the goal and review questions of our study (Section 3.2) are presented.

### 3.1 Overview

Based on our past experience in SLM and SLR studies, e.g., [34-39], and also using the well-known guidelines for conducting SLR and SLM studies in SE (e.g., [40-43]), we developed our SLM process, as shown in Figure 1. We discuss the SLM planning and design phase (its goal and RQs) in the next section. Section 4 to 6 then present each of the follow-up phases of the process.

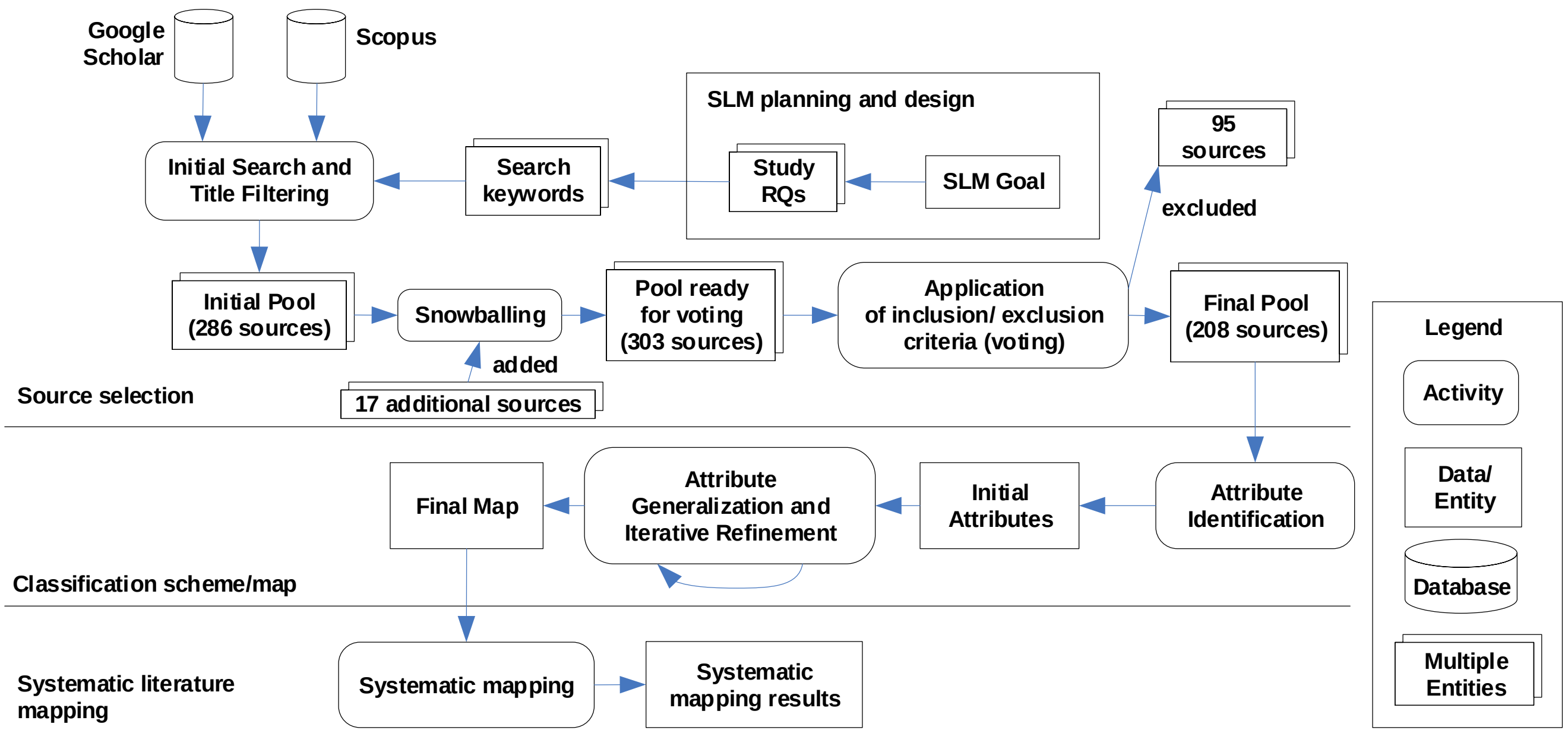

**Figure 1-An overview of the applied SLM process (represented as a UML activity diagram)**

## 3.2 GOAL AND REVIEW QUESTIONS

The goal of this study is to systematically map (classify), review and synthesize the body of knowledge in the area of software testability, to find out the recent trends and directions in this field, and to identify opportunities for future research, from the point of view of researchers and practitioners. Based on the above goal, we raise the following review questions (RQs) grouped into three categories:

<u>Group 1-Common to all SLM studies:</u>

- **RQ 1.1- Classification of studies by contribution types**: What are the different contributions of different sources? How many sources present approach, methods, tools or frameworks for handling testability?
- **RQ 1.2- Classification of studies by research method types**: What type of research methods have been used in the sources in this area? Some of the studies presented solution proposals or weak empirical studies while others presented strong empirical studies.

<u>Group 2-Specific to the domain (i.e., testability):</u>

- **RQ 2.1- Type of approaches for treating testability**: How is testability treated in the paper? For example, does the given paper present an approach for measurement or an approach improvement of testability?
- **RQ 2.2- Factors affecting testability**: What factors / attributes impact testability? It is important to understand what characteristics of a Systems Under Test (SUT) affects (lowers or increases) its testability.
- **RQ 2.3- Techniques for improving testability**: What are the techniques for improving testability? As discussed in Section 1, in our industrial software testing projects and interactions with the industry in the last 15 years (e.g., [4-6]), we have often observed that many companies struggle in improving software testability in their contexts. Thus, this RQ intended to synthesize the list of those techniques to benefit practitioners.

<u>Group 3-Specific to empirical studies:</u>

- **RQ 3.1- Research questions investigated in the empirical studies**: What are the research questions raised and studied in the empirical studies? Answering this RQ will assist us and readers (e.g., younger researchers) in exploring potential interesting future research directions.
- **RQ 3.2- Number and sizes of SUTs (examples) in each paper**: How many SUTs (example systems) are discussed in each paper, and how large are those systems? One would expect that each paper applies the proposed testing technique to at least one SUT. Some papers take a more comprehensive approach and apply the proposed testing technique to more SUTs.
- **RQ 3.3-Domains of SUTs**: What are the domains of the SUT(s) studied in each paper? The testability approaches proposed in some papers can be applied to any (generic) type of software, but some papers apply their idea on systems in specific domains, e.g., real-time or embedded systems, e.g., [P3].

<u>Group 4-Demographic information:</u>

- **RQ 4.1-Affiliation types of the study authors**: What are the affiliation types of the authors? We wanted to know the extent to which academics and practitioners are active in this area.
- **RQ 4.2- Highly-cited papers**: What are the highly-cited papers in this area? This RQ could help practitioners with deciding which sources to start reading first.

## 4 SEARCHING FOR AND SELECTION OF SOURCES

Let us recall from our SLM process (Figure 1) that the first phase of our study is article selection. For this phase, we followed the following steps in order:

- Source selection and search keywords (Section 4.1)
- Application of inclusion and exclusion criteria (Section 4.2)
- Finalizing the pool of articles and the online repository (Section 4.3)

### 4.1 SELECTING THE SOURCE ENGINES AND SEARCH KEYWORDS

In our review and mapping, we followed the standard process for performing systematic literature review (SLR) and systematic literature mapping (SLM) studies in software engineering. We performed the searches in both the Google Scholar database and Scopus (www.scopus.com), both widely used in review studies and bibliometrics papers, e.g., [44, 45]. The reason that we used Scopus in addition to Google Scholar was that several sources have mentioned that: "*it [Google Scholar] should not be used alone for systematic review searches*" [46] as it may miss to find a subset of papers.

All the authors did independent searches using the search strings, and, during this search phase, the authors already applied inclusion/exclusion criterion for including only those which explicitly addressed the study's topic. Our exact search string used in both search engines was: `"software testability" OR "testable software"`.

In terms of timeline, the searches were conducted during winter and spring 2017. Data extraction from the primary studies and their classifications were conducted during the summer 2017. We wanted to include all available papers on this topic, thus we did not restrict the papers' year of publication (e.g., only those published since 2000).

To ensure making our paper search and selection efforts efficiency, while doing the searches using the keywords, we also conducted title filtering to ensure that we would add to our candidate paper pool only directly- or potentially-relevant papers. After all, it would be meaningless to add an irrelevant paper to the candidate pool and then remove it. Our first inclusion/exclusion criterion (discussed in Section 4.2) was used for this purpose (i.e., Does the source focus on software testability?). For example, Figure 2 shows a screenshot of our search activity using Google Scholar in which directly- or potentially-relevant papers are highlighted by red boxes. To ensure efficiency of our efforts, we only added such related studies to the candidate pool.

Another issue was the stopping condition when searching using the Google Scholar. As Figure 2 shows, Google Scholar provided a very large number of hits using the above keyword as of this writing (more than 2 million records). Going through all of them was simply impossible for us. To cope with this challenge, we utilized the relevance ranking of the search engine (Google's PageRank algorithm) to restrict the search space. The good news was that, as per our observations, relevant results usually appeared in the first few pages and as we go through the pages, relevancy of results decreased. Thus, we checked the first n pages (i.e., somewhat a search "saturation" effect) and only continued further if needed, e.g., when at least one result in the $n^{th}$ page still was relevant (if at least one paper focused on testing embedded software). Similar heuristics have been reported in several other review studies, guideline and experience papers [47-51]. At the end of our initial search and title filtering, our candidate pool had 303 papers (as shown in our SLM process in Figure 1).

Also, as both Scopus and Google Scholar were used, there were chances of duplications in the pool. We only added a candidate paper to the paper pool if it was not already in the candidate pool.

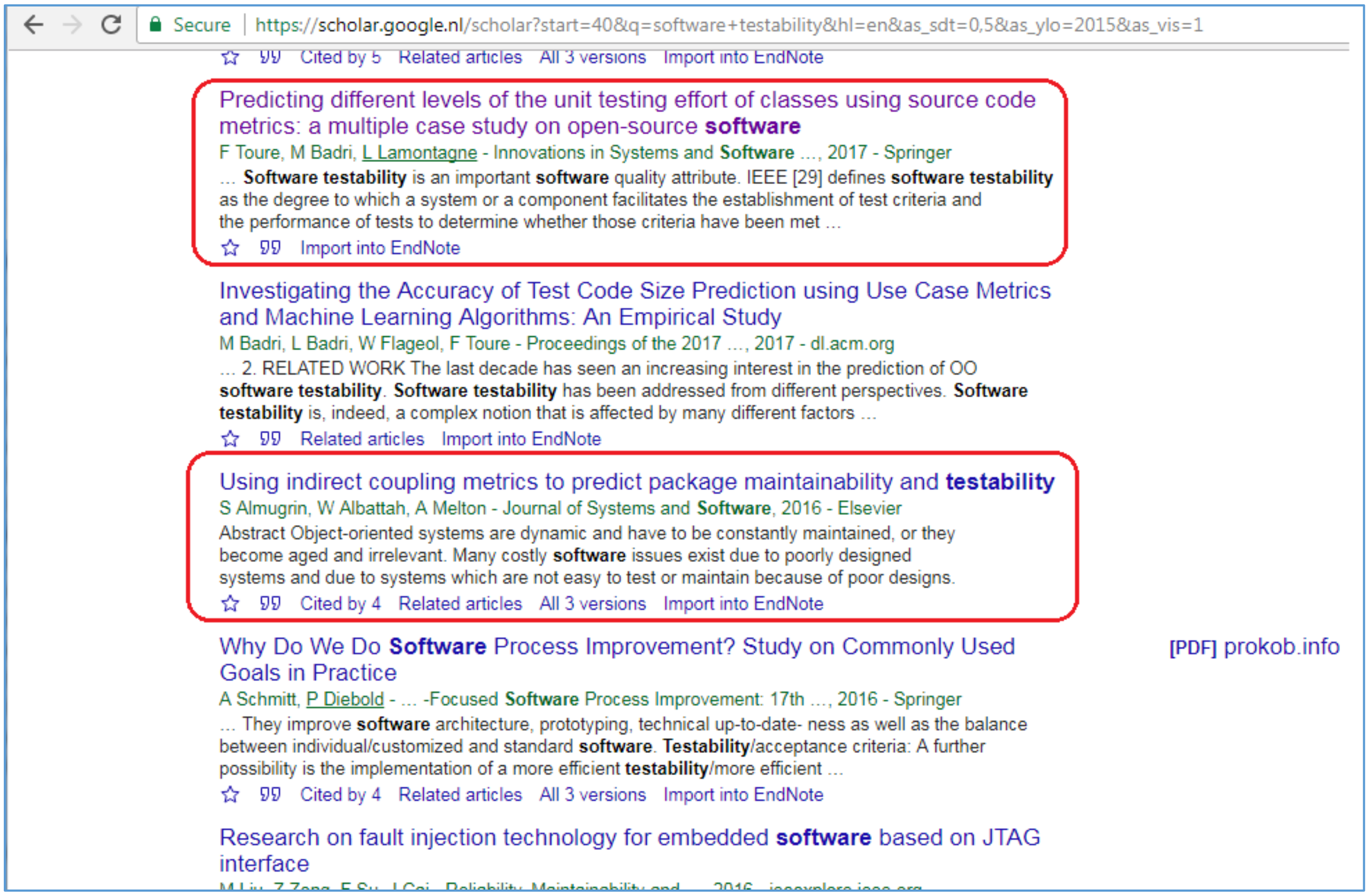

**Figure 2- A screenshot from the search activity using Google Scholar (directly- or potentially-relevant papers are highlighted by red boxes)**

To ensure including all the relevant sources as much as possible, we conducted forward and backward *snowballing* [41], as recommended by systematic review guidelines, on the set of papers already in the pool. Snowballing, in this context, refers to using the reference list of a paper (backward snowballing) or the citations to the paper to identify additional papers (forward) [41]. Snowballing provided 17 additional papers. Some examples of the papers found during snowballing are the followings. [P20] and [P172] were found by "forward" snowballing from [P9]. [P190] was found and added by backward snowballing of [P61]. Note that the citations in the form of [Pn] refer to the IDs of the primary studies (papers) reviewed in our study. They are available in the study's online spreadsheet (www.goo.gl/boNuFD) [52].

Note that, during our search, our goal was to add only primary studies to the pool. In our search, we also found 13 survey/review studies [20-32] which we did not add to the pool, but we noted them down and already discussed them in Section 2.2.

After compiling an initial pool of 303 "candidate" papers, a systematic voting (as discussed next) was conducted among the authors, in which a set of defined inclusion/exclusion criteria were applied to derive the final pool of the primary studies.

### 4.2 APPLICATION OF INCLUSION/EXCLUSION CRITERIA AND VOTING

We carefully defined the inclusion and exclusion criteria to ensure including all the relevant sources and not including the out-of-scope sources. The inclusion criteria were as follows:

- Does the paper focus on software testability?
- Does the paper include a relatively sound validation?
- Is the source in English and can its full-text be accessed on the internet?

The answer for each question could be binary: either Yes (value=1) or No (value=0). Our voting approach was as follows. One of the researchers voted on all the candidate papers w.r.t. the three criteria mentioned before. The other two researchers then peer reviewed all those votes. We used a consensus approach where disagreements were discussed until consensus was reached.

We included only the sources which received 1's for all the three criteria, and excluded the rest. Application of the above criteria led to exclusion of 95 sources, details for which can also be found in the study's online spreadsheet. For example, we excluded [53] since it did not report any validation of the proposed ideas. It was actually a position paper presenting work in progress without any validation.

### 4.3 Final pool of the primary studies

As mentioned above, the references for the final pool of 208 papers can be found in an online spreadsheet [52]. Once we finalized the pool of papers, we wanted to assess the growth of this field by the number of published papers each year. For this purpose, we depict in Figure 3 the annual number of papers (by their publication years). Note that, as discussed in Section 4.1, since we searched for the papers during winter and spring 2017, the number of papers for 2017 is partial and thus low (only 2 papers). Also, Figure 3 shows that the annual number of papers in this area had reached its highest in year 2009 (17 papers) and since that time, the trend has declined to less than 10 papers in a year. This could denote that interest and activity in this area have dropped slightly in latest years, and the reason for this needs further investigations.

To put things in perspective, we also compare the trend with data from five other SLM/SLR studies: (1) a SLM on web application testing [54], (2) a SLM on testing embedded software [12], (3) a SLM on Graphical User Interface (GUI) testing [55], (4) a survey on search-based testing (SBST) [56], and (5) a survey on mutation testing [57]. Note that the data for the other areas are not until year 2017, since the execution and publication timelines of those survey papers are in earlier years, e.g., the survey on mutation testing [57] was published in 2011 and thus only has the data until 2009. But still, the figure provides a comparative view of the growth of these six sub-areas of software testing.

As one can see in Figure 3, research on software testability has not been as active as in the other three areas above, notably search-based testing (SBST) and mutation testing. There may be various reasons for this observation, for instance, this subject of software testing has received less attention from the research community compared to SBST and mutation testing because it is more abstract and broad and less focused on test generation and its evaluation. However, growth of the software testability field is comparable (similar) to that of the GUI and web testing fields, as their corresponding curves have quite similar trends as shown in Figure 3. It is not easy to predict the future level of attention by researchers on testability. However, as testability is a critical issue in practice and also an active topic under discussion in the software industry, (see for instance [58, 59]). Therefore, the authors of this study hope that researchers take up and address especially industrial challenges on testability. It is a key objective of this mapping study to support industry relevant research on software testability.

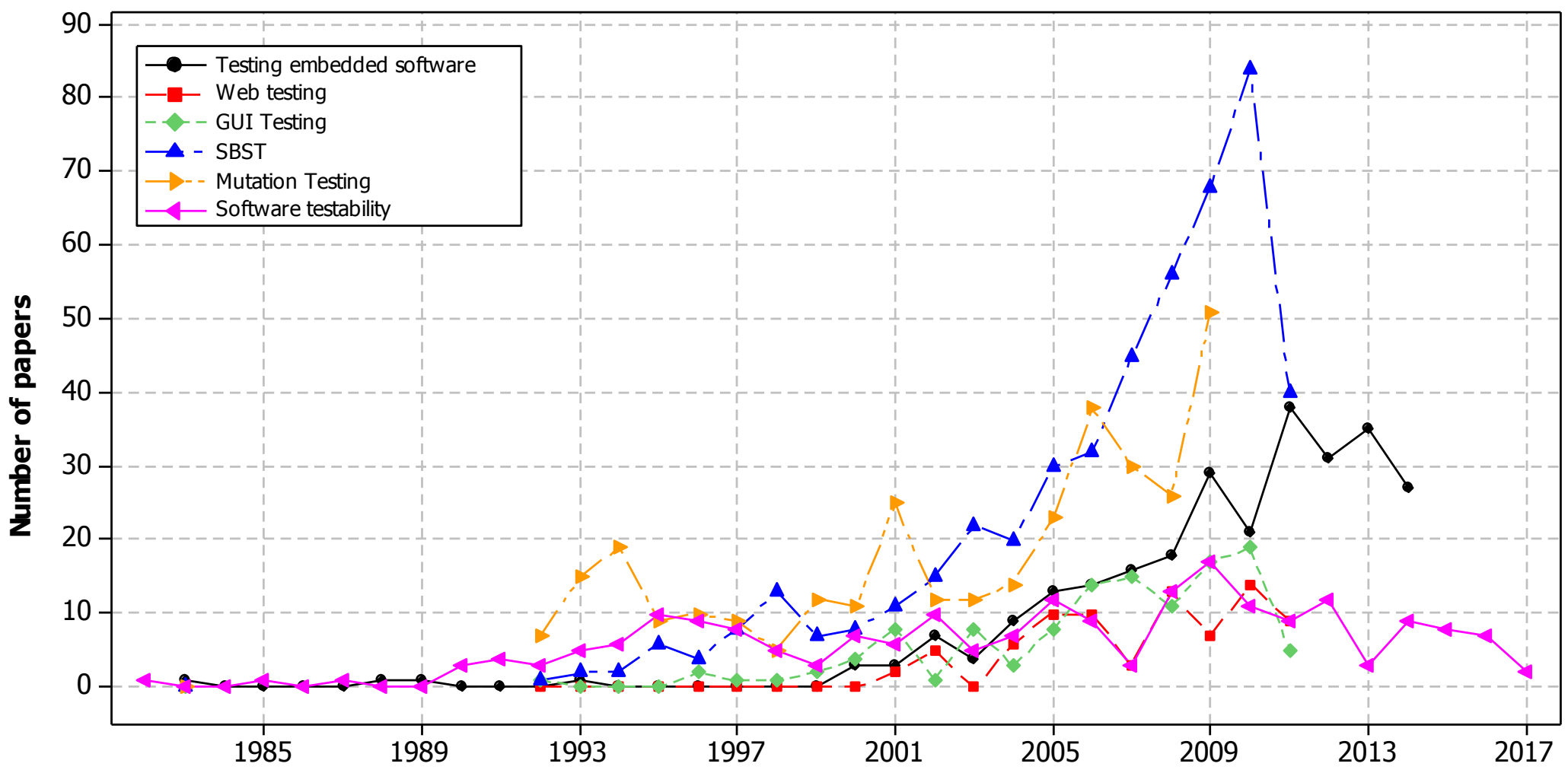

**Figure 3-Growth of the "software testability" field and comparison that growth from five other SLM/SLR studies**

## 5 Development of the systematic map and data-extraction plan

To answer each of the SLM's review questions, we developed a systematic map and then extracted data from papers to classify them using the map. In this section we first discuss the development of the systematic map (Section 5.1), then the conceptual model behind the classification (Section 5.2) and finally the data extraction process and data synthesis (Section 5.3).

### 5.1 Development of the systematic map

To develop our systematic map, we analyzed the studies in the pool and identified the initial list of attributes. As shown in Figure 1, we then used attribute generalization and iterative refinement, when necessary, to derive the final map.

As studies were identified as relevant to our study, we recorded them in our online spreadsheet to facilitate further analysis. Our next goal was to categorize the studies in order to begin building a complete picture of the research area and to answer the study RQs. We refined these broad interests into a systematic map using an iterative approach.

Table 3 shows the final classification scheme that we developed after applying the process described above. In the table, column 2 is the list of RQs, column 3 is the corresponding attribute/aspect. Column 4 is the set of all possible values for the attribute. Column 5 indicates for an attribute whether multiple selections can be applied. For example, for RQ 1.1 (contribution type), the corresponding value in the last column is 'M' (Multiple), indicating that one study can contribute more than one contribution type (e.g. method, tool, etc.). In contrast, for RQ 1.2 (research-method type), the corresponding value in the last column is 'S' (Single), indicating that one source can be classified under only one research-method type.

Contribution type and research type classifications in Table 3 were done similar to our past SLM and SLR studies, e.g., [34-38], and also using the well-known guidelines for conducting SLR and SLM studies, e.g., [40-43]. Among the research types, the least rigorous type is "Solution proposal" in which a given study only presents a simple example only (or proof of concept). Empirical evaluations are grouped under two categories: weak empirical studies (validation research) and strong empirical studies (evaluation research). The former is when the study does not pose hypothesis or research questions and does not conduct statistical tests (e.g., using t-test). We considered an empirical evaluation "strong" when it has considered these aspects. Explanations (definitions) of experience studies, philosophical studies, and opinion studies are provided in Peterson et al.'s guideline paper [42].

As discussed above, to derive the categories for all attributes/aspects in the systematic map (Table 3), we use attribute generalization and iterative refinement, e.g., for factors affecting testability (see RQ 2.2), we added the categories as we were finding them in the papers. For any category that appeared in at least five papers, we created a category in the corresponding set, otherwise, we added them in the "Other" categories.

**Table 3: Systematic map developed and used in our study**

| Group | RQ | Attribute/Aspect | Categories/metrics | (M)ultiple/ (S)ingle |
|---|---|---|---|---|
| Group 1-Common to all SLM studies | 1.1 | Contribution type | {Approach (method, technique), framework, tool, metric, model, process, guidelines to improve testability, empirical results only, other} | M |
| | 1.2 | Research-method type | {Solution proposal (simple examples only), weak empirical study (validation research), strong empirical study (evaluation research), experience studies, philosophical studies, opinion studies, other} | S |
| Group 2-Specific to the domain (testability) | 2.1 | Type of approaches for treating testability | {Requirement testability, design for testability, measurement / estimation, improvement, other} | M |
| | 2.2 | Factors affecting testability | {Observability, controllability, reliability, maintainability, efficiency, availability, flexibility, reusability, understandability, cohesion / coupling, inheritance , usability, other} | M |
| | 2.3 | Techniques for improving testability | {Adding assertions (for increasing chances of revealing defects), Improving observability, Improving controllability, Testability transformation, Refactoring, Architecture and test interfaces supporting testability, Dependency analysis (coupling, etc.), Other} | |
| Group 3-Specific to empirical studies and also the case-study of each paper | 3.1 | Research questions raised and studied in the empirical studies | Numbers of RQs and their texts | M |
| | 3.2 | Number of SUTs (examples) | Integer value, as indicated in the paper | S |
| | 3.3 | Domain of SUT (or example) | {Generic, real-time systems, embedded systems, communication systems and protocols, other} | M |
| Group 4- Demographic information | 4.1 | Affiliation types of the study authors | {A: Academic, I: Industry, C: collaboration} | S |
| | 4.2 | Active companies | Name(s) of the company (ies) involved in the paper | M |

### 5.2 Data extraction process and data synthesis

Once the systematic map (classification scheme) was ready, each of the researchers extracted and analyzed data from the subset of the sources (assigned to her/him). We included traceability links on the extracted data to the exact phrases in the sources to make explicit why the classification was performed in a specific way.

Figure 4 shows a snapshot of our online spreadsheet that we used to enable collaborative work and classification of sources with traceability links (as comments). In this snapshot, classification of sources w.r.t. RQ 1.1 (Contribution type) is shown and one researcher has placed the exact phrase from the source as the traceability link to facilitate peer reviewing and also quality assurance of data extractions.

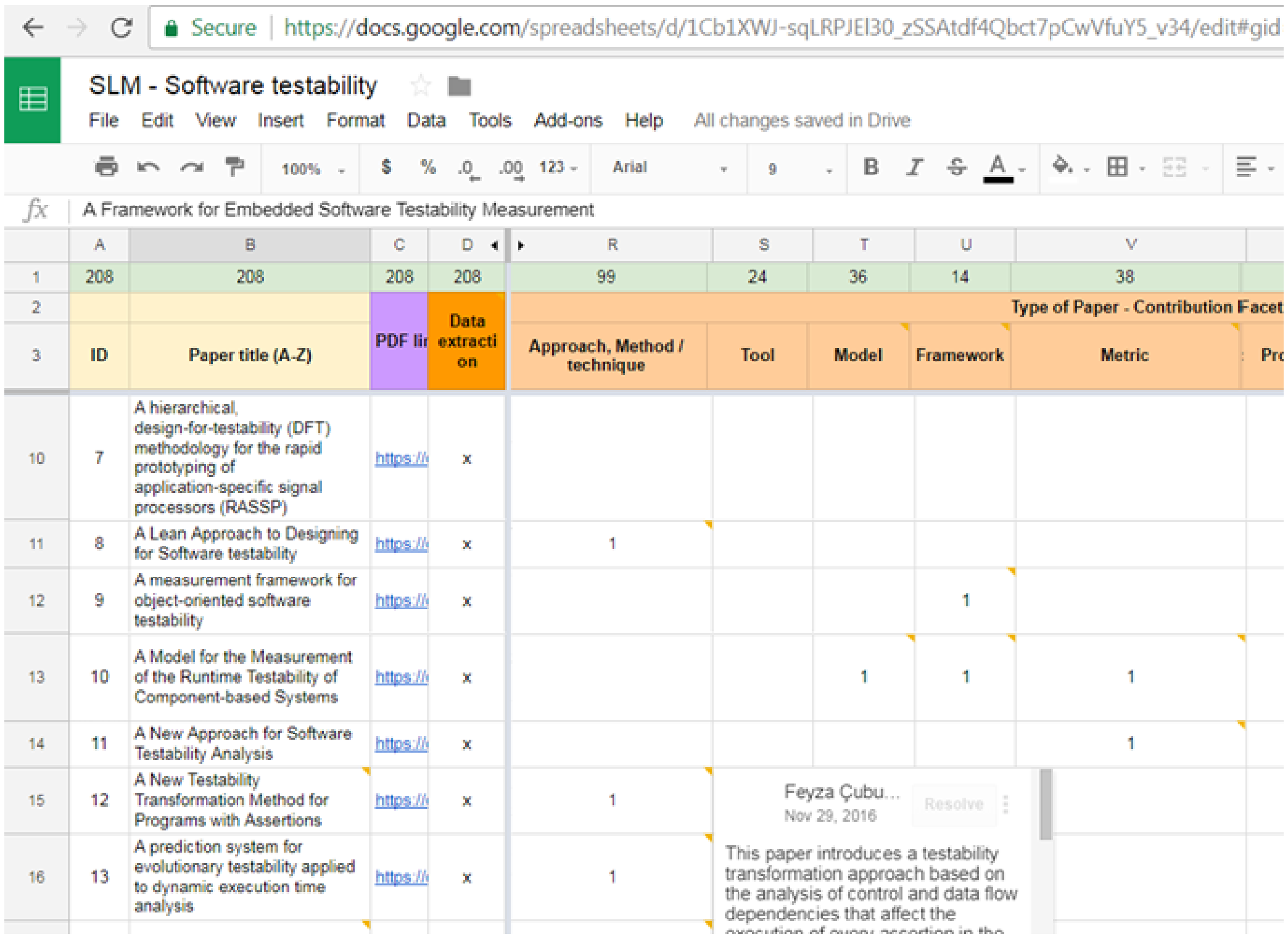

**Figure 4- A screenshot from the online repository of papers (goo.gl/MhtbLD).**

After all researchers finished data extractions, we conducted systematic peer reviewing in which researchers peer reviewed the results of each other's analyses and extractions. In the case of disagreements, discussions were conducted. This was done to ensure quality and validity of our results. Figure 5 shows a snapshot of how the systematic peer reviewing was done.

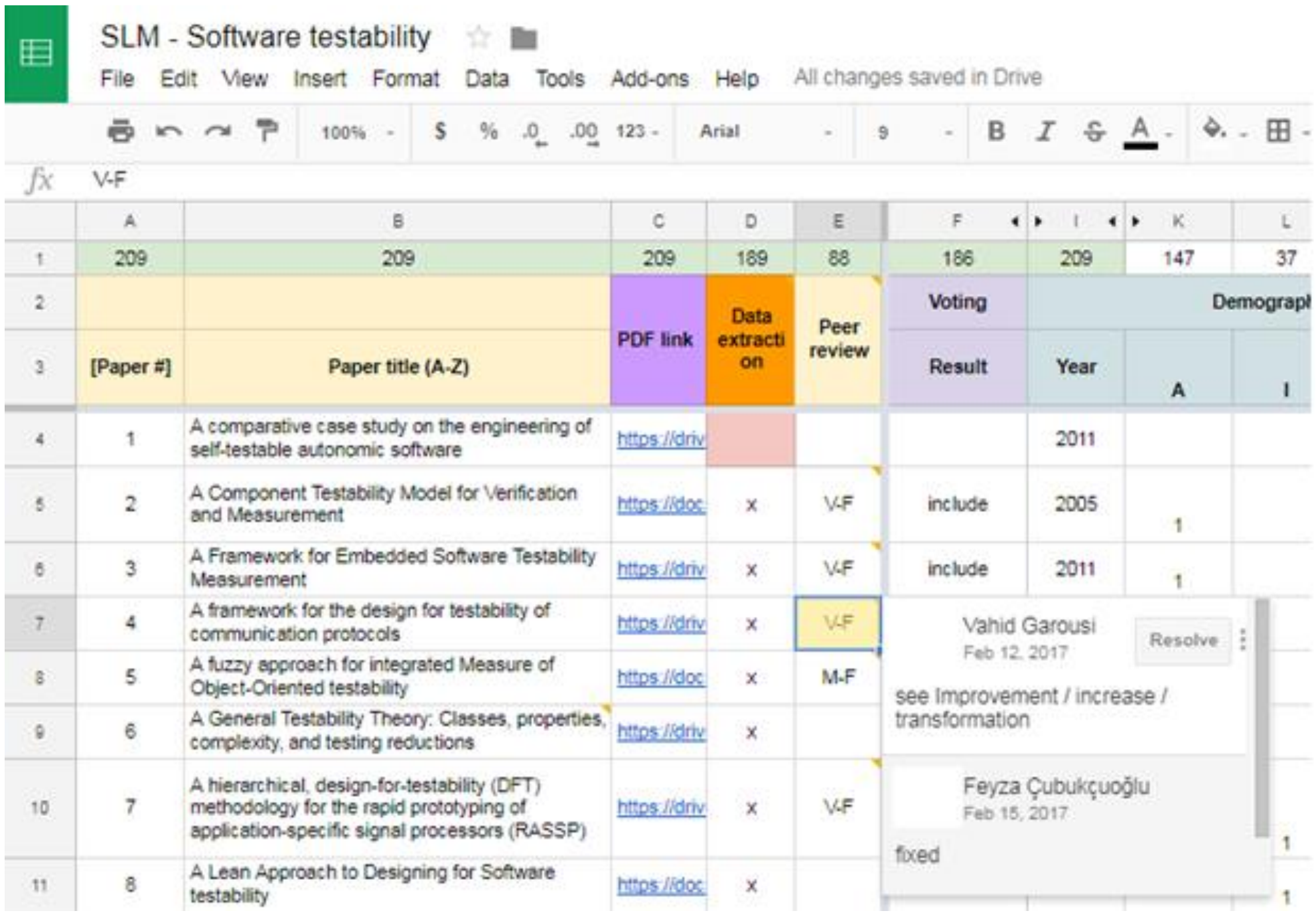

**Figure 5- A snapshot showing how the systematic peer reviewing was orchestrated and conducted**

## 6 RESULTS

This section presents results of the study's review questions (RQs). We present each group of RQs in a separate subsection, i.e., Group 1 (types of contributions and research methods) in Section 6.1, Group 2 (approaches for treating testability, factors affecting testability, techniques for improving testability) in Section 6.2, Group 3 (research questions, number and domain of SUTs) in Section 6.3, and finally Group 4 (demographic information) in Section 6.4.

### 6.1 GROUP 1- COMMON TO ALL SLM STUDIES

In this section we present the results of RQ 1.1 (classification of studies by contribution types) as well as of RQ 1.2 (classification of studies by types of research methods).

#### 6.1.1 RQ 1.1: Classification of studies by contribution types

Figure 6 shows the classification of studies by contribution types (facets). Note that as we discussed in the structure of the systematic map (Table 3), since each study could have multiple contribution types, it could thus be classified under more than one category in Figure 6. We discuss below a summary of each category by referring to few example papers in that category.

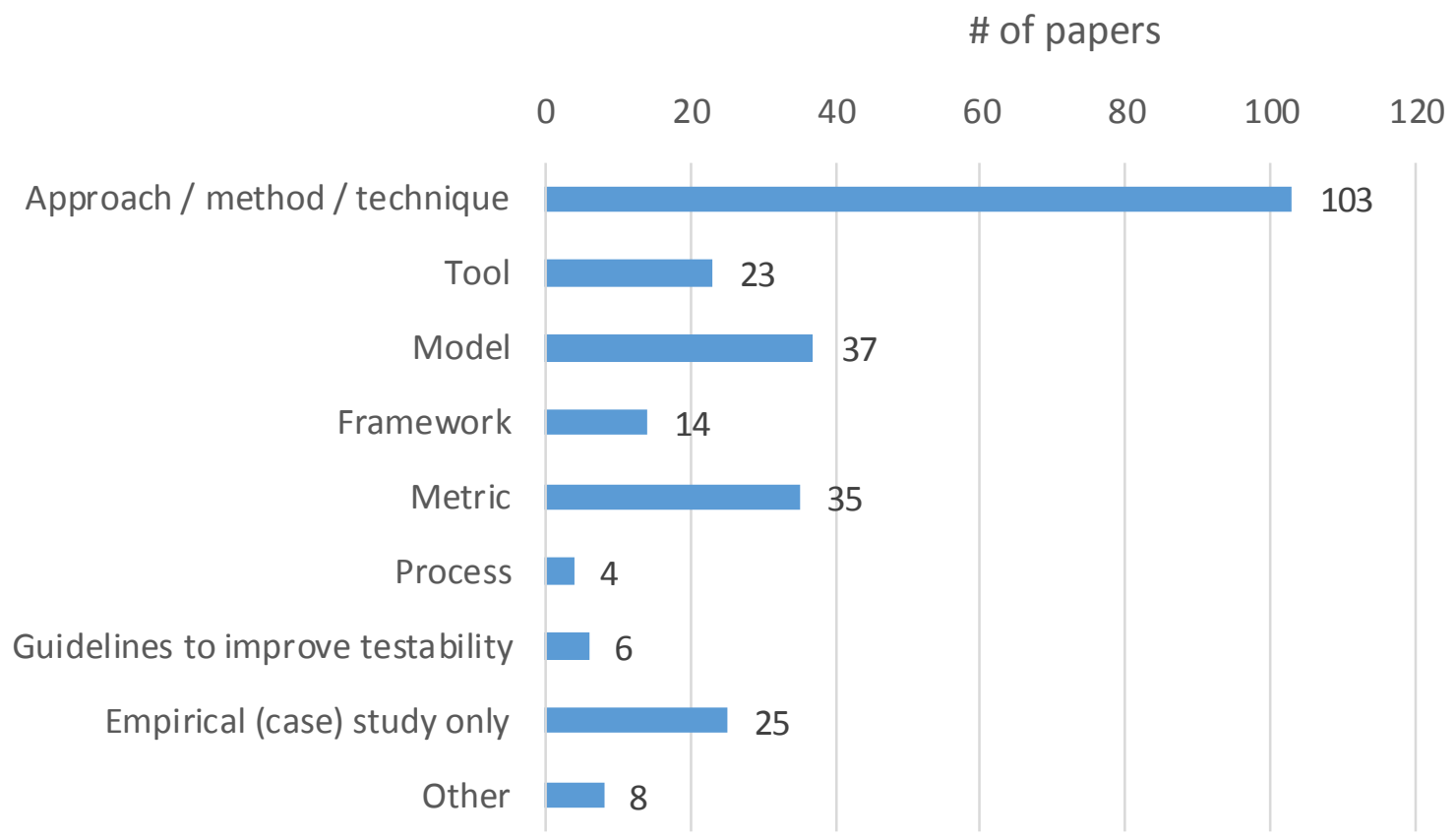

**Figure 6- Classification of studies by contribution types**

Since approaches, methods, and techniques are similar concepts, we grouped them together. 103 papers (about 48% of the pool) contributed approaches/methods/techniques to deal with testability. As one can see in Figure 6, this group is the category highest number of papers from the pool classified by the type of contribution.

23 papers (10% of the pool) presented tools to deal with testability. It is imperative that often, in the software engineering community, authors of some papers decide to "automate" the presented approach/method/technique by developing and presenting a tool. For example, in [P50], the authors implemented the described Java bytecode transformation approach as part of an evolutionary test generation tool named EvoSuite. [P57] presented a testability framework for OO software named COTT(Controllability and Observability Testing Tool). COTT helps the user to instrument OO software to gather the controllability and observability information. During testing, the tool facilitates creation of "difficult-to-reach" states required for testing of difficult- to-test conditions and observation of internal details of test execution. For improving testability with assertion insertion, [P83] presented a tool named C-Patrol. For measuring design testability of UML class diagrams, [P96] presented a tool named JTracor. We should note that, for a typical reader (researcher or practitioner), it is always helpful if the presented tools are available for download, but we did not check that aspect for each tool.

37 papers (18% of the pool) presented models to deal with testability. For example, [P5] presented a fuzzy model for an integrated software testability measure that takes into account the effect of various OO metrics, such as Depth of Inheritance Tree (DIT), Coupling Between Objects (CBO), and Response For a Class (RFC). [P10] presented a qualitative model for the measurement of the runtime testability of component-based systems. For software testability analysis, [P53] proposed a cost model to conduct cost-effectiveness analysis and to find out the percentage of cost reduction when improving testability. This model can help better understand the trade-offs between time, performance and cost when improving testability. [P97] proposed a prediction model for measuring evolutionary testability of real-time software. [P101] presented a metric-based testability model for OO design named MTMOOD. The model is based on the relationship of design properties with testability and their weightings in accordance with their anticipated influence and importance on testability.

14 papers (7% of the pool) presented (conceptual) frameworks. Note that this category is different than the “approaches” or “tools” category since as frameworks are usually in “conceptual” level and not in the level of tools, and also as per the terminology used by the authors of the studies. Here are some example papers in this category. [P9] proposed a generic and extensible measurement framework for OO software testability. The authors identified the design attributes that have an impact on testability directly or indirectly, by having an impact on testing activities and sub-activities. They also described the cause-effect relationships between these attributes and software testability based on a thorough review of the literature and their own testing experience. [P27] proposed a testability framework for OO software, which enables runtime constraint checking against implementation code and, in turn, improves the testability of a software system.

35 papers (17% of the pool) presented metrics to deal with testability. For example, [P5] proposed a fuzzy metric for OO software called Testability Index (TI) which is based on these metrics: Depth of Inheritance Tree (DIT), Coupling Between Objects (CBO), and Response For a Class (RFC). [P36] introduces five new factors (metrics) which can affect testability which take into account developer and development process: Years of coding experience, Time for development, Previous development of similar projects, Confidence in programming language, Effort needed for a task. [P52] proposed two new testability measures called Testability Vector (TV) and unified testability measure (UTM) for testability assessment of finite state machine (FSMs) in the context of communications protocols.

4 papers (2% of the pool) presented processes to deal with testability. For example, [P7] described a highly-automated, hierarchical design-for-testability process that spans the entire life cycle. The process was developed by Lockheed-Martin and another company named Self-Test Services for the US Department of Defense (DoD). [P7] mentioned that the process contributed to the DoD‘s goals of improvement in cycle time, design quality and life-cycle costs. [P166, P167] proposed processes for testability analysis of data-flow software and reactive software, respectively.

6 papers (3% of the pool) presented guidelines to improve testability. Papers in this category did not present other contribution types, but only present general heuristics. [P62, P63, P64] are among those papers. Entitled "Practicing Testability in the Real World", a paper authored by a Microsoft employee, [P120] presents real-life examples that the author has encountered in his career which exhibit how testability considerations could have made the testing job simpler. [P120] then presents a heuristics-based checklist consisting of the following groups: (1) Simplicity: The simpler a component, the less expensive it is to test, (2) Observability: Exposing state (visibility and transparency), (3) Control: Can you exercise every nook and cranny of the component?, and (4) Knowledge of expected results: Is the observed behavior correct?

The primary contribution of 25 papers (11% of the pool) were empirical (case) studies in this area. Note that if a paper had presented other contribution types (e.g., approach) and had an accompanying case study, we did not include it in this category, but only included those paper which presented empirical studies and results “only”. Most of these papers have the explicit term ”empirical study“ in their titles. The following five papers are a few examples in this category:

- A comparative case study on the engineering of self-testable autonomic software [P1]
- An empirical comparison of a dynamic software testability metric to static cyclomatic complexity [P31]
- An empirical study into class testability [P32]
- An empirical study on the effects of code visibility on program testability [P33]
- An empirical study on the usage of testability information to fault localization in software [P34]

Observing that more than 10% of the pool are focused on empirical studies in this area was seen by us as a good sign which shows the special attention of researchers in this area to empirical investigations.

8 papers (5% of the pool) presented 'Other' types of contributions. For example, [P107] proposed an embedded software architecture for improving testability of in-vehicle multimedia software. [P113] discussed factors that affect the testability of communication software. [P153] proposed a specification language (named Spill) for writing testable requirements specifications

#### 6.1.2 RQ 1.2: Classification of studies by research method types

In SLM studies, e.g., [34-38], it is also common to classify primary studies by their types of research methods. As the structure of the systematic map (Table 3) showed, based on established review guidelines [40-43], that classification includes these categories: (1) Solution proposal (simple examples only), (2) weak empirical study (validation research), (3) strong empirical study (evaluation research), (4) experience studies, (5) philosophical studies, (6) opinion studies, and (7) other research methods.

Figure 7 shows the cumulative trend of mapping of studies by research facet. As one can see, almost equally-divided portions of papers (63, 53 and 60 out of 208 papers) present solution proposals (by examples), weak empirical studies, and strong empirical studies. 20, 0 and 12 papers, in order, are experience papers, philosophical and opinion papers, respectively.

As a point of reference and for comparison, we show in Figure 8 the same chart from our other recent SLM study on testing embedded software [12]. By comparing the two charts (Figure 7 and Figure 8), we can see that the area of software testability, as a whole, puts more importance on conducting empirical studies than the area of testing embedded software. In the latter, the share of papers presenting solution proposals is higher.

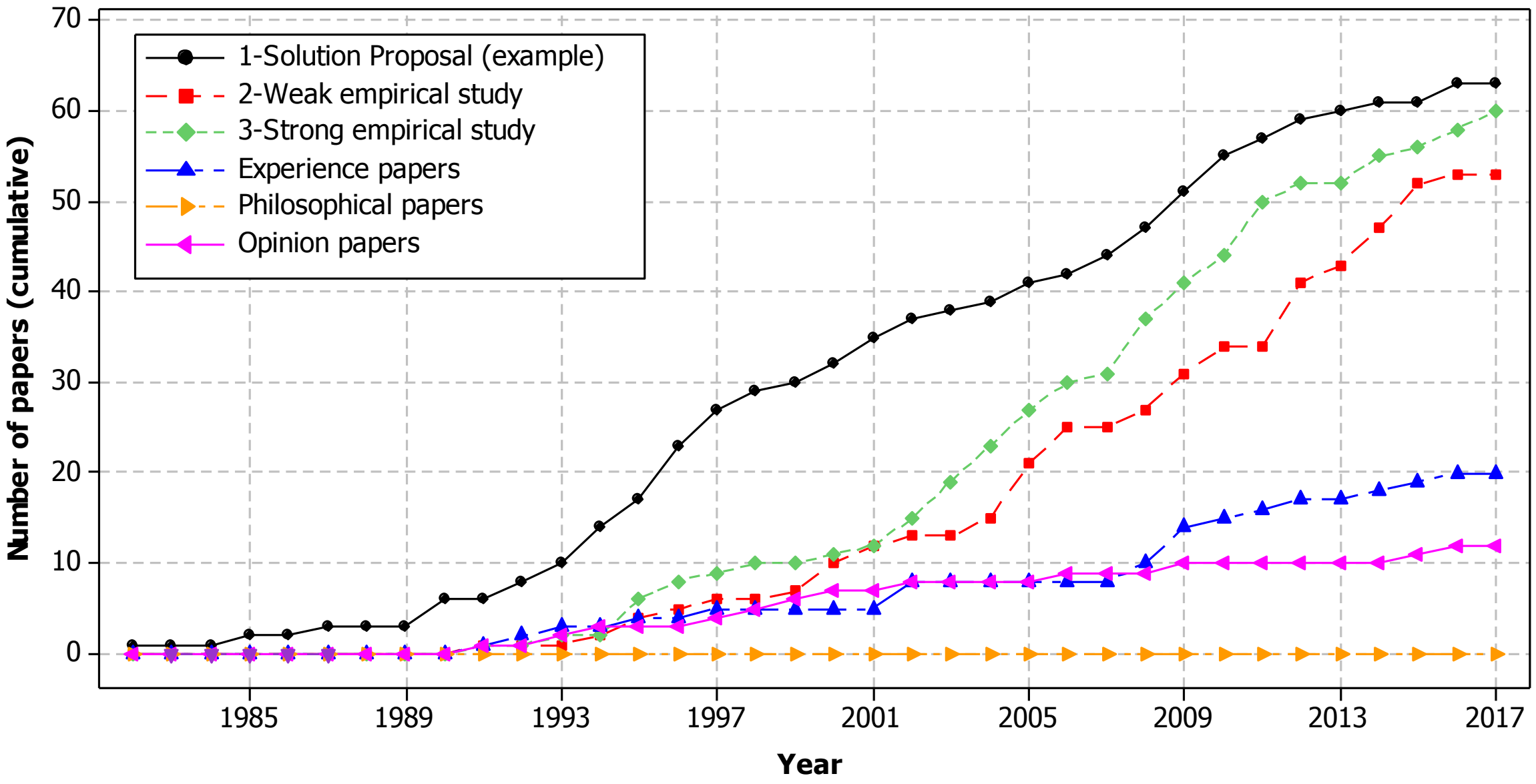

**Figure 7-Cumulative trend of mapping of studies by research-method types (total=208 papers)**

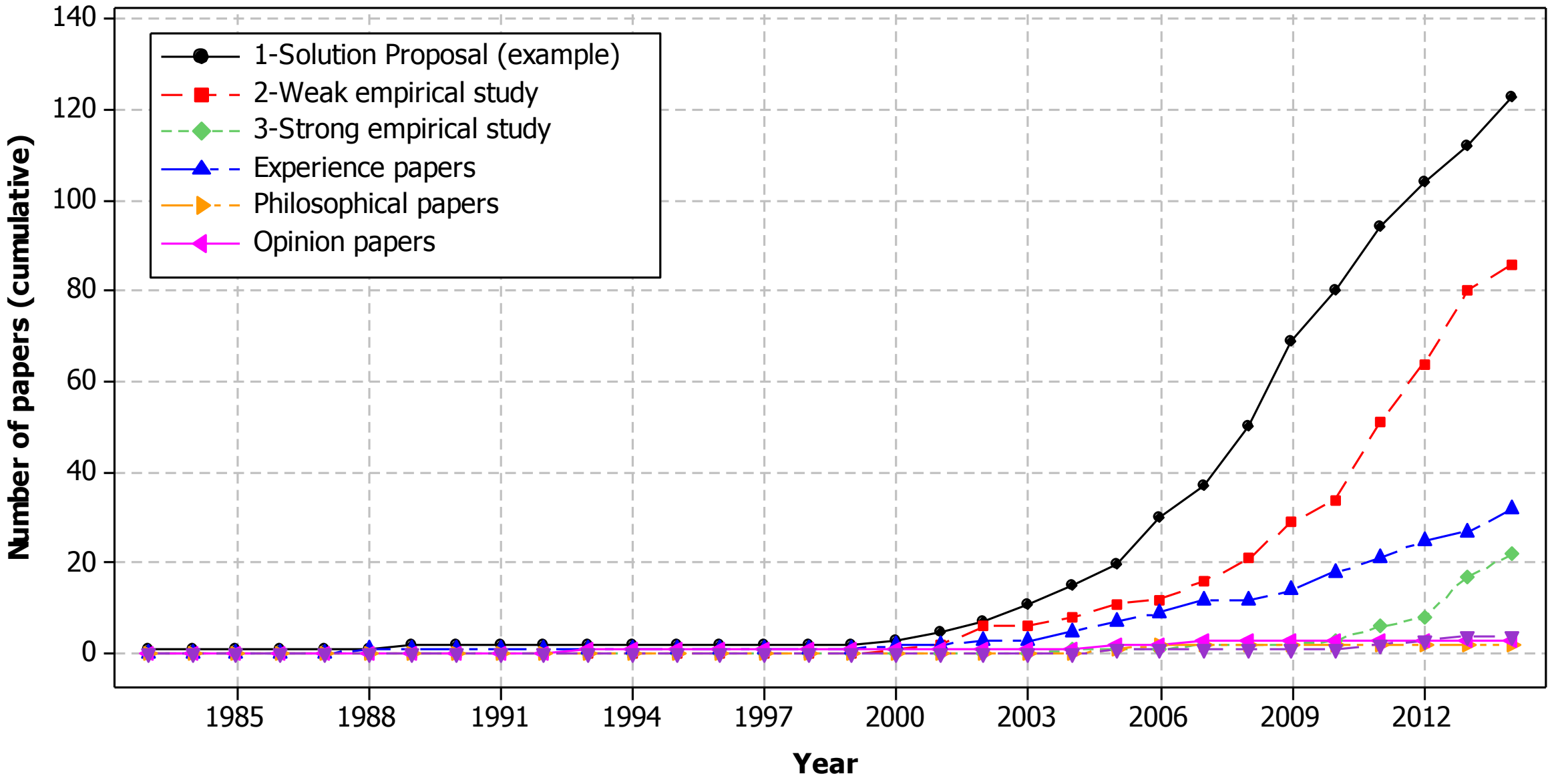

**Figure 8-Cumulative trend of mapping of studies in other SM study (testing embedded software [12]) by research-method types (total=272 papers)**

Since 'strong' empirical studies are the most rigorous studies in this context, we provide a few examples of those sources. [P20] presented a study of the relationship between class testability and runtime properties. The study raised and addressed two research questions:

- RQ1: Is dynamic coupling of a class significantly correlated with the class testability of its corresponding test class/unit?
- RQ2: Are key classes significantly correlated with the class testability of their corresponding test classes/units?"

[P72] conduced an empirical analysis for investigating the effect of control flow dependencies on testability of classes. [P73] reported an empirical evaluation of a nesting testability transformation for evolutionary testing.

12 papers were "opinion" papers, for example [P81] presented some heuristics of software testability based on author's opinion (a practitioner). The opinion papers had no empirical assessment. 20 paper were pure "experience" papers in which practical experience about testability were reported, for example [P120] talked about practicing testability in the "real world" in Microsoft Corporation. It discussed about the typical thought process in a test engineer's mind and present key insights into why practicing testability is hard.

## 6.2 GROUP 2-SPECIFIC TO THE DOMAIN (TESTABILITY)

In this section we present the results of RQ 2.1 (types of approaches for treating testability), RQ 2.2 (factors affecting testability) as well as of RQ 2.3 (techniques for improving testability).

### 6.2.1 RQ 2.1- Types of approaches for treating testability

To characterize how testability is addressed throughout different stages of the software development life cycle (SDLC), we synthesized and derived from the pool of papers the process model shown in Figure 9. We can see that testability is an activity which is not only done in the testing phase, but should be considered and conducted in all phases of the SDLC, from requirements engineering to design and to maintenance.

Based on this process model, we categorized how each paper has treated testability. We discuss next each of the categories below and mention a few example papers under each category.

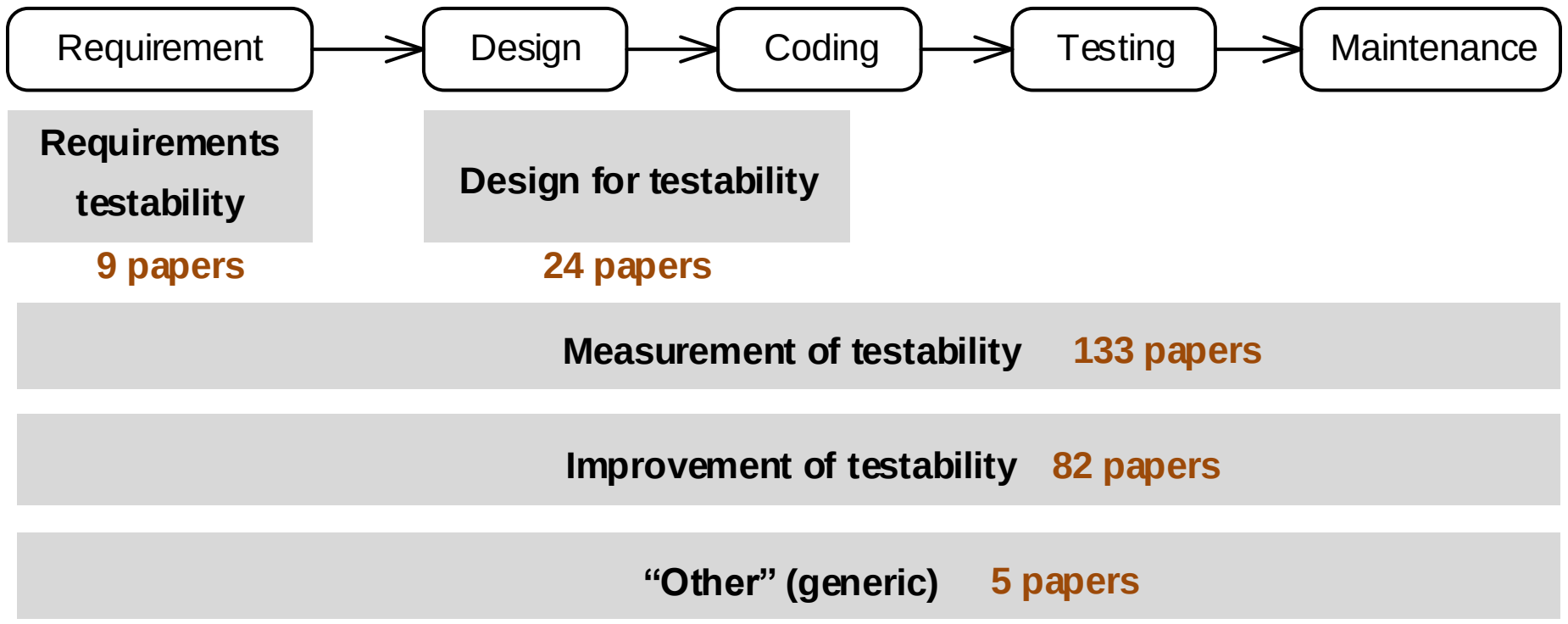

**Figure 9-Addressing testability in various stages of SDLC**

<u>Requirements testability (9 of the 208 papers):</u>

Requirements testability is the degree to which the requirements of a system can easily be tested. A testable requirement is the one that is specified in such a way that test suites and oracles can be easily derived from it. This issue should be considered as early as in the requirements phase, since going back to requirements and changing (improving) them to make them testable is often costly.

Surprisingly, very few studies have focused on the issue of requirements testability, although it is also considered in the IEEE recommended practice for software requirements specifications 830-1998 [60]. For example, [P96] presented an approach for measuring requirement quality to predict testability. The authors selected a set of document readability metrics that characterize the understandability and quality of requirements and assessed whether they characterize testable requirements. The seven selected metrics are well-known in the readability literature [61] and were: (1) Average Grade Level, (2) Flesch Kincaid Reading Ease, (3) Flesch Kincaid Grade Level, (4) Gunning Fog Score, (5) Simple Measure of Gobbledygook (SMOG Index), (6) Coleman Liau Index, and (7) Automated Readability Index. They then developed a model for requirement testability using machine learning and statistical analysis, and assessed whether that model of requirement testability can be learned and applied to other requirements.

[P135] focused on requirement testability in the context of safety-critical automotive software. The authors defined a requirement Ri as a logical expression Li: <Object X> shall <Action Y> [applied to] <Subject Z>. The requirement is mapped onto Object X which performs Action Y onto Subject Z. Testability of Ri is a property of Ri that this logical expression Li can be verified. They suggested that, to fulfill testability, the requirement has to consist of the object, the action and the subject and the object, the action and the subject must be identifiable within and present in the system. With these conditions valid, the requirement can be verified, and is testable.

[P143] entitled "*Sizing software with testable requirements*" is another interesting study, which is written by a practitioner. The work proposed testable requirements as a new software measurement paradigm. By applying it in industrial context, the paper stated that testable requirements is an intuitive, flexible measure that is a useful tool for communicating issues to users and management.

[P158] considered requirements testability in the context of offshore outsourcing (global software engineering). The authors argued that offshore outsourcing requires a set of testable requirements at the heart of the quality assurance of a legal agreement between the contractor and the client. They provided semantics for a model for testable requirements in that context.

[P159] examined testability use-cases by specifying them using the Abstract State Machine Language (ASML), which is an executable specification language developed at Microsoft Research. The authors then demonstrated the advantages of the approach by describing how to generate test cases and test oracles from use-cases specifications in ASML.

[P205] provides a case study of requirements testability for the validation of Airbus systems. The experiments showed that testability analysis can ease system validation activities. Design for testability (24 papers):

It is important to design a system in a way to facilitate its testing. Design for testability (DfT) is usually conducted in the design and coding phases of the SDLC, as shown in Figure 9. A bad design (e.g., falling in the trap of design anti-patterns [P88]) can negatively impact testability of a system or its units in the testing phase. Testability anti-pattern was also the focus of another work [P130] which defined testability anti-pattern as "*a design approach known to make test difficult and/or to increase the number of test cases to be executed*".

The authors of [P21] used system-dynamics modelling for studying the relationships between modularity and testability. The authors focused on the costs and benefits of modularizations w.r.t. testability. Using simulation data from system-dynamics, the authors found that adopting appropriate modularity and proper testing architecture can enhance efficiency and effectiveness of system testing.

[P22] reports a study on DfT in the context of component-based embedded software in two large-scale companies in the European telecom industry. Based on the interviews and technical documentation, differences and benefits of different DfT approaches (focusing on observability and controllability) were discussed. The paper presented several recommendations for DfT, e.g., (1) Especially in the case of embedded systems, a good host test environment enables high testability. When this environment matches the target system as much as possible, efficient host testing is possible; (2) Including test-support functionality in the system allows for more effective testing, including analysis of long running tests and deployed systems.

The paper [P62], written by a practitioner, provided practical suggestions on DfT. Based on the author's experience, it provided examples of testability features used in testing various software products, including event logging, assertions, diagnostics, resource monitoring, test points, fault injection hooks and features to support software installation and configuration. It also explored how testability issues could affect GUI test automation and what kinds of provisions testers must make when testability is lacking. The paper concluded with a discussion of how testers can work with product developers to get important testability features built into software products. [P64] is an influential 1994 paper entitled "*Design for testability in object-oriented systems*" (cited 378 times as of this writing) by Robert Binder, an expert in the field. The author presented important concepts and different architectures for DfT. It concluded with saying that: "*Nearly all the techniques and technology for achieving high testability are well established, but require financial commitment, planning, and conscious effort*".

[P108] focused on building testable software components, and introduced the concept of testable (Java) "beans", and proposed a way to construct a testable bean based on a testable architecture and well-defined built-in interfaces.

Measurement (estimation) of testability (133 papers):

A large ratio of papers (64%) proposed approaches or metrics for quantitative or qualitative measurement, estimation, and prediction of testability. As shown in Figure 9, in principle, such measurement can be done in any phase of SDLC. We discuss some example papers next

A runtime testability metric was proposed in [P10]. An empirical analysis of the lack-of-cohesion-metric (LCOM) for predicting testability of OO classes was reported in [P29]. A set of factors and metrics related to software developers that may affect testability were proposed in [P36], e.g., years of coding experience and previous experience in development of similar projects.

[P20] reported a case study of the relationship between class testability and runtime properties such as dynamic coupling metrics (e.g., import coupling and export coupling). The case study was conducted on four open-source projects. Testability was measured by the size of the test LOC.

The authors of [P29] conducted a case study using a metric-based testability model for object-oriented programs, presented in [P101]. The study explored empirically the capability of the model to assess testability of classes at the code level. The authors investigated testability from the perspective of unit testing and required testing effort. The empirical study was based on collected from two Java software systems for which JUnit test cases exist. In order to evaluate the capability of the model to predict testability of classes (characteristics of corresponding test classes), the authors used statistical tests using correlation.

[P101] presented a metric-based testability model for object-oriented design named MTMOOD. Based on empirical data and regression analysis, the study reported that the following quantitative formula: Testability = -0.08 * Encapsulation + 1.12 * Inheritance + 0.97 * Coupling.

[P72] reported an empirical analysis for investigating the effect of control-flow dependencies on testability of classes, in unit level testing. The results provided the evidence that there exist a significant relationship between control-flow dependencies and testability of classes.

[P85] presented an approach for improving the testability of object-oriented software through software contracts. Software contracts were instrumented in a class and test cases were designed for this class using the path testing technique and then it is compared with the class without instrumenting the software contracts. The study found that the instrumentation of software contracts reduces the number of test cases and hence improves the testability.

[P100] presented a set of metrics for testability assessment of web application. The set included these metrics: (1) IDP (number of elements with ID attribute), (2) TWI (number of workflow interruptions): number of situations, when an operation must be carried out outside the SUT by human user, (3) WIR (Workflow interruption ratio), and (4) TDI (number of difficult elements): number of various types of elements in SUT difficult to locate and handle by automated tests.

[P106] presented a tool named MuAspectJ for generating mutants to support measuring the testability of AspectJ programs. Testability was also measured by the tool. [P110] presented a method called Testable Object-Oriented Programming (TOOP) for building testability into objects during coding or compiling, so that the succeeding processes in test generation and implementation can be simplified. [P123] reported a case study on measuring testability of the Eclipse project.

The authors of [P129] argued that, for models used in model-based testing, evaluation of their testability is an important issue. The paper presented a quality management approach for the evaluation of software models based on a combination of the Goal Question Metric (GQM) and quality models. The approach also used the concept of "information need" in models. The quality model broken model testability into three aspects: syntactic quality (correspondence between the software model and its language definition), semantic quality (correspondence between the software model and the domain), and understandability (the capability of the software model to be understood). <u>Improvement of testability (82 papers):</u>

Many papers offered suggestions to increase testability. Some of the papers in this category improved testability by refactoring or testability "transformation". Such a transformation changes the structure of a unit under test (e.g., a Java method or a finite-state-machine) usually by increasing its observability and controllability, e.g., [P24, P43].

[P45] proposed a debugging framework to control the values of the inputs for a function under test, and as a result, it improved test controllability and, by opening the observation points in the debugging framework, tracking the changes of the variables under evaluation was made possible, which helped to improve the test observability. Another share of papers in this category suggested ideas to improve testability by adding assertions. For example, [P18] inserted assertions in the SUT to build self-checking capabilities inside a component, which would help to detect runtime faults easier (if they happen to manifest). As shown in Figure 9, in principle, similar to measurement of testability, testability improvement can also be done in any phase of SDLC, when the need is "felt" to improve testability.

Our RQ 2.3 will focus in depth on techniques for improving testability (Section 6.2.3). Thus, we do not discuss further works in this section.

<u>Need for attention to testability in all phases of the SDLC:</u>

Furthermore, we can imply from Figure 9 that one should put attention to testability in all phases of the SDLC. But as it is the case for other software quality attributes, the earlier, the better, i.e., it will be wiser if a development team considers (and improves) testability from as early as requirements and design phases, since costs of improving testability later on will usually be higher, similar to the notion of technical debt in software engineering [62]. In fact, there is tool support for this purpose. For instance, some recent tools such as *Sonar* have proposed plug-ins [63] to quantitatively show the testability measure of a given code-base in a technical debt model called "The SQALE Pyramid" [64].

### 6.2.2 RQ 2.2- Factors affecting testability

As the next review question, we synthesized and classified the factors affecting or influencing testability, as reported in the papers. These factors are also used to measure the degree of testability. Figure 10 shows the list and histogram of those factors. As discussed in Section 5.1, we developed this classification by iterative refinement, i.e., adding new factors to this list as we found them in the pool of papers.

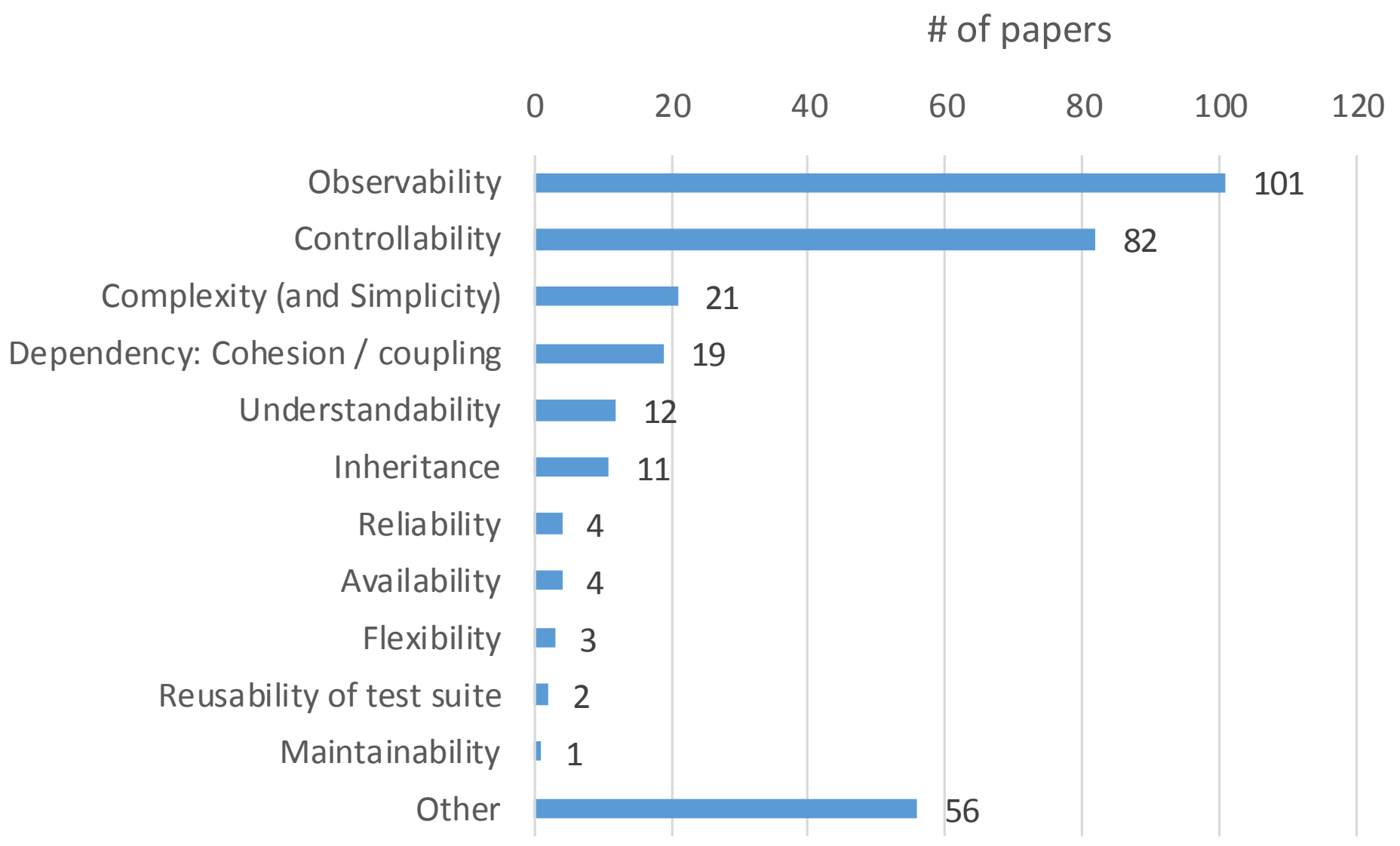

**Figure 10-Histogram of the factors influencing testability, as reported in the papers**

The two most often mentioned factors affecting testability are observability (mentioned in 101 papers) and controllability (82 papers).

*Observability:*

Observability determines how easy it is to observe the behavior of a program in terms of its outputs, effects on the environment, and other hardware and software components. It focuses on the ease of observing outputs. Binder [P64] stated that: "*if users cannot observe the output, they cannot be sure how a given input has been processed*". Observability directly influences testability, since if it is not easy to observe the behavior of a program in terms of its outputs, testing will be more challenging.

To increase observability, when conducting integration testing, [P90] suggests to collect definition-use pairs between the last definition in a method and the first use in other methods. [P110] suggests that full observability of an object under test can be implemented by inserting a probe instrument such as write or print in the code.

[P116] focuses on the testability of distributed real-time systems. In such a context, observability is important for determining whether the SUT performs correctly or not, due to two aspects: (1) the test engineer must be able to observe every significant event generated by the environment, and to determine the correct ordering and timing of events, and (2) it is necessary to observe the actions and outputs of the system during testing, but without disturbing its time behavior.

[P120] entitled "*Practicing testability in the real world*", is a paper from Microsoft. The study was motivated by a shortage of bugs in the company requesting a change in a feature because it will become simpler to test it. The author present experiences while applying testability concepts and provided guidelines to ensure testability consideration during feature planning and design. It is based on a model named SOCK (Simplicity, Observability, Control and Knowledge of expected results). The paper argues that the simpler a component, the less expensive it is to test. Observability related to exposing state (visibility and transparency). Control related to whether the tester could exercise every aspect of the component. Knowledge of expected results also impacts testing.

[P127] entitled "*Putting assertions in their place*" advocates for a middle ground between no assertions at all (a common practice) and the theoretical ideal of assertions at every location. The proposed compromise was to place assertions only at

locations where testing is unlikely to uncover software faults. Testability measurement, used in the paper, was able to identify locations where testing is unlikely to be effective. The paper defines assertions as *"a method for increasing observability in software by increasing the dimensionality of the output space"*.

[P151] entitled "*Software testability: The new verification*" is the most cited paper in the pool (392 times as of this writing). [P151] relates software observability to hardware observability in the context of IC (integrated circuit) design. It reports that IC design engineers often used the observability notion, which is closely related to software observability. In the hardware domain, observability is the ability to view the value of a particular node embedded in a circuit. In such a context, the principal obstacle in testing large-scale ICs is the inaccessibility of the internal signals. One method of increasing observability in that context is to increase the chip's pin count, letting the extra pins carry out additional internal signals that can be checked during testing. For software, when modules contain local variables, one loses the ability to see information in the local variables during functional testing, which could become a major issue for object-oriented systems. To remedy this, testers can apply a notion similar to increasing the pin count in a chip, by increasing the amount of data-state information that is checked during unit testing.

*Controllability:*

There are several slightly-different definitions for controllability. It is the degree to which it is possible to control the state of the component under test as required for testing. Another definition is that, it determines how easy it is to provide a program with the needed inputs to exercise a certain condition or path, in terms of values, operations, and behaviors. If a system or component under test supports various ways of supplying inputs to exercise it, it tends to provide better controllability and thus better testability. Binder [P64] defined the significance of controllability as follows: "*if users cannot control the inputs, they cannot be sure what caused a given output*". Controllability directly influences testability, since if it is not easy to control the state of the component under test as required for testing and as a consequence testing various conditions, paths, operations, and behaviors would be challenging.

[P25] presents an analysis technique to increase testability of object-oriented components. A component's bytecode was analyzed to develop a control and data flow graph, which is then used to improve component testability by increasing both controllability and observability. To increase controllability, the definition and uses information were applied to collect definition-use pairs of variable between last-definitions and first-uses. The definition–use pairs of a variable were applied to increase controllability by supporting test case generation to cover all the necessary tests.

[P52] focuses on finite state machines (FSM) used in the context of communication software. By using formal methods, it defined a controllability degree metric for FSMs which was based on reachability sets.

[P56] focuses on constructing self-testable software components. One of the ideas was to use instrumentation, which is extra software introduced into the class under test to increase its controllability and observability. The instrumentation, also referred to as built-in test (BIT) capabilities, comprised of: assertions, a reporter method and a BIT access control.

[P81] discusses controllability in two different context: (1) environmental controllability, in the context of project-related testability, which is the degree to which one can control all potentially relevant experimental variables in the environment surrounding our tests; and (2) intrinsic controllability, in the context of intrinsic testability, which relates to being able to providing any possible input and invoking any possible state, combination of states, or sequence of states on demand, easily and immediately.

[P160] proposes a quantitative measure for controllability of a module M based on data-flow path F as follows: $CO(M)=T(I)/C(I)$, where $T(M)$ is the maximum information quantity that module M receives from inputs of flow F and $C(M)$ is the total information quantity that module M would receive if isolated.

[P171] links controllability to several internal OO features (encapsulation, coupling cohesion and polymorphism) as follows. Encapsulation promotes controllability. Coupling makes controllability difficult. Cohesion helps improving it. Polymorphism further reduces controllability.

*Complexity (and Simplicity):*

Software complexity is a term that encompasses numerous properties of a piece of software, all of which affect internal interactions. Simplicity is seen as the opposite of complexity, which is defined as the degree to which a software artifact has a single, well-defined responsibility. [P17] found that simplicity positively influences design for testability.

[P31] reports an empirical comparison of a dynamic software testability metric with static cyclomatic complexity and found a positive correlation between them.

In addition, [P37] also found that a highly complex class is likely to require more unit testing effort than a low complexity class. If a class is highly complex (for example, its methods have highly complex control flow structures and/or the interaction patterns between methods and attributes are highly complex), it is likely that more effort is required. The code used to examine whether a test case execution is successful may also be more complex. Therefore, more effort is likely to be required to unit test a highly complex class.

<u>*Dependency (Cohesion / coupling):*</u>

Cohesion refers to the degree to which the elements inside a module belong together. It is a measure of the strength of relationship between the methods and data of a class and some unifying purpose or concept served by that class. Coupling is the degree of interdependence between software modules; a measure of how closely connected two routines or modules are; the strength of the relationships between modules.

Both cohesion and coupling were mentioned to be influencing testability. Cohesion in general has been found to enhance testability. On the one hand, cohesiveness among methods within a class is desirable, since it usually means decreasing the number of methods, the size of the class and hence the testing effort [P15]. On the other hand, low cohesion usually means poor design or poor organization of a class and thereby increasing the complexity of a class and the likelihood of errors. Therefore, the more cohesive the methods within a class are, the higher the testability of that class. The lack of cohesion in methods results in lowering the testability of the class [P15].

Coupling in general decreases testability. [P15] reports that coupling between classes through message passing is detrimental to testability. [P9] mentioned that test stubbing is required for unit testing and integration testing whenever the unit to be tested or integrated is dependent on other units that are not yet tested or even coded. Coupling between such units drives the stubbing effort as increasing coupling will likely lead to additional features in the stubs, and will thus decreases testability. However, a certain type of coupling, the one using inheritance among classes is encouraged since it enhances testability.

[P37] also reports similar findings. It states that a class with low cohesion is likely to require more unit testing effort than a class with high cohesion. Weak cohesion indicates that a class may serve several unrelated goals, which suggests inappropriate design. Encapsulating unrelated attributes/methods is likely to lead to more complex tests. Therefore, more effort is likely to be required to unit test a class with low cohesion. A class with high coupling is likely to require more unit testing effort than a class with a low coupling. On the one hand, a class having a high import coupling depends on many externally provided services. When unit testing such a class, many stubs have to be developed to emulate the behavior of any externally provided services that are not yet available. On the other hand, a class with high export coupling has a large influence on the system: many other classes rely on it. Such a class will likely receive extra attention during testing and hence is also likely to require a larger test suite. Therefore, more effort is likely to be required to unit test a class with high coupling.

[P40] referred to dependency (a type of coupling) among objects/units as a testability anti-pattern.

<u>*Understandability:*</u>

Understandability is the degree to which a software artifact is documented or self-explaining. [P2] discusses that component understandability, which refers to how well the generated component user manual, API specification are easy for user operations and testing. Thus, it influences testability.

<u>*Inheritance:*</u>

In general, inheritance also decreases testability. [P9] found that the higher the size of the inheritance hierarchy, the more expensive it is to test due to the dependences between the child classes. In other words, testing the interface between the super and sub classes may be more expensive as a result of inheritance, as there would be a need for additional test cases and possibly a need for modifications of test oracles for each subclass.

[P15] found that the depth of a class in an inheritance tree affects the testability of the class. Usually, the deeper the class is in an inheritance tree, the greater the number of methods it will inherit. This makes it more complex to test and maintain it. If a class is near the root of an inheritance tree, there are more chances where it will be tested since every time a test case goes through its offspring, it will also go through the class itself. Therefore the nearer the class is to the root of an inheritance tree, the higher the testability of a class while the deeper the class is in an inheritance tree, the lower its testability is.

Similar findings were reported by [P37]. A class with many ancestors, many descendants, deep inheritance hierarchy, many overridden methods and/or many overloaded methods is likely to require more unit testing effort than a class with few ancestors, few descendants, shadow inheritance hierarchy, few overridden methods and few overloaded methods. If a class has more ancestors or is situated deep in an inheritance hierarchy, it will likely inherit more attributes and methods from its ancestors. More effort is hence likely to be required to cover the inherited attributes/methods and the test code used to examine whether a test case execution is successful may also be more complex.

*Reliability of tests:*

[P21] defined an interesting aspect for testability: reliability of the test, which is defined as the confidence in the outcome of the test, or the probabilistic accuracy of the test in having the correct outcome.

*Availability:*

Availability is a characteristic of a system, which aims to ensure an agreed level of operational performance, usually uptime, for a higher than normal period.

[P2] found that document availability, which refers to the availability of component artifacts, including requirements specification, API specification, and user reference manual, influences testability.

[P10] presents a model for the measurement of the runtime testability of component-based systems. It found that if a component has high availability requirements, runtime testing under such as requirement cannot be performed, as it decrease high availability. Thus, its runtime testability would decrease. Runtime testability was defined in [P10] in two ways, i.e., (1) the degree to which a system or a component facilitates runtime testing without being extensively affected; (2) the specification of which tests are allowed to be performed during runtime without extensively affecting the running system.

*Flexibility:*

Entitled "*Flexibility: a key factor to testability*", [P78, P177] highlighted the importance of unit or system flexibility and its impact on testability. Furthermore, the main research question explored in [P173] was: How flexibility affects the testability of framework-based applications?

*Reusability of test suite:*

[P32] and [P182] mentioned that high reusability and structure of the test suite positively impacts testability.

*Maintainability:*

Entitled " *Modifiability: a key factor to testability*", [P105] highlighted the importance of unit or system modifiability and its impact on testability.

*Other factors affecting testability:*

Furthermore, 56 papers mentioned additional other factors. Here are some examples of that factors: unit size [P9]; statefulness [P10]; isolateability, test automatability [P17]; software process capability [P29]; modularity [P33]; built-in test capabilities, and test support environment [P64]; fault-proneness [P76]; manageability and supportability [P120] quality of the test suite [P122]; and self-documentation [P169].

*Other attempts to classify testability and factors influencing it:*

We should also note that other attempts to classify testability and factors influencing it have been presented in the community. For example, when presenting heuristics of software testability, a practitioner named James Bach classified testability into five categories (www.satisfice.com/tools/testable.pdf): (1) intrinsic testability: internal characteristics of the SUT which affect testability (e.g., observability and controllability); (2) project-related testability: issues such as information availability (have we got all information we want or need to test well?) or tool availability (are we provided with all the tools we need to test well?); (3) subjective testability: issues such as testing skill (our ability to test in general obviously makes testing easier) and product knowledge (knowing a lot about the product, including how it works internally, profoundly improves our ability to test it); (4) value-related testability: what we want from the product, e.g., oracle reliability: we benefit from oracles that can be trusted to work in different test executions and in many conditions; and (5) epistemic testability: the gap between what we know and what we need to know about the status of the product. For example, test automation is excellent in providing us with the illusion of increased epistemic testability, e.g.: "*Every night, we run 10,000 tests in less than an hour!*" while it can actually decrease testability: "*Bob spends four hours every day processing the results from test automation!*".

Kedemo used James Bach's classification to develop an extended model and referred to it as the three *dimensions* of testability [65]: (1) product dimensions: code and environment, team and vision; (2) tester dimensions: skills and knowledge, mental state; and (3) context drivers such as: risk, resources, development paradigm. Each of these drivers has positive or negative impact on testability, e.g., more resources would generally lead to better testability. More project risk, arising from the test activity or faults, would generally lead to low testability. Intrinsic testability factors discussed earlier correspond to the factors under the "product" group in this model.

Finally, a few other authors provided checklists. For instance, entitled "*Practicing testability in the real world*", [P120] presented a testability checklist including the following questions:

- Does it take you a considerable amount of time and code to "setup" before testing your actual component and "cleanup" later?
- Does your test fail intermittently due to a component external to the component that you are testing?
- Could you make your test execution simpler and easier by directly working with the component?
- Can your component be initialized or started in isolation by itself?
- Can you exactly point out what failed?
- Does your component follow established design patterns?

#### 6.2.3 RQ 2.3- Techniques for improving testability

Albeit the high importance of testability, it is common that testability is often not considered from the very beginning of software development. Thus, it is essential to investigate ways to improve it when a test team finds out that testability of a unit or a system is low. As discussed in Section 6.2.1, overall 82 papers in the pool addressed testability improvement issues. As already shown in Figure 9, testability improvement can be done in any phase of SDLC, when the need is "felt" to improve testability. After extracting the techniques for improving testability as reported in each paper, we synthesized and grouped them as shown in Figure 11. We found six recurring types of approaches for improving testability and placed the rest in the "other" category as shown in Figure 11. We discuss them next.

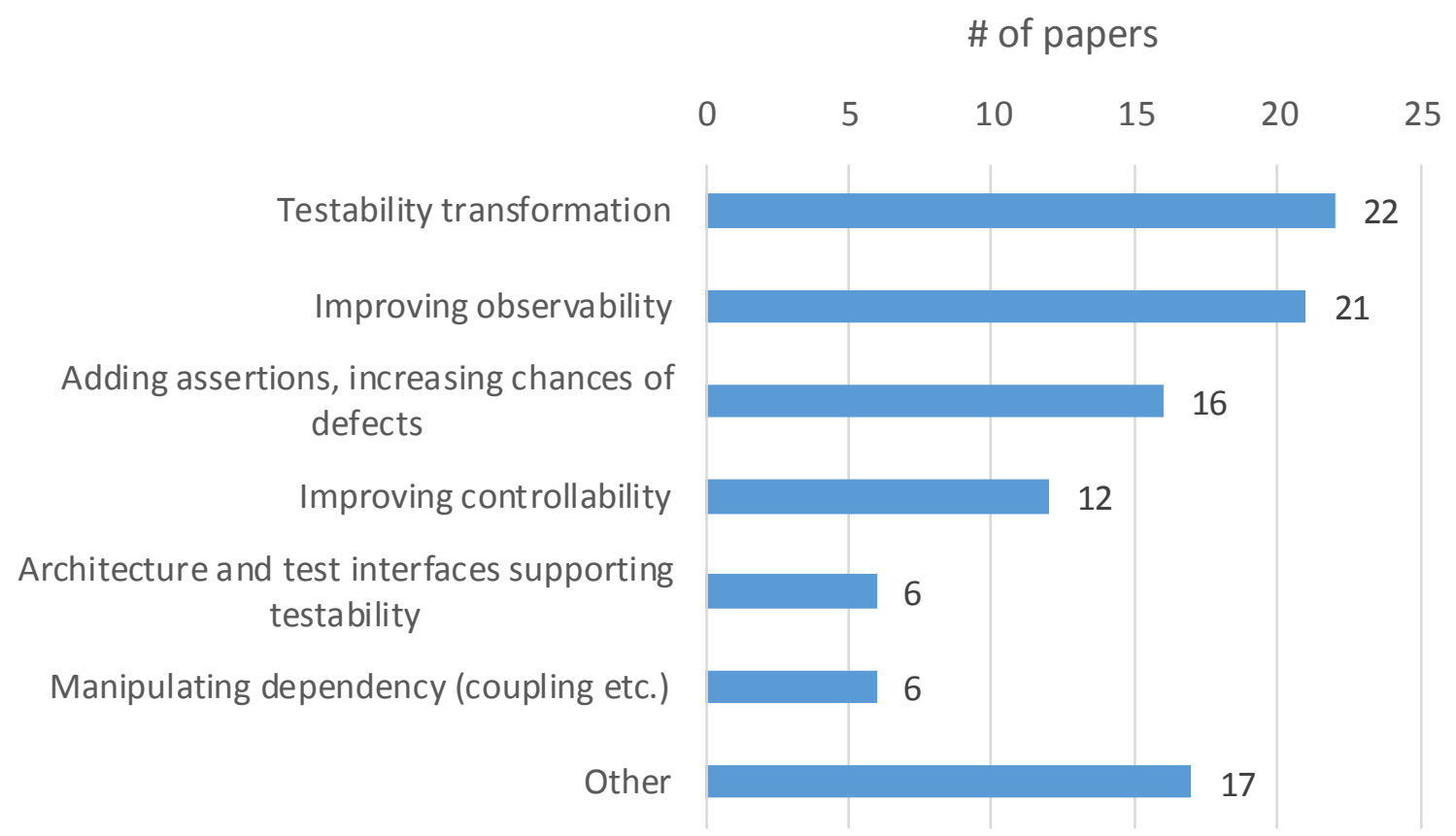

**Figure 11- Techniques for improving testability**

The most common way to improve testability is by testability transformation (proposed in 22 papers). Such a transformation changes the structure of a unit under test (e.g., a Java method or a finite-state-machine) to make it more testable. For instance, [P172] presented transformations that make it easier to generate test data. To address the challenge of evolutionary testing in the presence of loops, a certain type of testability transformation was conducted in [P75] to help evolutionary test-data generation. Furthermore, a prominent subclass of testability transformations are testability refactoring, e.g., automated code refactoring [P47], refactoring to reduce polymorphism and complexity [P87], and refactoring to reduce class interactions (coupling) [P96].

Improving observability and improving controllability are among the widely discussed techniques for improving testability (discussed in 21 and 12 papers, respectively). This is in alignment with findings of RQ 2.2 as we found that the two most often mentioned factors affecting testability are observability (mentioned in 101 papers) and controllability (86 papers), in Section 6.2.2. Here are some examples of those papers: [P14] proposed a method for improving the observability of specifications written in the Business Process Execution Language (BPEL). For improving observability in [P25], Java components' byte-code was analyzed to derive control and data flow graphs, which is then used to increase components testability by increasing both controllability and observability. The approach in [P42] extracted execution traces to facilitate test-case selection and thus testability. A debugging framework was presented in [P45], which was then used for tracking the changes of variables under evaluation, which, in turn helped improve the test observability.

For improving controllability, the approach in [P24] eliminated state variables from guard conditions of state-machines and determined which functions to call in order to satisfy guards with state variables. In [P49], to test unstructured programs to improve test data generation, the approach produced single-entry, single-exit control flows.

The other widely discussed technique for improving testability is adding assertions to the unit under test, which usually increases the chances of revealing faults if there are any. For instance, [P18, P134] used assertions to build self-testing (also called "built-in test") capabilities inside a component. [P27] utilized run-time constraint checking for improving testability. [P85, P93] improving the testability of object-oriented software through software contracts, e.g., method preconditions, method post-conditions, and class invariants. [P86] used distribution and complexity of past faults for smarter testing and for improving testability. [P200] was another paper in this category and had this title: "*Using assertions to make untestable software more testable*". Voas points out that the placement of "*assertions is one relatively simple trick for improving testability*" [P149]. [P93] proposed an instrumentation approach to add, to the code, assertions based on specification contracts.

To improve testability, 6 papers presented specific architecture and/or test interfaces supporting testability. For instance, [P54] use macros and test interfaces for increasing testability and then compared them to other techniques for increasing testability. [P107] presented an Embedded Software Design Architecture for Improving the Testability of In-vehicle Multimedia Software.

Furthermore, 6 other papers presented approaches based on manipulating dependency (coupling etc.). For instance, [P59, P201] utilized data-dependency analysis to help test-data generation. [P62] utilized Dependency-Injection, also known as "Inversion of Control" (IoC), which is as a technique for loose coupling, and in turn improving testability. [P175] removed dependency among components. [P181] decreased coupling among objects.

Finally, 17 papers presented other means to improve testability. For instance, isolation of untestable components [P137] or of source code that is likely to hide faults [P151] were suggested. [P82] entitled "*Improving architecture testability with patterns*" exploited the information described at Architectural Patterns to drive the definition of tests. As a result, that approach can assist developers in finding relatively shorter and cheaper test paths to high dependable software. The approach in [P95] is based on avoiding object interactions and concurrent accesses to shared objects. For design for testability of communication software, [P113] suggested to implement protocols in a deterministic way. In "*Practicing testability in the real world*", [P120] suggested simplicity of the component and knowledge of expected results (oracle).

## 6.3 Group 3-Specific to empirical studies and also the case-study of each paper

In this section we present the results of RQ 3.1 (research questions raised and studied in the empirical studies), RQ 3.2 (number of SUTs in each paper) as well as of RQ 3.3 (domains of SUTs).

### 6.3.1 RQ 3.1-Research questions investigated in the empirical studies

By this review question, we analyzed the types of RQs studied in the papers. It helps us to get an overview of the investigated research questions on testability and to explore potential interesting future research directions.

For that purpose, we extracted the list of RQs raised in studies of the pool. 37 papers from the paper pool include explicit research questions. Overall, 110 RQs are raised in the papers with explicit RQs. We classified the raised RQs according to Easterbrook et al. [66] into the following types: (1) existence questions, (2) descriptive and classification questions, (3) descriptive-comparative questions, (4) frequency and distribution questions, (5) descriptive-process questions, (6) relationship questions, (7) causality questions, (8), causality-comparative interaction questions, as well as (9) design questions. Table 4 shows the overall numbers of RQs per each type and also representative examples together with the source references. Most stated RQs (52) fall into the category of descriptive-process questions investigating how things actually work.

**Table 4: Number of research questions and examples per research question type**

| Research Question Type | Number | Examples for Research Questions |
|---|---|---|
| Existence | 10 | *Does a fault exist?* [P140] |
| Descriptive and classification | 12 | *What are the properties of "easily testable" programs?* [P182] |
| Descriptive-comparative | 6 | *How is the testability of fault prone code compared to non-fault prone code?* [P193] |
| Frequency and distribution | 9 | *What is the probability this code will fail if it is faulty?* [P145] |
| Descriptive-process | 52 | *How can component testability be increased?* [P68] |
| Relationship | 15 | *Does each single concurrent program metric correlate with the testability score?* [P122] |
| Causality | 2 | *Can software product line testability be controlled by variability binding time?* [P104] |
| Causality-comparative interaction | 2 | *Can partial least square regression models predict the effort involved in unit testing effort more accurately than multiple linear regression models?* [P37] |
| Design | 2 | *How to construct testable software components with a systematic approach?* [P108] |

### 6.3.2 RQ 3.2-Number and sizes of SUTs (examples) in each paper

One would expect that each paper applies the proposed testing technique to at least one SUT or example. Some papers take a more comprehensive approach and apply the proposed testing technique to more SUTs. Overall, 182 out of the 208 papers, i.e., 88%, refer to a SUT or example. From those 182 papers (see Figure 12), most papers, i.e., 109 (60%) refer to exactly one SUT/example. 27 (15%) refer to two SUTs/examples and 11 (6%) to three SUTs/examples. So about 80% of the papers referring to an SUT/example, refer to one, two or three SUTs/examples. Four and five SUTs/examples are also well represented, i.e., in 9 (5%) and 5 (3%) papers, respectively. So about 90% refer to up to five SUTs/examples. Overall, most papers referring to an SUT or an example (88%) and from these papers most to exactly one SUT/example (109, 60%) and about 90% to five or less SUTs/examples. So the expectation that most papers refer to an SUT/example (88%) and most of them to one or just a few SUTs/examples to provide the respective contribution holds.

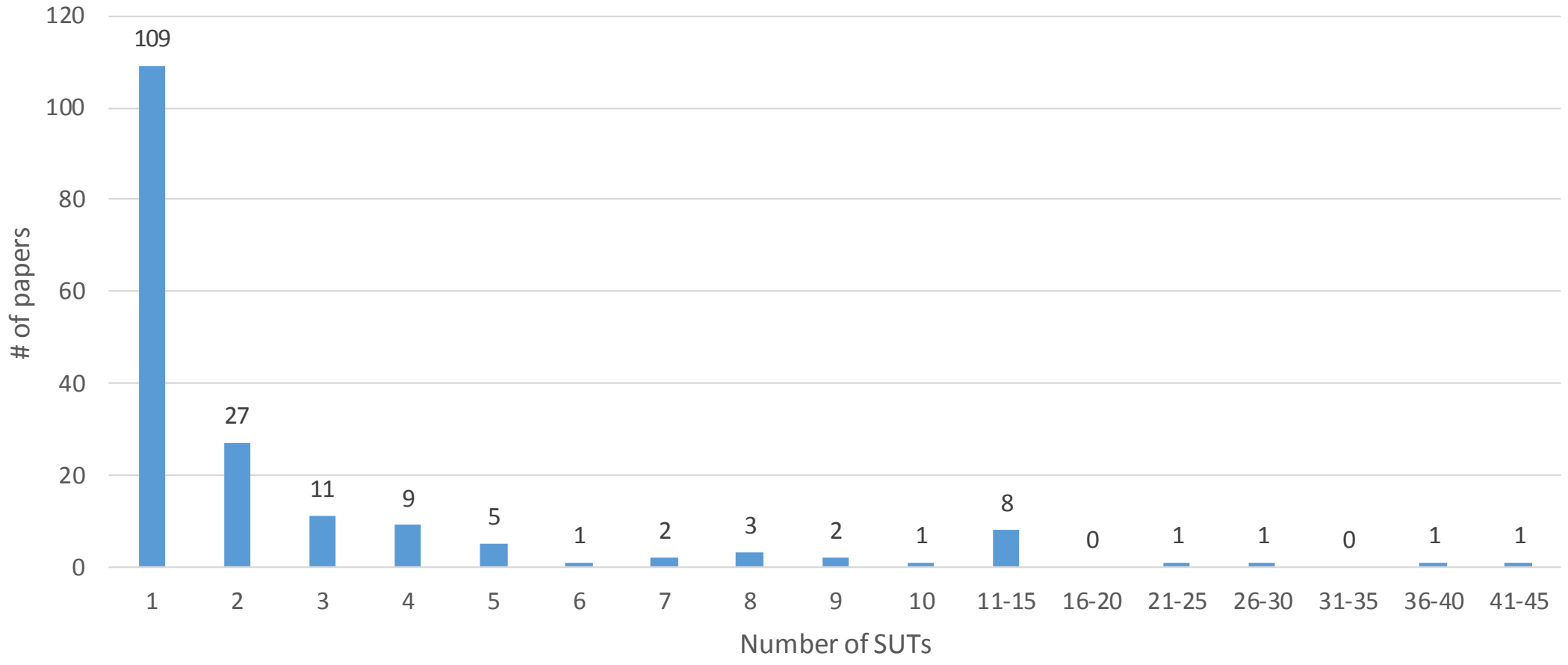

**Figure 12- Number of Systems Under Test (SUT) or examples in each paper**

We also extracted the size measured by the lines of code (LOC) of the SUTs (examples) used in the papers. Figure 13 shows the distribution of that data. As one can see, a large ratio of the papers have worked on rather small SUTs (up to 12.5 KLOC). In terms of large SUTs, [P186] conducted an empirical study of testability on 40 SUTs which in total add up to 919,096 LOC (i.e., about 920 KLOC).

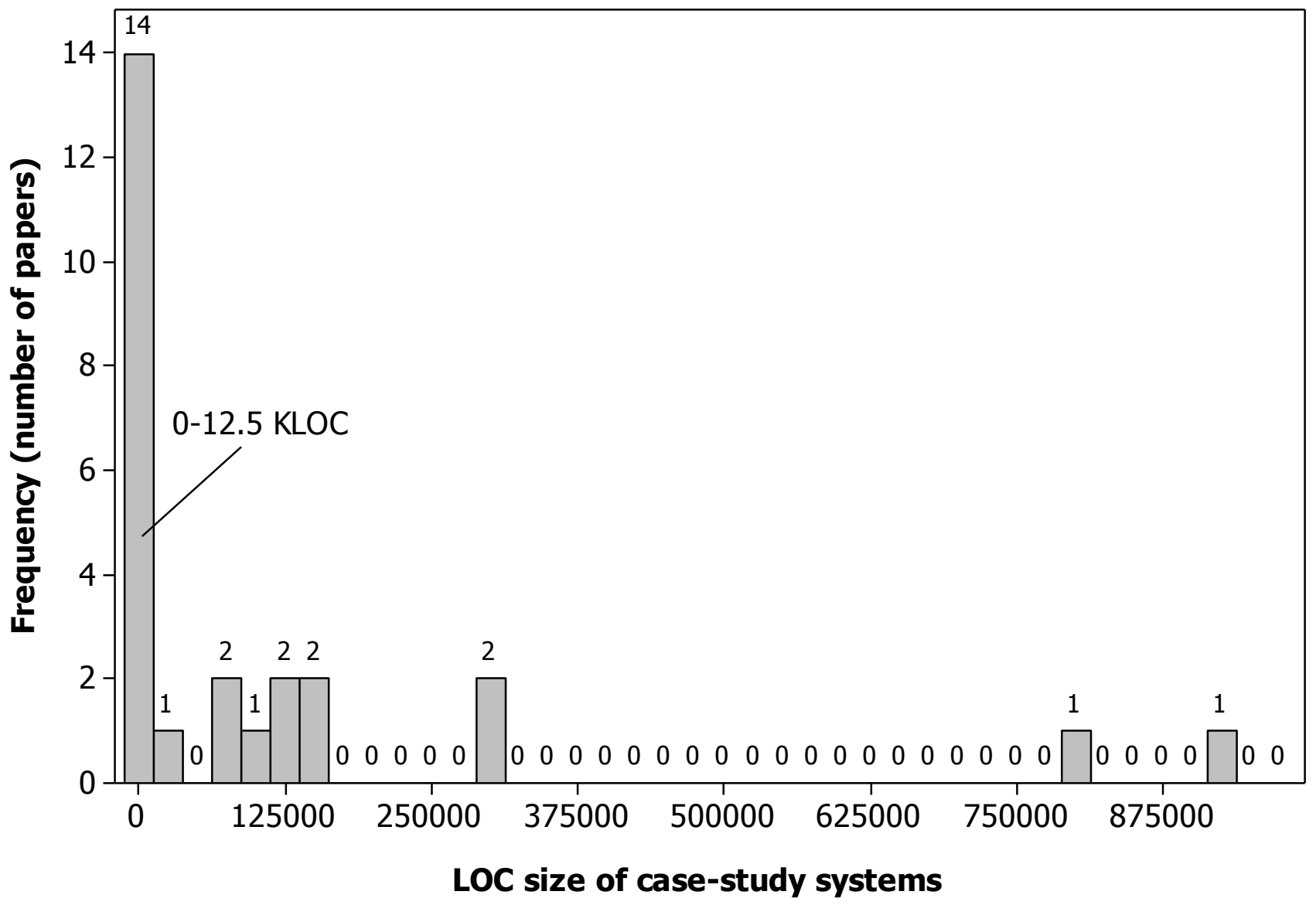

**Figure 13- Distribution of the size measured in lines of code (LOC) of Systems Under Test (SUT) in the papers**

#### 6.3.3 RQ 3.3-Domains of SUTs

As mentioned before, 182 papers refer to an SUT or an example. Most of them (131, 72%) are not specific to a domain, but address generic issues, e.g., related to the usage of assertions [P200], object-oriented programming [P70], or testability estimation [P172]. In addition, 24 (13%) address embedded systems, 20 (11%) real-time systems, 6 (3%) communication systems and protocols, and finally 4 (2%) web applications. In summary, we can say that most papers cover generic issues (about 70%), but system-specific issues (related to embedded, real-time, and communication systems) are also covered (about 30%), whereas web-specific issues are only rarely covered.

### 6.4 GROUP 4- DEMOGRAPHIC INFORMATION

In this section we present the results of RQ 4.1 (affiliation types of the study authors) and RQ 4.2 (highly-cited papers).

#### 6.4.1 RQ 4.1-Affiliation types of the study authors

To answer this research question we have broken down the papers based on the affiliation types of the study authors, i.e., industrial, academic or collaborative. In Figure 14, we show (as a stack chart) the number of studies per year published solely by academic researchers (those working in universities and research centers), solely by practitioners (also including corporate research centers), or as collaborative work. As one can see, the attention level on this topic has risen and fallen over the years. Note that our study was conducted during the first six months of 2017, thus we decided to include the papers published until end of 2016. The peak year in terms of number of papers was 2009 in which 15 papers were published on this topic.

In terms of breakdown of papers by affiliation types of the authors, 147 papers were authored solely by academic researchers, 36 papers by practitioners only, and 25 papers as joint (collaborative) work between researchers and practitioners. Among the active companies in conducting research and publishing papers in this area, we could see various company names such as: Lockheed Martin [P7], Boeing [P8], Microsoft [P120], and Siemens [P181].

To put the annual trend of papers in this area in perspective, we show, also in Figure 14, a similar chart reported by another recent survey study in the area of testing embedded software [12]. It is interesting to observe that while for the area of testing embedded software, the number of papers is generally increasing, for papers in the area of software testability, there are ups and downs, perhaps denoting a continuous change in challenges and/or interest in this area over the years. Also, by comparing the two charts, we can say that the area of software testability is slightly "older" than the area of testing embedded software.

Literature on software testability

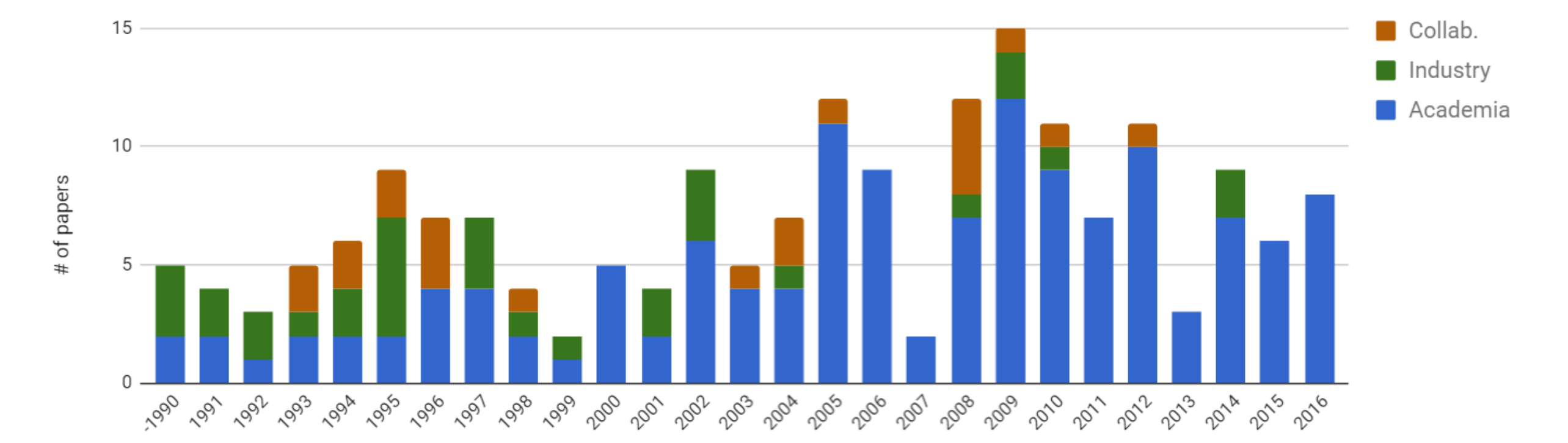

Literature on testing embedded software

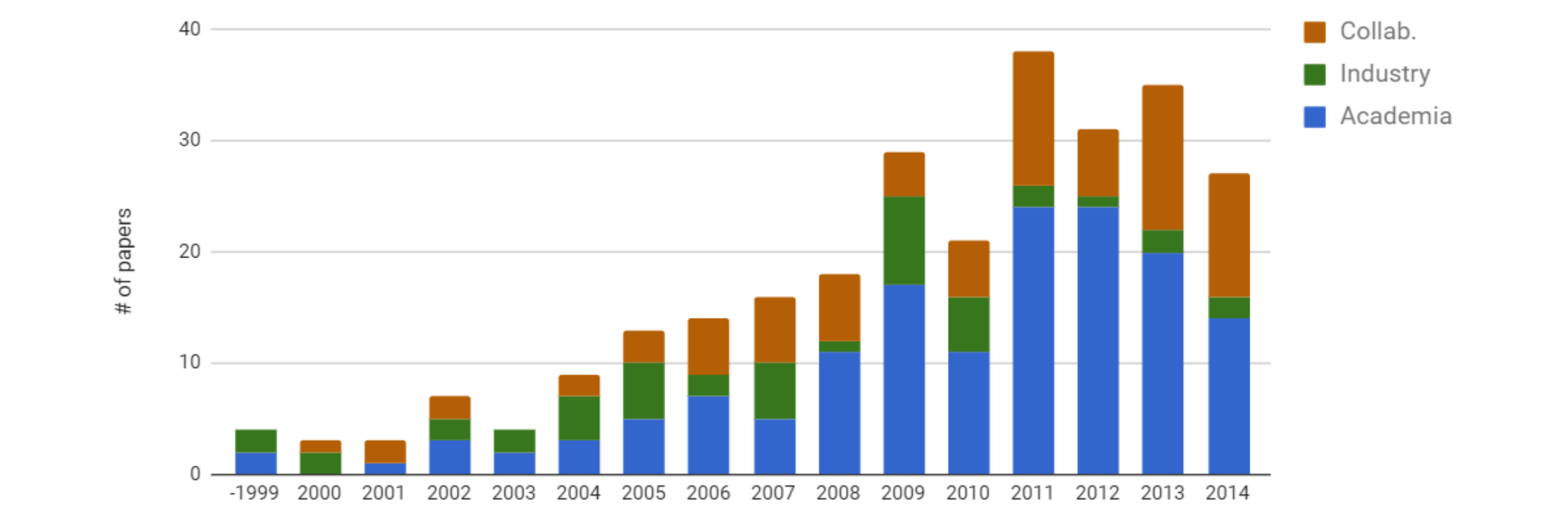

**Figure 14-(Top): Growth of the field and affiliation types of the study authors across years in this area. (Bottom): The same chart for another survey study on topic of testing embedded software [12]**

### 6.4.2 RQ 4.2-Highly cited papers

To help practitioners with deciding which sources to start reading first, we assessed the popularity of papers to find the most popular sources. In the academic world, citations are the de-facto metric for this purpose and have been used in many studies, e.g., a recent IEEE Software paper [67]. We list in Table 5 and Table 6 the top 5 highly-cited papers based on two metrics: (1) absolute citation values, and (2) normalized citations (average citations per year). Both of these metrics are widely used in bibliometric studies, e.g., [44, 45]. We can see that four papers appear in both top-5 lists.

While we believe all studies in the pool are interesting and could be useful for practitioners, they may want to consider starting their review of the literature by reading first the five top-cited papers in this pool. Note that the citation numbers in Table 5 were extracted from the Google Scholar system on Dec. 29, 2017. One can immediately see from the title of the highly cited papers that they address generic fundamental aspects of testability, i.e., how to define, measure, design for, achieve or apply testability.

**Table 5- Ranking the of five most cited (popular) sources, by absolute citation values**

| Rank | Paper title | Year of publication | Num. of citations | Average citations per year | ID in the pool |
|---|---|---|---|---|---|
| 1 | On testing non-testable programs | 1982 | 540 | 15.4 | [P111] |
| 2 | Software testability: the new verification | 1995 | 392 | 17.5 | [P151] |
| 3 | Design for testability in object-oriented systems | 1994 | 377 | 16.0 | [P64] |
| 4 | Testability of software components | 1991 | 288 | 10.8 | [P182] |
| 5 | Testability transformation | 2004 | 216 | 16.3 | [P184] |

**Table 6- Ranking the of five most cited (popular) sources, by normalized citations**

| Rank | Paper title | Year of publication | Num. of citations | Average citations per year | ID in the pool |
|---|---|---|---|---|---|
| 1 | Software testability: The new verification | 1995 | 386 | 17.55 | [P151] |
| 2 | Testability transformation | 2004 | 212 | 16.31 | [P182] |
| 3 | Design for testability in object-oriented systems | 1994 | 369 | 16.04 | [P64] |
| 4 | On testing non-testable programs | 1982 | 540 | 15.43 | [P111] |
| 5 | An empirical study into class testability | 2006 | 124 | 11.27 | [P32] |

We briefly discuss next the top-5 papers based on the absolute number of citations. The most cited and seminal paper "*On testing non-testable programs*" highlighted the significance of testability more than three decades ago (as early as in 1982). The second most cited paper "*Software testability: the new verification*" by Jeffrey Voas and Keith Miller [P151] defined how to design for testability and how to measure testability by sensitivity analysis, which estimates for a particular location in a program the probability of failure that would be induced in the program by a single fault.

The third most cited paper "*Design for testability in object-oriented systems*" by Robert Binder [P64] defined software testability for object-oriented software and identifies six main facets of testability, i.e., characteristics of the representation, characteristics of the implementation, built-in test capabilities, the test suite, the test support environment, and the software process in which testing is conducted. Each facet has sub-facets represented in a fishbone diagram. For instance, the facet representation has the sub-facets requirements, specification, traceability, and separation of concerns. The sub-facet separation of concerns is influenced by user interface, control strategy, collaboration packaging, and architectural packaging.

The fourth most cited paper "*Testability of software components*" by Roy Freedman [P182] defines and evaluates the concept of domain testability. A domain testable program is observable and controllable. In addition, metrics are defined, i.e., translation and redesign time, to assess the level of effort required to modify a program so that it becomes domain testable.

The fifth most cited paper "*Testability transformation*" by Mark Harman, Lin Hu, Rob Hierons, Joachim Wegener, Harmen Sthamer, André Baresel, and Marc Roper [P184] introduces testability transformation. It is a source-to-source transformation that aims to improve the ability of a given test generation method to generate test data for the original program. The paper illustrates the theory of testability transformation with an example application to evolutionary testing.

## 7 DISCUSSIONS

In this section we first summarize the research findings and discuss implications (Section 7.1), then we present benefits of the review (Section 7.2) and finally we discuss potential threats to validity of this review (Section 7.3).

### 7.1 SUMMARY OF RESEARCH FINDINGS

In this section we summarize the research findings of each RQ and discuss implications.

<u>Group 1-Common to all SLM studies:</u>

- **RQ 1.1-Classification of studies by contribution types**: Most papers (103) actually contribute testability approaches, methods or techniques. Model and metrics are also well represented and provided in 37 and 35 papers, respectively. Empirical studies only and tools are provided by 25 and 23 papers, respectively. Frameworks are provided by 14 papers, whereas guidelines to improve testability (6) and processes (4) are rather rare. Finally, 8 papers contain other contributions.
- **RQ 1.2-Classification of studies by research method types**: Almost equally-divided portions of papers (63, 53 and 60 out of 208 papers) present solution proposals (by examples), weak empirical studies, and strong empirical studies. 20, 0 and 12 papers, in order, are experience papers, philosophical and opinion papers, respectively.

<u>Group 2-Specific to the domain (i.e., testability):</u>

- **RQ 2.1- Type of approaches for treating testability**: In 9 papers requirements testability is addressed; design for testability is addressed in 24 papers; measurement of testability is addressed in 133 papers; and finally improvement of testability in 82 papers. Approaches to quantitative and qualitative measurement of testability are very prominent and address all phases of development.
- **RQ 2.2- Factors affecting testability**: The two most often mentioned factors affecting testability are observability (mentioned in 101 papers) and controllability (82 papers). In addition, specific aspects of software quality are frequently reported to affect testability. Especially, complexity and simplicity (21 papers), dependency: cohesion / coupling (19 papers), and understandability (12 papers) are among other factors affecting testability. Furthermore, 56 papers mentioned additional other factors, for instance unit size, potential for test automation, modularity, fault-proneness, or the degree to which a requirement is readable.
- **RQ 2.3- Techniques for improving testability**: Common ways to improve testability are testability transformation (proposed in 22 papers), improving observability (21), adding assertions (16), and improving controllability (12). Furthermore, specific architecture and/or test interfaces supporting testability (6) and manipulating dependency (6) are proposed. Finally, 17 papers present other means to improve testability, for instance, isolation of untestable components or of source code that is likely to hide faults.

<u>Group 3-Specific to empirical studies and those with case studies:</u>

- **RQ 3.1-Research questions investigated in the empirical studies**: 37 papers from the paper pool include explicit research questions. Overall, 110 research questions are raised in the papers with explicit research questions. Most stated research questions (52) fall into the category of descriptive-process questions investigating how things actually work.
- **RQ 3.2-Number and sizes of SUTs (examples) in each paper**: Most papers refer to a SUT or example (182 papers, 88%) and from those papers most to exactly one SUT/example (109), about 80% of the papers to three or less SUT/example and about 90% to five or less SUTs/examples. Most papers work on rather small SUTs of up to 12.5 KLOC.
- **RQ 3.3-Domains of SUTs**: Most of the 182 papers referring to an SUT or example are not domain-specific, but address generic issues, e.g., related to the usage of assertions, object-oriented programming, or testability estimation. In addition, 24 address embedded systems, 20 real-time systems, 6 communication systems and protocols, whereas web-specific issues are only investigated in 4 papers.

Group 4-Demographic information:

- **RQ 4.1-Affiliation types of the study authors**: Papers authored solely by academic researchers are dominating (147), but also papers authored solely by practitioners (36) as well as jointly by researchers and practitioners (25) are well represented. This shows that testability is a topic not only of academic, but also practical relevance and also actively investigated by practitioners.
- **RQ 4.2- Highly-cited papers**: The five most highly cited papers all address generic fundamental aspects of testability, i.e., how to define, measure, design for, achieve or apply testability. For instance, the most cited and seminal paper "Software testability: the new verification" by Voas and and Miller [P151] defines how to design for testability and how to measure testability by sensitivity analysis.

## 7.2 POTENTIAL BENEFITS OF THIS REVIEW

We discuss next the potential benefits of our mapping study for practitioners and researchers.

### 7.2.1 Benefits for practitioners

Recall from beginning of this paper that this review study was conducted based on a real need that we had in our industrial project. The authors and their collaborators have already started to benefit from the results of this review. In our ongoing collaborations with several industry partners in Turkey, Austria and the Netherlands in the area of software testing, our colleagues and we did not have an adequate overview of the literature and this review provided that. Thanks to our review study, we are currently assessing several existing testability assessment techniques based on the review at hand for possible adoption/extension in our ongoing industry-academia collaborations.

To further assess the benefits of this review, we asked two active test engineers (one in the Netherlands and one in Austria who is involved mostly in the insurance software domain) to review this review paper and the online spreadsheet of papers, and let us know what they think about their potential benefits. Their general opinion was that a review paper like this article is a valuable resource.

One of the practitioners provided the following feedback and also mentioned some ideas for further work in this area: "*Software projects in the insurance domain are developed iteratively over a period of 3 to 4 years. For instance, new GUI are added [to the existing software applications]. In order to efficiently test the applications, their testability has to be improved from iteration to iteration. A literature survey [such as this paper] could benefit our projects. In order to assess the testability of a design, methods and tools have to be selected. Many tools are available on the market. However, a survey of practical methods and metrics are still needed to select an appropriate approach. Interestingly, only a few number of articles on requirements testability were found. In the company that I work for, it [requirements testability] is a key issue in all projects. Papers dealing with skills of requirements analysts and an efficient review process are needed [to ensure requirements testability]. Also empirical studies on the efficiency of review processes [for the purpose of requirements testability] are needed. The results of experiments, assessing the benefit of investments in reviews of requirements specifications and their benefit to test-case design, could foster support for those activities by [higher-level] managers*".

Our other industry contact mentioned: "*According to my experience, problems and risks with testability are seen on all levels of testing, from unit tests that are hard to write because of a low cohesion and tight coupling through to testers having to resort to the user interface to indirectly test business logic that is implemented in lower levels, yet has no direct access hooks. In order to ensure that applications have high levels of testability, we [as a community] need to advocate the notion of testability throughout the entire SDLC. We can only succeed in doing this if we are aware of what testability is, how it can be measured and how it can be improved. I feel that many practitioners are unaware of the large number of the resources available in this area, written by both researchers and practitioners. This instantly confirms the need for an index on the existing literature concerning testability, if we want to prevent organizations reinventing the wheel over and over again.*"

Since the body of knowledge in this area is extremely large, this review would serve practitioners as a repository to access that large body of knowledge. For example, a practitioner can review Figure 9 in which we classified the pool of 208 papers by how they address testability in different stages of the SDLC. Testability is often considered late in the SDLC and, in that case, it is most of the time not easy to improve testability. Thus, Figure 9 and our online spreadsheet supports practitioners to consider testability as early as possible in the SDLC and pay attention to requirements testability. Other approaches for treating testability, such as design for testability and measurement (estimation) of testability, are also important and our mapping provides links to studies under those groups.

Our synthesis of techniques for improving testability (Figure 11) will also benefit practitioners since systematic approach for increasing testability are often needed in practice. A practitioner looking for advice in such a problem can use our survey to find the studies which for example have improved testability by improving observability.

Furthermore, practitioners often complain that approaches developed by researchers have only been tried on small examples or small-scale systems [68]. However, as we explored in RQ 3.2, the sizes of SUTs (examples) in some papers of the pool were quite large and thus practitioners can review those papers to see how testability approaches have been applied to large-scale systems, and could potentially be applied in their own context as well.

Last, but not the least, we explored the domains of SUTs (in Section 6.3.3) and found that various domains are represented in the pool, e.g., embedded systems, real-time systems, communication systems and protocols, and web applications. Thus, practitioners working especially in those domains and are looking for advice on testability, may first review the studies from the pool performed in their own domain.

Finally, one practitioner who referred to this study already benefited from it when developing a goal-question-metrics approach for measuring and improving testability of requirements [69]. The mapping study pointed to the relevant related work and other examples where the Goal Question Metric (GQM) approach has been applied to measure the testability of requirements.

#### 7.2.2 Benefits for researchers

In addition to practitioners, our mapping study provides various benefits for researchers. Note that this study is a systematic literature mapping (classification), and thus it provides a map (general picture) of this area for new or established researchers, who want to pursue further research in the area of testability. It goes without saying that reviewing a large set of 208 papers would be challenging and a survey paper like this can serve as an ideal starting point.

In a paper entitled "*What Makes Good Research in Software Engineering?*" [70], Mary Shaw discussed the importance of asking "interesting" RQs. Important of raising (asking) good and interesting RQs have also been discussed extensively in other fields, e.g., [71-73]. We synthesized and categorized the list of all 110 RQs raised and studied in the empirical studies in the pool (Section 6.3.1). We believe that researchers would benefit from them in formulating even more interesting and relevant RQs for future studies in the area of testability.

Furthermore, studying highly-cited papers (Section 6.4.2) would be beneficial for both researchers and practitioners.

Finally, one author of this mapping study, who is a researcher, himself already benefited from the study at hand when developing a framework for measuring testability of non-functional properties [74]. The mapping study provided a pointer to relevant related work, showed that a framework for measuring testability of non-functional properties does not exist so far, and provided a comprehensive overview of input factors for measuring testability.

### 7.3 Potential threats to validity

The main issues related to threats to validity of this literature review are inaccuracy of data extraction, and incomplete set of studies in our pool due to limitation of search terms, selection of academic search engines, and researcher bias with regards to exclusion/inclusion criteria. In this section, these threats are discussed in the context of the four types of threats to validity based on a standard checklist for validity threats presented in [75]: internal validity, construct validity, conclusion validity and external validity. We discuss next those validity threats and the steps that we have taken to minimize or mitigate them.

Internal validity: The systematic approach that has been utilized for source selection is described in Section 4. In order to make sure that this review is repeatable, search engines, search terms and inclusion/exclusion criteria are carefully defined and reported. Problematic issues in selection process are limitation of search terms and search engines, and bias in applying exclusion/inclusion criteria.

Limitation of search terms and search engines can lead to incomplete set of primary sources. Different terms have been used by different authors to point to a similar concept. In order to mitigate risk of finding all relevant source, formal searching using defined keywords has been done followed by manual search in references of initial pool and in web pages of active researchers in our field of study. For controlling threats due to search engines, not only we have included comprehensive academic databases such as Google Scholar. Therefore, we believe that adequate and inclusive basis has been collected for this study and if there is any missing publication, the rate will be negligible.

Applying inclusion/exclusion criteria can suffer from researchers' judgment and experience. Personal bias could be introduced during this process. To minimize this type of bias, joint voting was applied in source selection and only source with scores passing our set threshold were selected for this study. Also, to minimize human error/bias, we conducted extensive peer reviewing to ensure quality of the extracted data.

To perform data extraction, the papers were divided among the three authors. It would not have been efficient and is even not required to independently extract all the data from all the papers by each of the three authors. However, the authors performed quality assurance of each other's results and jointly discussed the extracted data to mitigate the threat of divided data extraction. According to the authors' experience dividing the data extraction task among authors is common in SM and SLR studies in software engineering and the only way for efficient data extraction if the paper pool is large. Furthermore, the quality of the extracted data is high if quality assurance and respective feedback cycles as in our setting are established.

Construct validity: Construct validities are concerned with issues that to what extent the object of study truly represents theory behind the study [75]. Threats related to this type of validity in this study were suitability of RQs and categorization scheme used for the data extraction. We carefully defined the RQs in our team of three researchers and linked them to a suitable data extraction scheme.

Conclusion validity: Conclusion validity of a literature review study is asserted when correct conclusions are reached through rigorous and repeatable treatment. In order to ensure reliability of our treatments, an acceptable size of primary sources was selected and terminology in defined schema reviewed by authors to avoid any ambiguity. All primary sources are reviewed by at least two authors to mitigate bias in data extraction. Each disagreement between authors was resolved by consensus among researchers. Following the systematic approach and described procedure ensured replicability of this study and assured that results of similar study will not have major deviations from our classification decisions.

External validity: External validity is concerned with to what extent the results of our literature review can be generalized. The study provides a comprehensive view on testability (overall 208 sources are included) taking both the academic and industrial perspective into account (see Figure 14). The collected sources contain a significant proportion of academic and industrial work, which forms an adequate basis for concluding results useful for both academia and applicable in industry. Also, note that our findings in this study are mainly within the field of testability. We have no intention to generalize our results beyond this subject. Therefore, few problems with external validity are worthy of substantial attention.

## 8 CONCLUSIONS AND FUTURE WORK

To identify the state-of-the-art and the state-of-the-practice in the area of software testability and to find out what we know about testability, we conducted and presented in this article a systematic literature mapping. Our article aims to benefit the readers (both practitioners and researchers) in providing the most comprehensive survey of the area of testability.

By classifying the entire body of knowledge, this survey paper "mapped" the body of knowledge on software testability. We systematically classified a large set of 208 papers and investigated several review questions under four groups. The first group investigated the contribution as well as the research method types. The second group investigated the approaches for treating testability, factors affecting testability, as well as techniques for improving testability. The third group investigated the studies research questions, the number of SUTs or examples in each paper, as well as the domains of the SUTs or examples. Finally, the fourth group investigated the affiliation types of the authors as well as the highly cited papers.

The area of software testability has been comprehensively studied by researchers but also practitioners. Most papers actually contribute testability approaches, methods or techniques, relate to a concrete SUT or example and provide solution proposals (by examples), weak empirical studies or strong empirical studies. As approaches to treat testability measurement of testability and improvement of testability are most frequently addressed in the papers. The two most often mentioned factors affecting testability are observability and controllability. Common ways to improve testability are testability transformation, improving observability, adding assertions, and improving controllability.

By summarizing what we know in this area, this paper provides an "index" to the vast body of knowledge in this area. Practitioners and researchers, who are interested in reading each of the classified studies in depth, can conveniently use the online Google spreadsheet at www.goo.gl/boNuFD to navigate to each of the papers.

Our future work includes using the findings of this SLM in our industry-academia collaborative projects as well as the development of a testability framework for testing non-functional properties like performance or security. We also encourage practitioners to report their concrete challenges in the area of testability so that researchers can work on and solve those challenges. This would strengthen the bridge between industry and academia in this area [76].

Also, let us note that we raised in this mapping study only a carefully selected number of RQs focusing – besides general RQs on the type of research and contribution as well as demography of the papers - on the domain (i.e., testability): types of approaches for treating testability, factors affecting testability, and techniques for improving testability. Future work comprises the analysis of additional aspects and RQs related to testability, e.g., challenges of testability and its measurement, and context issues (factors) such as: human aspects like the skill and experience level of testers (testability may depend on the testers' skill in testing), architecture, programming language, testability in the scope of test automation and DevOps, development process (e.g. Agile or Waterfall). Furthermore, this first effort of ours was only a mapping (classification) study. A natural follow-up work in this direction would be an in-depth SLR to systematically review some of the most important aspects, e.g., we only classified the types of approaches for treating testability (in Section 6.2.1), but it will be important to review and cross-compare various existing approaches for treating testability. We think that an an SLR based on this mapping study will provide further insight into what the most important challenges about testability are. Such information would be particularly beneficial to academics as a source for future research.

## ACKNOWLEDGEMENTS

The authors would like to thank Bas Dijkstra and Armin Beer for providing their opinions on benefits of this mapping study. Michael Felderer was partly funded by the Knowledge Foundation (KKS) of Sweden through the project 20130085: Testing of Critical System Characteristics (TOCSYC).

## 9 REFERENCES

The References section is divided into two parts: (1) Citations to the 208 sources reviewed in the literature review; and (2) Other (regular) references cited throughout the paper.

### 9.1 SOURCES REVIEWED IN THE LITERATURE REVIEW

[P1] T. M. King, A. A. Allen, Y. Wu, P. J. Clarke and A. E. Ramirez, "A Comparative Case Study on the Engineering of Self-Testable Autonomic Software", IEEE International Conference and Workshops on Engineering of Autonomic and Autonomous Systems, Las Vegas, 2011, pp. 59-68.

[P2] J. Gao and M. C. Shih, "A Component Testability Model for Verification And Measurement", Annual International Computer Software and Applications Conference, 2005, pp. 211-218 Vol. 1.

[P3] J. Fu, B. Liu, and M. Lu, "A Framework for Embedded Software Testability Measurement, in Information and Automation, Springer, 2011, p. 105-111.

[P4] S.T. Vuong, A.A.F. Loureiro, and S.T. Chanson, "A Framework For The Design for Testability of Communication Protocols", International Workshop on Protocol Test Systems, Pau, France, 1993.

[P5] V. Gupta, K. Aggarwal, and Y. Singh, "A Fuzzy Approach for Integrated Measure of Object-Oriented Software Testability", Journal of Computer Science, 2005. 1(2): p. 276-282.

[P6] [77] I. Rodríguez, L. Llana and P. Rabanal, "A General Testability Theory: Classes, Properties, Complexity, and Testing Reductions", IEEE Transactions on Software Engineering, vol. 40, no. 9, pp. 862-894

[P7] R. Sedmak and J. Evans, "A Hierarchical, Design-For-Testability (DFT) Methodology for The Rapid Prototyping of Application-Specific Signal Processors (RASSP)", Proceedings of IEEE International Test Conference, Washington, DC, 1995, pp. 319-327.

[P8] A. L. Alwardt, N. Mikeska, R. J. Pandorf and P. R. Tarpley, "A Lean Approach to Designing for Software Testability", IEEE Systems Readiness Technology Conference. (AUTOTESTCON), Anaheim, CA, 2009, pp. 178-183

[P9] S. Mouchawrab, L.C. Briand, and Y. Labiche, "A Measurement Framework for Object-Oriented Software Testability", Information and Software Technology, 2005. 47(15): p. 979-997.

[P10] A. González, É. Piel and H. G. Gross, "A Model for the Measurement of the Runtime Testability of Component-Based Systems," International Conference on Software Testing, Verification, and Validation Workshops, Denver, CO, 2009, pp. 19-28

[P11] L. Zhao, "A New Approach for Software Testability Analysis", Proceedings of the International Conference on Software Engineering, pp. 985-988, 2006.

[P12] A.M. Alakeel, "A New Testability Transformation Method for Programs with Assertions", International Journal of Computer Science and Network Security, 2016, 16(7): p. 137

[P13] H.-G. Groß, "A Prediction System for Evolutionary Testability Applied to Dynamic Execution Time Analysis", Information and Software Technology, 2001, 43(14): p. 855-862.

[P14] S. Salva and I. Rabhi, "A Preliminary Study on BPEL Process Testability", International Conference on Software Testing, Verification, and Validation Workshops, Paris, 2010, pp. 62-71.

[P15] B. W. N. Lo and Haifeng Shi, "A Preliminary Testability Model for Object-Oriented Software", International Conference Software Engineering: Education and Practice, 1998, pp. 330-337.

[P16] T.M. King, A. Ramirez, P.J. Clarke, and B. Quinones-Morales, "A Reusable Object-Oriented Design to Support Self-Testable Autonomic Software", Proceedings of ACM Symposium on Applied Computing, pp. 1664-1669, 2008

[P17] R. Poston, J. Patel, and J.S. Dhaliwal, "A Software Testing Assessment to Manage Project Testability", European Conference on Information Systems, 2012

[P18] C.R. Rocha and E. Martins, "A Strategy to Improve Component Testability without Source Code", Proceedings of International Workshop on Software Quality, 2004, 58: p. 47-62.

[P19] T. H. Tsai, C. Y. Huang and J. R. Chang, "A Study of Applying Extended PIE Technique to Software Testability Analysis", IEEE International Computer Software and Applications Conference, Seattle, WA, 2009, pp. 89-98

[P20] A. Tahir, S. MacDonell, and J. Buchan, "A Study of the Relationship Between Class Testability and Runtime Properties", International Conference on Evaluation of Novel Approaches to Software Engineering, 2014.

[P21] M. Efatmaneshnik and M. Ryan, "A Study of the Relationship between System Testability and Modularity", INSIGHT Journal, 2017, 20(1): p. 20-24.

[P22] [78] T. Kanstrén, "A Study on Design for Testability in Component-Based Embedded Software", International Conference on Software Engineering Research, Management and Applications, Prague, 2008, pp. 31-38.

[P23] A. Tahir, "A Study on Software Testability and the Quality of Testing In Object-Oriented Systems", University of Otago, PhD thesis, 2016.

[P24] A. Kalaji, R. M. Hierons and S. Swift, "A Testability Transformation Approach for State-Based Programs", International Symposium on Search Based Software Engineering, Windsor, 2009, pp. 85-88.

[P25] S. Kansomkeat and W. Rivepiboon, "An Analysis Technique to Increase Testability of Object-Oriented Components", Software Testing, Verification and Reliability, 2008, 18(4): p. 193-219.

[P26] Jin-Cherng Lin and Szu-Wen Lin, "An Analytic Software Testability Model", Proceedings of the Asian Test Symposium, 2002, pp. 278-283.

[P27] N. Pan and E. Song, "An Aspect-Oriented Testability Framework", Proceedings of the ACM Research in Applied Computation Symposium, 2012.

[P28] P.K. Singh, O.P. Sangwan, A.P. Singh, and A. Pratap, "An Assessment of Software Testability using Fuzzy Logic Technique for Aspect-Oriented Software", International Journal of Information Technology and Computer Science, 2015, 7(3): p. 18.

[P29] A. Kout, F. Toure, and M. Badri, "An Empirical Analysis of a Testability Model for Object-Oriented Programs", ACM SIGSOFT Software Engineering Notes, 2011, 36(4): p. 1-5.

[P30] L. Badri, M. Badri, and F. Toure, "An Empirical Analysis of Lack of Cohesion Metrics for Predicting Testability of Classes", International Journal of Software Engineering and Its Applications, 2011, 5(2): p. 69-86.

[P31] J.M. Voas, K.W. Miller, and J.E. Payne, "An Empirical Comparison of a Dynamic Software Testability Metric to Static Cyclomatic Complexity", Transactions on Information and Communication Technologies, vol. 8, pp.15, 1994

[P32] M. Bruntink and A. van Deursen, "An Empirical Study into Class Testability", Journal of Systems and Software, 2006, 79(9): p. 1219-1232.

[P33] L. Ma, C. Zhang, B. Yu, and H. Sato, "An Empirical Study on the Effects of Code Visibility on Program Testability", Software Quality Journal, 2017, 25(3): p. 951-978.

[P34] A. Gonzalez-Sanchez, R. Abreu, H.-G. Gross, and A.J. van Gemund, "An Empirical Study on the Usage of Testability Information to Fault Localization in Software", Proceedings of the ACM Symposium on Applied Computing, 2011.

[P35] Jin-Cherng Lin, Szu-Wen Lin and Ian-Ho, "An Estimated Method For Software Testability Measurement", Proceedings IEEE International Workshop on Software Technology and Engineering Practice incorporating Computer Aided Software Engineering, London, 1997, pp. 116-123.

[P36] U. Garg and A. Singhal, "An Estimation of Software Testability Using Fuzzy Logic", International Conference Cloud System and Big Data Engineering, Noida, 2016, pp. 95-100.

[P37] Y. Zhou, H. Leung, Q. Song, J. Zhao, H. Lu, L. Chen, and B. Xu, "An in-depth investigation into the relationships between structural metrics and unit testability in object-oriented systems", Science China Information Sciences, 2012, 55(12): p. 2800-2815.

[P38] S. Khalid, S. Zehra, and F. Arif, "Analysis of Object Oriented Complexity and Testability Using Object Oriented Design Metrics", Proceedings of National Software Engineering Conference, 2010.

[P39] L. d. Bousquet, M. Delaunay, H. V. Do and C. Robach, "Analysis of Testability Metrics for Lustre/Scade Programs", International Conference on Advances in System Testing and Validation Lifecycle, Nice, 2010, pp. 26-31.

[P40] M. R. Shaheen and L. d. Bousquet, "Analysis of the Introduction of Testability Antipatterns during the Development Process", International Conference on Software Engineering Advances, Porto, 2009, pp. 128-133.

[P41] Y. Le Traon, F. Ouabdesselam and C. Robach, "Analyzing Testability On Data Flow Designs", Proceedings International Symposium on Software Reliability Engineering, San Jose, CA, 2000, pp. 162-173.

[P42] C. Mao, "AOP-based Testability Improvement for Component-based Software," Annual International Computer Software and Applications Conference, Beijing, 2007, pp. 547-552.

[P43] J.M. Voas and K.W. Miller, "Applying A Dynamic Testability Technique to Debugging Certain Classes of Software Faults", Software Quality Journal, 1993, 2(1): p. 61-75.

[P44] M. C. K. Yang, W. E. Wong and A. Pasquini, "Applying Testability to Reliability Estimation," Proceedings International Symposium on Software Reliability Engineering, Paderborn, 1998, pp. 90-99.

[P45] Q. Guo, J. Derrick, and N. Walkinshaw, "Applying Testability Transformations to Achieve Structural Coverage of Erlang Programs", Testing of Software and Communication Systems, 2009, p. 81-96.

[P46] M. Z. Meetei, "Aspect-oriented software for testability", 5th International Conference on BioMedical Engineering and Informatics, 2012, pp. 1326-1331.

[P47] M. Ó. Cinnéide, D. Boyle and I. H. Moghadam, "Automated Refactoring for Testability", IEEE nternational Conference on Software Testing, Verification and Validation Workshops, Berlin, 2011, pp. 437-443.

[P48] L. Zucconi and K. Reed, "Building Testable Software", ACM SIGSOFT Software Engineering Notes, 1996, 21(5): p. 51-55.

[P49] R.M. Hierons, M. Harman, and C. Fox, "Branch-Coverage Testability Transformation for Unstructured Programs", The Computer Journal, 2005, 48(4): p. 421-436.

[P50] Y. Li and G. Fraser, "Bytecode Testability Transformation", International Symposium on Search Based Software Engineering, 2011.

[P51] S. Kansomkeat, J. Offutt, and W. Rivepiboon, "Class-Component Testability Analysis", Proceedings of WSEAS International Conference on Software Engineering, 2006.

[P52] R. Dssouli, K. Karoui, K. Saleh, and O. Cherkaoui, "Communications Software Design for Testability: Specification Transformations and Testability Measures". Information and Software Technology, 1999, 41(11): p. 729-743.

[P53] J.-R. Chang, C.-Y. Huang, C.-J. Hsu, and T.-H. Tsai, "Comparative Performance Evaluation of Applying Extended PIE Technique to Accelerate Software Testability Analysis", Journal of Systems Science, 2012, 43(12): p. 2314-2333.

[P54] J. Metsä, M. Katara and T. Mikkonen, "Comparing Aspects with Conventional Techniques for Increasing Testability", International Conference on Software Testing, Verification, and Validation, Lillehammer, 2008, pp. 387-395.

[P55] M. Liangli, W. Houxiang and Li Yongjie, "Construct Metadata Model based on Coupling Information to Increase the Testability of Component-based Software", International Conference on Computer Systems and Applications, Amman, 2007, pp. 24-31.

[P56] E. Martins, C. M. Toyota and R. L. Yanagawa, "Constructing Self-Testable Software Components", International Conference on Dependable Systems and Networks, Goteborg, Sweden, 2001, pp. 151-160.

[P57] A. Goel, S. Gupta, and S. Wasan, "Cott–a Testability Framework for Object-Oriented Software Testing", Int. Jounal Comput. Sci, 2008, 3(1): p. 813-820.

[P58] Z. Al-Khanjari, M. Woodward, and H.A. Ramadhan, "Critical Analysis of the Pie Testability Technique", Software Quality Journal, 2002, 10(4): p. 331-354.

[P59] B. Korel, M. Harman, S. Chung, P. Apirukvorapinit, R. Gupta and Q. Zhang, "Data Dependence Based Testability Transformation In Automated Test Generation", IEEE International Symposium on Software Reliability Engineering, Chicago, IL, 2005, pp. 10 pp.-254.

[P60] J. Bainbridge, "Defining Testability Metrics Axiomatically", Software Testing, Verification and Reliability, 1994, 4(2): p. 63-80.

[79] A. Lundberg, "Dependency Injection Frameworks: An Improvement to Testability", MSc thesis, Umeå University, 2015.

[80] B. Pettichord, "Design for Testability", Pacific Northwest Software Quality Conference, 2002.

[81] H. P. E. Vranken, M. F. Witteman and R. C. Van Wuijtswinkel, "Design for Testability in Hardware Software Systems", IEEE Design & Test of Computers, 1996, vol. 13, no. 3, pp. 79-86.

[82] R.V. Binder, "Design for Testability in Object-Oriented Systems", Communications of the ACM, 1994, 37(9): p. 87-101.

[83] M. Kim, S.T. Chanson, and S. Yoo, "Design for Testability of Protocols Based on Formal Specifications", Protocol Test Systems VIII. Springer, 1996, p. 252-264.

[84] M. Nakazato, S. Ohtake, M. Inoue and H. Fujiwara, "Design for Testability of Software-Based Self-Test for Processors", Asian Test Symposium, Fukuoka, 2006, pp. 375-380.

[85] H. König, A. Ulrich, and M. Heiner, "Design For Testability: A Step-Wise Approach to Protocol Testing", Testing of Communicating Systems, Springer, 1997, p. 125-140.

[86] Y.-T. Lin, "Design of Software Components with Increased Testability", MSc thesis, University of Delft, 2004.

[87] K.L. Bellman and C. Landauer, "Designing Testable, Heterogeneous Software Environments", Journal of Systems and Software, 1995.,29(3): p. 199-217.

[88] J.E. Payne, R.T. Alexander, and C.D. Hutchinson, "Design-For-Testability for Object-Oriented Software", Object Magazine, 1997, 7(5): p. 34-43.

[89] T. M. Khoshgoftaar, R. M. Szabo and J. M. Voas, "Detecting Program Modules With Low Testability", Proceedings of International Conference on Software Maintenance, Opio, 1995, pp. 242-250.

[90] M. Badri and F. Toure, "Empirical Analysis for Investigating the Effect of Control Flow Dependencies on Testability of Classes", International Conference on Software Engineering and Knowledge Engineering, 2011.

[91] P. McMinn, D. Binkley, and M. Harman, "Empirical Evaluation of a Nesting Testability Transformation for Evolutionary Testing", ACM Transactions on Software Engineering and Methodology, 2009. 18(3): p. 11.

[92] F. Lammermann, A. Baresel, and J. Wegener, "Evaluating Evolutionary Testability for Structure-Oriented Testing With Software Measurements", Applied Soft Computing, 2008, 8(2): p. 1018-1028.

[93] A. Baresel, D. Binkley, M. Harman, and B. Korel, "Evolutionary Testing In The Presence of Loop-Assigned Flags: A Testability Transformation Approach", ACM SIGSOFT Software Engineering Notes, pp. 108-118, 2004.

[94] L. Badri, M. Badri, and F. Toure, "Exploring Empirically the Relationship Between Lack of Cohesion and Testability in Object-Oriented Systems", International Conference on Advanced Software Engineering and Its Applications, pp, 78-92, 2010.

[95] J.M. Voas, "Factors That Affect Software Testability", NASA Langley Technical Report, 1991.

[96] M. Khan and R. Srivastava, "Flexibility: A Key Factor to Testability", International Journal of Software Engineering and Applications, 2015. 6(1): p. 89.

[97] F. Martensson, H. Grahn, And M. Mattsson, "Forming Consensus on Testability in Software Developing Organizations", Conference on Software Engineering Research and Practice, Sweden, 2005.

[98] Y. Le Traon and C. Robach, "From Hardware to Software Testability", Proceedings of IEEE International Test Conference , Washington, DC, 1995, pp. 710-719.

[99] J. Bach, "Heuristics of software testability", white paper, version 2.3, http://www.satisfice.com/tools/testability.pdf, 2015,.

[100] R. Coelho, U. Kulesza, and A. Von Staa, "Improving Architecture Testability With Patterns", ACM SIGPLAN Conference on Object-Oriented Programming, Systems, Languages, and Applications, pp. 114-115, 2005.

[101] Hwei Yin and J. M. Bieman, "Improving Software Testability With Assertion Insertion", International Test Conference, Washington, DC, 1994, pp. 831-839.

[102] M.C. Benton and N.M. Radziwill, "Improving Testability and Reuse by Transitioning to Functional Programming", arXiv preprint arXiv:1606.06704, 2016.

[103] Y. Singh and A. Saha, "Improving the Testability of Object Oriented Software Through Software Contracts", ACM SIGSOFT Software Engineering Notes, 2010, 35(1): p. 1-4.

[104] S. Khatri, R.S. Chhillar, and V. Singh, "Improving The Testability of Object-Oriented Software During Testing and Debugging Processes", arXiv preprint arXiv:1308.3320, 2013.

[105] I. Cabral, M.B. Cohen, and G. Rothermel, "Improving the Testing and Testability of Software Product Lines", Proceedings of international conference on Software product lines, pp. 241-255, 2010.

[106] N. Alshahwan, M. Harman, A. Marchetto and P. Tonella, "Improving Web Application Testing Using Testability Measures", IEEE International Symposium on Web Systems Evolution, Edmonton, AB, 2009, pp. 49-58.

[107] D. Deveaux, P. Frison and J. M. Jezequel, "Increase Software Trustability with Self-Testable Classes in Java", Proceedings Australian Software Engineering Conference, Canberra, ACT, 2001, pp. 3-11.

[108] S. Kansomkeat, J. Offutt, and W. Rivepiboon, "Increasing Class-Component Testability", IASTED Conf. on Software Engineering, 2005.

[109] E. A. Esker, J. P. Martin, W. R. Simpson and J. W. Sheppard, "Integrated Design for Testability and Automatic Testing Approaches", IEEE Conference on Systems Readiness Technology, San Antonio, TX, 1990, pp. 509-514.

[110] Z. Al-Khanjari and M. Woodward, "Investigating The Partial Relationships Between Testability and The Dynamic Range-To-Domain Ratio", Australasian Journal of Information Systems, 2003. 11(1).

[111] L.C. Briand, Y. Labiche, and H. Sun, " Investigating the Use of Analysis Contracts To Improve The Testability of Object-Oriented Code", Software: Practice and Experience, 2003, 33(7): p. 637-672.

[112] R. Bache and M. Müllerburg, "Measures of Testability as a Basis for Quality Assurance", Software Engineering Journal, 1990, 5(2): p. 86-92.

[113] B. Baudry, Y. L. Traon, G. Sunye and J. M. Jezequel, "Measuring and Improving Design Patterns Testability", International Workshop on Enterprise Networking and Computing in Healthcare Industry , 2003, pp. 50-59.

[114] B. Baudry and Y. Le Traon, "Measuring Design Testability of a UML Class Diagram", Information and Software Technology, 2005, 47(13): p. 859-879.

[115] H.-G. Gross, "Measuring Evolutionary Testability of Real-Time Software", PhD thesis, University of Glamorgan, 2000.

[116] J. H. Hayes, W. Li, T. Yu, X. Han, M. Hays and C. Woodson, "Measuring Requirement Quality to Predict Testability", IEEE International Workshop on Artificial Intelligence for Requirements Engineering, Ottawa, 2015, pp. 1-8.

[117] M. Kumar, P. Sharma and H. Sadawarti, "Measuring Testability of Aspect Oriented Programs", International Conference on Computer Applications and Industrial Electronics, Kuala Lumpur, 2010, pp. 345-349.

[118] J. Fu, M. Lu, S. Yang, and Z. Li, "Method to analyzing software testability affecting factors based on testability tree", Software Technology and Engineering, World Scientific, 2009, p. 206-209.

[119] R.A. Khan and K. Mustafa, "Metric Based Testability Model for Object Oriented Design (MTMOOD) ", ACM SIGSOFT Software Engineering Notes, 2009. 34(2): p. 1-6.

[120] M. Bures, "Metrics for Automated Testability of Web Applications", Proceedings of International Conference on Computer Systems and Technologies, pp. 83-89, 2015.

[121] A. Gonzalez-Sanchez, E. Piel, H. G. Gross and A. J. C. v. Gemund, "Minimising the Preparation Cost of Runtime Testing Based on Testability Metrics", IEEE Annual Computer Software and Applications Conference, Seoul, 2010, pp. 419-424.

[122] M. Jaring, R. L. Krikhaar and J. Bosch, "Modeling Variability and Testability Interaction in Software Product Line Engineering", Seventh International Conference on Composition-Based Software Systems, Madrid, 2008, pp. 120-129.

[123] R. Srivastava and M. Khan, "Modifiability: A Key Factor To Testability", International Journal of Advanced Information Science and Technology, Vol.3, No.6, pp. 85-94, 2014.

[124] A. Jackson and S. Clarke, "MuAspectJ: Mutant Generation to Support Measuring the Testability of AspectJ Programs", Technical Report (TCD-CS-2009-38), ACM,2009.

[125] B. O. Obele and D. Kim, "On an Embedded Software Design Architecture for Improving the Testability of In-vehicle Multimedia Software", IEEE International Conference on Software Testing, Verification and Validation Workshops, Cleveland, OH, 2014, pp. 349-352.

[126] J. Gao, K. Gupta, S. Gupta, and S. Shim, "On Building Testable Software Components", International Conference on COTS-Based Software Systems, pp. 108-121, 2002.

[127] Y. Wang, D. Patel, G. King, G. Staples, M. Ross, and M. Fayad, "On Built-In Test Reuse in Object-Oriented Framework Design", ACM Computing Surveys, 2000, 32(1es): p. 7.

[128] Y. Wang, G. King, M. Ross, and G. Staples, "On Testable Object-Oriented Programming", ACM SIGSOFT Software Engineering Notes, 1997. 22(4): p. 84-90.

[129] E.J. Weyuker, "On Testing Non-Testable Programs", The Computer Journal, 1982. 25(4): p. 465-470.

[130] N. Yevtushenko, A. Petrenko, R. Dssouli, K. Karoui, and S. Prokopenko, "On the Design for Testability of Communication Protocols", Protocol Test Systems, Springer, 1996, p. 265-280.

[131] S. T. Chanson, A. A. F. Loureiro and S. T. Vuong, "On the Design for Testability of Communication Software", Proceedings of IEEE International Test Conference, Baltimore, MD, 1993, pp. 190-199.

[132] L. White and H. K. N. Leung, "On the Edge Regression testability", IEEE Micro, 1992, vol. 12, no. 2, pp. 81-84.

[133] M. Badri, A. Kout and L. Badri, "On the Effect of Aspect-Oriented Refactoring on Testability of Classes: A Case Study", International Conference on Computer Systems and Industrial Informatics, Sharjah, 2012, pp. 1-7.

[134] W. Schutz, "On the Testability of Distributed Real-Time Systems", Proceedings Symposium on Reliable Distributed Systems, Pisa, 1991, pp. 52-61.

[135] A. Bertolino and L. Strigini, "On the Use of Testability Measures for Dependability Assessment", IEEE Transactions on Software Engineering, 1996, vol. 22, no. 2, pp. 97-108.

[136] M. Harman, "Open Problems in Testability Transformation", IEEE International Conference on Software Testing Verification and Validation Workshop, Lillehammer, 2008, pp. 196-209.

[137] J. M. Voas, K. W. Miller and J. E. Payne, "PISCES: A Tool for Predicting Software Testability", Proceedings of Symposium on Assessment of Quality Software Development Tools, New Orleans, LA, 1992, pp. 297-309.

[138] V. Chowdhary, "Practicing Testability in the Real World", International Conference on Software Testing Verification and Validation, Denver, CO, 2009, pp. 260-268.

[139] F. Toure, M. Badri, and L. Lamontagne, "Predicting Different Levels of the Unit Testing Effort of Classes Using Source Code Metrics: A Multiple Case Study on Open-Source Software", Innovations in Systems and Software Engineering, 2017, p. 1-32.

[140] T. Yu, W. Wen, X. Han and J. H. Hayes, "Predicting Testability of Concurrent Programs", IEEE International Conference on Software Testing, Verification and Validation, Chicago, IL, 2016, pp. 168-179.

[141] Y. Singh and A. Saha, "Predicting Testability of Eclipse: A Case Study", Journal of Software Engineering, 2010, 4(2): p. 122-136.

[142] T. M. Khoshgoftaar, E. B. Allen and Z. Xu, "Predicting Testability of Program Modules Using A Neural Network", Proceedings IEEE Symposium on Application-Specific Systems and Software Engineering Technology, Richardson, TX, 2000, pp. 57-62.

[143] J. Vincent, G. King, P. Lay, and J. Kinghorn, "Principles of Built-In-Test for Run-Time-Testability In Component-Based Software Systems", Software Quality Journal, 2002, 10(2): p. 115-133.

[144] Jin-Cherng Lin, Pu-Lin Yeh and Shou-Chia Yang, "Promoting the Software Design for Testability Towards A Partial Test Oracle," Proceedings IEEE International Workshop on Software Technology and Engineering Practice incorporating Computer Aided Software Engineering, London, 1997, pp. 209-214.

[145] J. M. Voas and K. W. Miller, "Putting Assertions in Their Place," Proceedings of IEEE International Symposium on Software Reliability Engineering, Monterey, CA, 1994, pp. 152-157.

[146] B. Baumgarten and H. Wiland, "Qualitative Notions of Testability", in Testing of Communicating Systems, Springer, 1998, p. 349-364.

[147] H. Voigt and G. Engels, "Quality Plans for Measuring Testability of Models", International Conference on Quality Engineering in Software Technology, pp, 353-370 2008, p. 1-16.

[148] M.R. Shaheen and L. du Bousquet, "Quantitative Analysis of Testability Anti-patterns on Open Source Java Applications", Proceedings of Workshop on Quantitative Approaches in Object-Oriented Software Engineering, 2008, p. 21.

[149] S. Sohn and P. Seong, "Quantitative Evaluation of Safety Critical Software Testability Based on Fault Tree Analysis and Entropy", Journal of Systems and Software, 2004, 73(2): p. 351-360.

[150] H.M. Sneed, "Reengineering for Testability", Proceedings of Workshops on Software Engineering, 2006.

[151] M. Harman, "Refactoring as Testability Transformation", IEEE International Conference on Software Testing, Verification and Validation Workshops, Berlin, 2011, pp. 414-421.

[152] E. Martins and V.G. Vieira, "Regression Test Selection for Testable Classes", European Dependable Computing Conference, pp. 453-470, 2005.

[153] V. Izosimov, U. Ingelsson, and A. Wallin, "Requirement Decomposition and Testability in Development of Safety-Critical Automotive Components", Computer Safety, Reliability, and Security, 2012, p. 74-86.

[154] M. Chandrasekharan, B. Dasarathy and Z. Kishimoto, "Requirements-Based Testing of Real-Time Systems: Modeling for Testability", Computer, 1985, vol. 18, no. 4, pp. 71-80.

[155] A. Gonzalez-Sanchez, E. Piel and H. G. Gross, "RiTMO: A Method for Runtime Testability Measurement and Optimisation", International Conference on Quality Software, Jeju, 2009, pp. 377-382.

[156] S. Singh, R. Kumar, and R. Singh, "RTM: A Relation based Testability Metric for Object Oriented Systems", International Journal of Computer Applications, 2013. 67(8), pp. 50-56.

[157] A. Gonzalez-Sanchez, E. Piel, H. G. Gross and A. J. C. van Gemund, "Runtime Testability in Dynamic High-Availability Component-Based Systems", International Conference on Advances in System Testing and Validation Lifecycle, Nice, 2010, pp. 37-42

[158] P. McMinn, "Search-Based Failure Discovery Using Testability Transformations to Generate Pseudo-Oracles", Proceedings of Annual conference on Genetic and evolutionary computation, 2009.

[159] Y. Le Treon, D. Deveaux and J. M. Jezequel, "Self-Testable Components: From Pragmatic Tests To Design-For-Testability Methodology", Proceedings Technology of Object-Oriented Languages and Systems, Nancy, 1999, pp. 96-107.

[160] J.M. Voas and K.W. Miller, "Semantic Metrics for Software Testability", Journal of Systems and Software, 1993, 20 (3): p. 207-216.

[161] P.B. Wilson, "Sizing Software with Testable Requirements", Systems Development Management, 2000, p. 34-10.

[162] W. Albattah, "Software Maintainability and Testability Predictions Using Package Cohesion", Kent State University, 2014.

[163] J. Voas, J. Payne, R. Mills, and J. McManus, "Software Testability: An Experiment in Measuring Simulation Reusability", ACM SIGSOFT Software Engineering Notes, 1995.

[164] H.H.M. Vo, "Software Testability Measure for SAE Architecture Analysis and Design Language (AADL)", MSc thesis, Clemson University, 2012.

[165] J. Fu, M. Lu and B. Liu, "Software Testability Measurement Based on Rough Set Theory," International Conference on Computational Intelligence and Software Engineering, Wuhan, 2009, pp. 1-4.

[166] J.M. Voas, "Software Testability Measurement for Assertion Placement and Fault Localization", Automated and Algorithmic Debugging workshop, 1995.

[167] J. Voas, "Software Testability Measurement for Intelligent Assertion Placement", Software Quality Journal, 1997, 6(4): p. 327-336.

[168] Pu-Lin Yeh and Jin-Cherng Lin, "Software Testability Measurements Derived From Data Flow Analysis", Proceedings of the Euromicro Conference on Software Maintenance and Reengineering, Florence, 1998, pp. 96-102.

[1] J. M. Voas and K. W. Miller, "Software Testability: The New Verification", IEEE Software, 1995, vol. 12, no. 3, pp. 17-28.

[169] K. Karoui, R. Dssouli and O. Cherkaoui, "Specification Transformations and Design for Testability", Global Telecommunications Conference, , London, 1996, pp. 680-685 vol.1.

[170] F. Kluźniak and M. Miłkowska, "Spill—A Logic Language for Writing Testable Requirements Specifications". Science of Computer Programming, 1997, 28(2): p. 193-223.

[171] K. R. Pattipati, S. Deb, M. Dontamsetty and A. Maitra, "START: System Testability Analysis and Research Tool", IEEE Aerospace and Electronic Systems Magazine, 1991, vol. 6, no. 1, pp. 13-20.

[172] J. M. Voas and K. W. Miller, "Substituting Voas's Testability Measure for Musa's Fault Exposure Ratio", IEEE International Conference on Communications, , Dallas, TX, 1996, pp. 230-234 vol.1.

[173] H. P. E. Vranken, M. P. J. Stevens, M. T. M. Segers and J. H. M. M. van Rhee, "System-Level Testability of Hardware/Software Systems", International Test Conference, Washington, DC, 1994, pp. 134-142.

[174] K. R. Pattipati, V. Raghavan, M. Shakeri, S. Deb and R. Shrestha, "TEAMS: Testability Engineering and Maintenance System", American Control Conference, 1994, vol.2. pp. 1989-1995

[175] J.-P. Corriveau, "Testable Requirements for Offshore Outsourcing", International Conference on Software Engineering Approaches for Offshore and Outsourced Development, pp. 27-43, 2007.

[176] W. Grieskamp, M. Lepper, W. Schulte and N. Tillmann, "Testable Use Cases in the Abstract State Machine Language", Proceedings Asia-Pacific Conference on Quality Software, Hong Kong, 2001, pp. 167-172.

[177] Thanh Binh Nguyen, M. Delaunay and C. Robach, "Testability Analysis Applied to Embedded Data-Flow Software", International Conference on Quality Software, 2003, pp. 351-358.

[178] T. B. Nguyen, M. Delaunay and C. Robach, "Testability Analysis for Software Components", International Conference on Software Maintenance, 2002, pp. 422-429.

[179] K. Drira, P. Azéma, and P. de Saqui Sannes, "Testability Analysis in Communicating Systems", Computer Networks, 2001, 36(5): p. 671-693.

[180] T. B. Nguyen, T. B. Dang, M. Delaunay and C. Robach, "Testability Analysis Integrated into Scicos Development Environment", IEEE International Conference on Computing & Communication Technologies, Hanoi, 2010, pp. 1-4.

[181] B. Baudry, Y. Le Traon and G. Sunye, "Testability Analysis of a UML Class Diagram", Proceedings IEEE Symposium on Software Metrics, 2002, pp. 54-63.

[182] Y. Le Traon and C. Robach, "Testability Analysis of Co-Designed Systems", Proceedings of the Asian Test Symposium, Bangalore, 1995, pp. 206-212.

[183] T.B. Nguyen, M. Delaunay, and C. Robach, "Testability Analysis of Data-Flow Software", Electronic Notes in Theoretical Computer Science, 2005, 116: p. 213-225.

[184] Thanh Binh Nguyen, C. Robach and M. Delaunay, "Testability Analysis of Reactive Software", International Workshop on Testability Assessment, IWoTA, 2004, pp. 15-25.

[185] K. Karoui and R. Dssouli, "Testability Analysis of the Communication Protocols Modeled by Relations", University of Montreal, Technical Report No. 1050, 1996.

[186] G. Cantone, A. Cimitile, and U. de Carlini, "Testability and Path Testing Strategies", Microprocessing and Microprogramming, 1987, 21(1-5): p. 371-381.

[187] P.K. Singh, O.P. Sangwan, A. Pratap, and A.P. Singh, "Testability Assessment of Aspect Oriented Software Using Multicriteria Decision Making Approaches", World Applied Sciences Journal, 2014, 32(4): p. 718-730.

[188] H. Singhani and P.R. Suri, "Testability Assessment of Object Oriented Software Using Internal & External Factor Model and Analytic Hierarchy Process", International Journal of Scientific & Engineering Research, Volume 6, pp. 1694-0784, 2015.

[189] M. Nazir and R.A. Khan, "Testability Estimation Model (TEMOOD)", Advances in Computer Science and Information Technology. Computer Science and Information Technology, 2012: p. 178-187.

[190] N. Goel and M. Gupta, "Testability Estimation of Framework Based Applications", Journal of Software Engineering and Applications, 2012, 5(11): p. 841-849.

[191] Z. Al-Khanjari and N. Kraiem, "Testability Guidance Using a Process Modeling", Journal of Software Engineering and Applications, 2013, 6(12): p. 645.

[192] S. Jungmayr, "Testability Measurement and Software Dependencies", Proceedings of the International Workshop on Software Measurement, 2002.

[193] D. Abdullah, R. Srivastava, and M. Khan, "Testability Measurement Framework: Design Phase Perspective", International Journal of Advanced Research in Computer and Communication Engineering, 2014, 3(11): p. 8573-8576.

[194] M. Abdullah and R. Srivastava, "Testability Measurement Model for Object Oriented Design (TMMOOD)", arXiv preprint arXiv:1503.05493, 2015.

[195] Y. Le Traon and C. Robach, "Testability Measurements for Data Flow Designs", Proceedings International Software Metrics Symposium, Albuquerque, NM, 1997, pp. 91-98.

[196] B. Lindström, J. Offutt and S. F. Andler, "Testability of Dynamic Real-Time Systems: An Empirical Study of Constrained Execution Environment Implications", International Conference on Software Testing, Verification, and Validation, Lillehammer, 2008, pp. 112-120.

[197] M. Bruntink, "Testability of Object-Oriented Systems: A Metrics-Based Approach", Master's Thesis, University of Amsterdam, 2003. 45.

[198] P. Khanna, "Testability of Object-Oriented Systems: An AHP-Based Approach for Prioritization of Metrics", International Conference on Contemporary Computing and Informatics, Mysore, 2014, pp. 273-281.

[199] R. S. Freedman, "Testability of software components", IEEE Transactions on Software Engineering, 1991, vol. 17, no. 6, pp. 553-564.

[200] W. T. Tsai, J. Gao, X. Wei and Y. Chen, "Testability of Software in Service-Oriented Architecture", Annual International Computer Software and Applications Conference, Chicago, IL, 2006, pp. 163-170.

[201] M. Harman et al., "Testability transformation", IEEE Transactions on Software Engineering, 2004, vol. 30, no. 1, pp. 3-16.

[202] D. Gong and X. Yao, "Testability Transformation Based On Equivalence of Target Statements", Neural Computing and Applications, 2012, 21(8): p. 1871-1882.

[203] P. McMinn, D. Binkley, and M. Harman, "Testability Transformation for Efficient Automated Test Data Search in the Presence Of Nesting", Proceedings of the UK Software Testing Workshop, 2005.

[204] M. Harman, A. Baresel, D. Binkley, R. Hierons, L. Hu, B. Korel, P. McMinn, and M. Roper, "Testability Transformation–Program Transformation to Improve Testability", Formal methods and testing, 2008: p. 320-344.

[205] K. Saleh, "Testability-Directed Service Definitions and Their Synthesis," International Phoenix Conference on Computers and Communication, Scottsdale, AZ, USA, 1992, pp. 674-678.

[206] M.R. Woodward and Z.A. Al-Khanjari, "Testability, Fault Size and the Domain-to-Range Ratio: An Eternal Triangle", ACM SIGSOFT Software Engineering Notes, 2000, 25(5): p. 168-172.

[207] C. Izurieta and J. M. Bieman, "Testing Consequences of Grime Buildup in Object Oriented Design Patterns", International Conference on Software Testing, Verification, and Validation, Lillehammer, 2008, pp. 171-179.

[208] S. Do Sohn and P.H. Seong, "Testing Digital Safety System Software with a Testability Measure Based on a Software Fault Tree", Reliability Engineering & System Safety, 2006. 91(1): p. 44-52.

[209] L. Ma, H. Wang and Y. Lu, "The Design of Dependency Relationships Matrix to improve the testability of Component-based Software", International Conference on Quality Software (QSIC'06), Beijing, 2006, pp. 93-98.

[210] M. Hays and J. Hayes, "The Effect of Testability on Fault Proneness: A Case Study of the Apache HTTP Server", IEEE International Symposium on Software Reliability Engineering Workshops, Dallas, TX, 2012, pp. 153-158.

[211] B. Baudry, Y. Le Traon, G. Sunye and J. M. Jezequel, "Towards a 'Safe' Use of Design Patterns To Improve OO Software Testability", Proceedings International Symposium on Software Reliability Engineering, 2001, pp. 324-329.

[212] Y. Le Traon and C. Robach, "Towards a Unified Approach to The Testability of Co-Designed Systems", International Symposium on Software Reliability Engineering, Toulouse, 1995, pp. 278-285.

[213] S. Ghosh, "Towards Measurement of Testability of Concurrent Object-Oriented Programs Using Fault Insertion: A Preliminary Investigation", IEEE International Workshop on Source Code Analysis and Manipulation, 2002, pp. 17-25.

[214] R. Dssouli, K. Karoui, A. Petrenko, and O. Rafiq, "Towards Testable Communication Software", Protocol Test Systems, Springer, 1996, p. 237-251.

[215] A. Rodrigues, P.R. Pinheiro, M.M. Rodrigues, A.B. Albuquerque, and F.M. Gonçalves, "Towards the selection of testable use cases and a real experience", International Conference on Communications in Computer and Information Science, 2009, p. 513-521.

[216] A. Tahir, S. G. MacDonell and J. Buchan, "Understanding Class-Level Testability Through Dynamic Analysis", International Conference on Evaluation of Novel Approaches to Software Engineering, Lisbon, Portugal, 2014, pp. 1-10.

[217] J. Voas and L. Kassab, "Using Assertions to Make Untestable Software More Testable", Software Quality Professional, 1999, 1(4): p. 31.

[218] M. Liangli, W. Houxiang and Li Yongjie, "Using Component Metadata based on Dependency Relationships Matrix to improve the Testability of Component-based Software", International Conference on Digital Information Management, Bangalore, 2007, pp. 13-18.

[219] J. Voas, L. Morell, and K. Miller, "Using Dynamic Sensitivity Analysis to Assess Testability", NASA Technical Reports, 1990.

[220] S. Almugrin, W. Albattah, and A. Melton, "Using Indirect Coupling Metrics to Predict Package Maintainability And Testability", Journal of Systems and Software, 2016, 121: p. 298-310.

[221] A. Bertolino and L. Strigini, "Using Testability Measures for Dependability Assessment", International Conference on Software Engineering, Seattle, Washington, USA, 1995, pp. 61-61.

[222] F. Doumbia, O. Laurent, C. Robach and M. Delaunay, "Using the Testability Analysis Methodology for the Validation of AIRBUS Systems", International Conference on Advances in System Testing and Validation Lifecycle, Porto, 2009, pp. 86-91.

[223] M.-C. Shih, "Verification and Measurement of Software Component Testability", 2004.

[224] D. Owen, T. Menzies and B. Cukic, "What makes finite-state models more (or less) testable?", Proceedings IEEE International Conference on Automated Software Engineering, 2002, pp. 237-240.

[225] D. Deveaux, G.S. Denis, and R.K. Keller, " XML Support to Design for Testability", European Conference on Object-Oriented Programming, 2000.

## 9.2 Bibliography

[1] J. M. Voas and K. W. Miller, "Software testability: the new verification," *IEEE Software,* vol. 12, no. 3, pp. 17-28, 1995.

[2] M. Bolton, "More About Testability," *http://www.developsense.com/blog/2014/07/very-short-blog-posts-20-more-about-testability/*, Last accessed: Nov. 25, 2017.

[3] R. V. Binder, "Design for testability in object-oriented systems," *Commun. ACM,* vol. 37, no. 9, pp. 87-101, 1994.

[4] V. Garousi, M. M. Eskandar, and K. Herkiloğlu, "Industry-academia collaborations in software testing: experience and success stories from Canada and Turkey," *Software Quality Journal,* pp. 1-53, 2016.

[5] V. Garousi and J. Zhi, "A survey of software testing practices in Canada," *Journal of Systems and Software,* vol. 86, no. 5, pp. 1354-1376, 2013.

[6] V. Garousi, A. Coşkunçay, A. B. Can, and O. Demirörs, "A survey of software testing practices in Turkey (Türkiye'deki yazılım test uygulamaları anketi)," in *Proceedings of the Turkish National Software Engineering Symposium "Ulusal Yazılım Mühendisliği Sempozyumu" (UYMS)*, 2013.

[7] M. Felderer and F. Auer, "Software Quality Assurance During Implementation: Results of a Survey in Software Houses from Germany, Austria and Switzerland," in *International Conference on Software Quality. Complexity and Challenges of Software Engineering in Emerging Technologies*, 2017, pp. 87-102.

[8] M. Felderer and R. Ramler, "A multiple case study on risk-based testing in industry," *International Journal on Software Tools for Technology Transfer,* vol. 16, no. 5, pp. 609-625, 2014.

[9] Z. Obrenovic, "Insights from the Past: The IEEE Software History Experiment," *IEEE Software,* vol. 34, no. 4, pp. 71-78, 2017.

[10] T. Dyba and T. Dingsoyr, "What Do We Know about Agile Software Development?," *Software, IEEE,* vol. 26, no. 5, pp. 6-9, 2009.

[11] T. Hall, H. Sharp, S. Beecham, N. Baddoo, and H. Robinson, "What Do We Know about Developer Motivation?," *Software, IEEE,* vol. 25, no. 4, pp. 92-94, 2008.

[12] V. Garousi, M. Felderer, Ç. M. Karapıçak, and U. Yılmaz, "What we know about testing embedded software," *IEEE Software, In press,* 2017.

[13] ISO and IEEE, "ISO/IEC/IEEE 24765:2010-Systems and software engineering - Vocabulary," 2010.

[14] ISO, "International standard ISO/IEC 9126-1: 2001: Software engineering -- Product quality -- Part 1: Quality model," 2001.

[15] US Department of Defence, "Military standard MIL-STD-2165: Testability program for electronic systems and equipment " 1985.

[16] ISO, "ISO/IEC 25010:2011-Systems and software engineering -- Systems and software Quality Requirements and Evaluation (SQuaRE) -- System and software quality models," 2011.

[17] ISTQB, "Standard Glossary of Terms Used in Software Testing, Version 3.1," *https://www.istqb.org/downloads/send/20-istqb-glossary/104-glossary-introduction.html*, Last accessed: Dec. 2017.

[18] ISO, "ISO standard 12207:2008-Systems and software engineering -- Software life cycle processes," 2008.

[19] ISO and IEEE, "ISO/IEC/IEEE 29148:2011(E) - ISO/IEC/IEEE International Standard - Systems and software engineering -- Life cycle processes -- Requirements engineering," 2011.

[20] J. A. Caroli, "A Survey of Reliability, Maintainability, Supportability, and Testability Software Tools," *Technical report ADA236148, US Defense Technical Information Center, http://www.dtic.mil/docs/citations/ADA236148*, Last accessed: Nov. 25, 2017.

[21] P. Nikfard, S. B. Ibrahim, B. D. Rohani, H. B. Selamat, and M. N. Mahrin, "An Evaluation for Model Testability approaches," *International journal of computers and technology,* no. 1, pp. 938-947, 2010.

[22] F. Jianping, L. Bin, and L. Minyan, "Present and future of software testability analysis," in *International Conference on Computer Application and System Modeling*, 2010, vol. 15, pp. V15-279-V15-284: IEEE.

[23] M. R. Shaheen and L. Du Bousquet, "Survey of source code metrics for evaluating testability of object oriented systems," *Technical Report RR-LIG-005, Inria France,* 2010.

[24] P. Malla and B. Gurung, "Adaptation of Software Testability Concept for Test Suite Generation: A systematic review," *Master thesis, Blekinge Institute of Technology,* 2012.

[25] P. Nikfard, M. K. Najafabadi, B. D. Rouhani, F. Nikpay, and H. B. Selamat, "An Empirical Study into Model Testability," in *International Conference on Informatics and Creative Multimedia*, 2013, pp. 85-92.

[26] P. R. Suri and H. Singhani, "Object Oriented Software Testability Survey at Designing and Implementation Phase," *International Journal of Science and Research,* vol. 4, no. 4, pp. 3047-3053, 2015.

[27] M. Huda, Y. Arya, and M. Khan, "Measuring Testability of Object Oriented Design: A Systematic Review," *International Journal of Scientific Engineering and Technology,* vol. 3, pp. 1313-1319, 2014.

[28] S. Srivastava, "Object Oriented Design for Testability: A Systematic Review," *International Journal of Advance Research in Computer Science and Management Studies,* vol. 2, no. 10, 2014.

[29] R. Srivastava, N. Dhanda, and S. Lavania, "Testability Estimation of Object Oriented Software: A Systematic Review," *International Journal of Advance Research in Computer Science and Management Studies,* pp. 360-367, 2014.

[30] M. M. Hassan, W. Afzal, M. Blom, B. Lindström, S. F. Andler, and S. Eldh, "Testability and software robustness: A systematic literature review," in *Euromicro Conference on Software Engineering and Advanced Applications*, 2015, pp. 341-348.

[31] P. R. Suri and H. Singhani, "Object Oriented Software Testability Metrics Analysis," *International Journal of Computer Applications Technology and Research,* vol. 4, no. 5, pp. 359-367, 2015.

[32] M. M. Hassan, W. Afzal, B. Lindström, S. M. A. Shah, S. F. Andler, and M. Blom, "Testability and software performance: a systematic mapping study," in *Proceedings of ACM Symposium on Applied Computing*, 2016, pp. 1566-1569: ACM.

[33] E. İ. Hanoğlu, F. N. Kılıçaslan, A. Tarhan, and V. Garousi, "Software testability: a systematic literature mapping," in *Proceedings of the Turkish National Software Engineering Symposium "Ulusal Yazılım Mühendisliği Sempozyumu" (UYMS)*, 2016, pp. 260-271.

[34] J. Zhi, V. Garousi, B. Sun, G. Garousi, S. Shahnewaz, and G. Ruhe, "Cost, benefits and quality of software development documentation: a systematic mapping," *Journal of Systems and Software,* vol. 99, pp. 175–198, 2015.

[35] V. Garousi, Y. Amannejad, and A. Betin-Can, "Software test-code engineering: a systematic mapping," *Journal of Information and Software Technology,* vol. 58, pp. 123–147, 2015.

[36] S. Doğan, A. Betin-Can, and V. Garousi, "Web application testing: a systematic literature review," *Journal of Systems and Software,* vol. 91, pp. 174-201, 2014.

[37] F. Häser, M. Felderer, and R. Breu, "Software paradigms, assessment types and non-functional requirements in model-based integration testing: a systematic literature review," presented at the Proceedings of the International Conference on Evaluation and Assessment in Software Engineering, 2014.

[38] M. Felderer, P. Zech, R. Breu, M. Büchler, and A. Pretschner, "Model-based security testing: a taxonomy and systematic classification," *Software Testing, Verification and Reliability,* 2015.

[39] M. Felderer and E. Fourneret, "A systematic classification of security regression testing approaches," *International Journal on Software Tools for Technology Transfer,* vol. 17, no. 3, pp. 305-319, 2015.

[40] K. Petersen, S. Vakkalanka, and L. Kuzniarz, "Guidelines for conducting systematic mapping studies in software engineering: An update," *Information and Software Technology,* vol. 64, pp. 1-18, 2015.

[41] C. Wohlin, "Guidelines for snowballing in systematic literature studies and a replication in software engineering," presented at the Proceedings of the 18th International Conference on Evaluation and Assessment in Software Engineering, London, England, United Kingdom, 2014.

[42] K. Petersen, R. Feldt, S. Mujtaba, and M. Mattsson, "Systematic mapping studies in software engineering," presented at the International Conference on Evaluation and Assessment in Software Engineering (EASE), 2008.

[43] B. Kitchenham and S. Charters, "Guidelines for Performing Systematic Literature Reviews in Software engineering," *Technical report, School of Computer Science, Keele University, EBSE-2007-01,* 2007.

[44] V. Garousi and M. V. Mäntylä, "Citations, research topics and active countries in software engineering: A bibliometrics study " *Elsevier Computer Science Review,* vol. 19, pp. 56-77, 2016.

[45] V. Garousi, "A bibliometric analysis of the Turkish software engineering research community," *Springer Journal on Scientometrics,* vol. 105, no. 1, pp. 23-49, 2015.

[46] N. R. Haddaway, A. M. Collins, D. Coughlin, and S. Kirk, "The Role of Google Scholar in Evidence Reviews and Its Applicability to Grey Literature Searching," *PLoS ONE,* vol. 10, no. 9, 2015.

[47] K. Godin, J. Stapleton, S. I. Kirkpatrick, R. M. Hanning, and S. T. Leatherdale, "Applying systematic review search methods to the grey literature: a case study examining guidelines for school-based breakfast programs in Canada," (in Eng), *Systematic Reviews,* vol. 4, pp. 138-148, Oct 22 2015.

[48] Q. Mahood, D. Van Eerd, and E. Irvin, "Searching for grey literature for systematic reviews: challenges and benefits," *Research Synthesis Methods,* vol. 5, no. 3, pp. 221-234, 2014.

[49] J. Adams *et al.*, "Searching and synthesising 'grey literature' and 'grey information' in public health: critical reflections on three case studies," *Systematic Reviews,* journal article vol. 5, no. 1, p. 164, 2016.

[50] V. Garousi and M. V. Mäntylä, "When and what to automate in software testing? A multivocal literature review," *Information and Software Technology,* vol. 76, pp. 92-117, 2016.

[51] V. Garousi, M. Felderer, and T. Hacaloğlu, "Software test maturity assessment and test process improvement: A multivocal literature review," *Information and Software Technology,* vol. 85, pp. 16–42, 2017.

[52] V. Garousi, M. Felderer, and F. N. Kılıçaslan, "Online dataset for: What we know about software testability: a systematic classification of the literature," www.goo.gl/boNuFD, Last accessed: Nov. 25, 2017.

[53] A. Sabane, "Improving System Testability and Testing with Microarchitectures," in *Working Conference on Reverse Engineering*, 2010, pp. 309-312.

[54] V. Garousi, A. Mesbah, A. Betin-Can, and S. Mirshokraie, "A systematic mapping study of web application testing," *Elsevier Journal of Information and Software Technology,* vol. 55, no. 8, pp. 1374–1396, 2013.

[55] I. Banerjee, B. Nguyen, V. Garousi, and A. Memon, "Graphical User Interface (GUI) Testing: Systematic Mapping and Repository," *Information and Software Technology,* vol. 55, no. 10, pp. 1679–1694, 2013.

[56] M. Harman, S. A. Mansouri, and Y. Zhang, "Search-based software engineering: Trends, techniques and applications," *ACM Comput. Surv.,* vol. 45, no. 1, pp. 1-61, 2012.

[57] Y. Jia and M. Harman, "An Analysis and Survey of the Development of Mutation Testing," *IEEE Transactions on Software Engineering,* vol. 37, no. 5, pp. 649-678, 2011.

[58] R. Binder, "GTAC 2010: What Testability Tells Us About the Software Performance Envelope," https://www.youtube.com/watch?v=1keyEiJHqPw, Last accessed: June 2018.

[59] M. Hevery, "Design Tech Talk Series Presents: OO Design for Testability," https://www.youtube.com/watch?v=acjvKJiOvXw, Last accessed: June 2018.

[60] IEEE., "IEEE Recommended Practice for Software Requirements Specifications 830-1998," http://standards.ieee.org/findstds/standard/830-1998.html, Last accessed: Nov. 25, 2017.

[61] K. Kirk, *Writing for Readability*. American Society for Training and Development, 2010.

[62] Z. Li, P. Avgeriou, and P. Liang, "A systematic mapping study on technical debt and its management," *Journal of Systems and Software,* vol. 101, pp. 193-220, 2015.

[63] SonarSource S.A., "Sonar SQALE 1.2 in screenshot," *https://blog.sonarsource.com/sonar-sqale-1-2-in-screenshot/*, Last accessed: Nov. 25, 2017.

[64] J. L. Letouzey and M. Ilkiewicz, "Managing Technical Debt with the SQALE Method," *IEEE Software,* vol. 29, no. 6, pp. 44-51, 2012.

[65] M. Kedemo, "Testability Awakens: Moving Testability into New Dimensions," *Testing Trapeze magazine,* pp. 31-37, December 2015.

[66] S. Easterbrook, J. Singer, M.-A. Storey, and D. Damian, "Selecting Empirical Methods for Software Engineering Research," in *Guide to Advanced Empirical Software Engineering*, F. Shull, J. Singer, and D. K. Sjøberg, Eds.: Springer London, 2008, pp. 285-311.

[67] F. Shull, T. Dybå, H. Sharp, and R. Prikladnicki, "Voice of Evidence: A Look Back," *IEEE Software,* vol. 34, no. 4, pp. 23-25, 2017.

[68] V. Garousi, K. Petersen, and B. Özkan, "Challenges and best practices in industry-academia collaborations in software engineering: a systematic literature review," *Information and Software Technology,* vol. 79, pp. 106–127, 2016.

[69] A. Beer and M. Felderer, "Measuring and improving testability of system requirements in an industrial context by applying the goal question metric approach," in *Proceedings of the International Workshop on Requirements Engineering and Testing*, 2018, pp. 25-32.

[70] M. Shaw, "What makes good research in software engineering?," *International Journal on Software Tools for Technology Transfer,* vol. 4, no. 1, pp. 1-7, 2002.

[71] A. Bryman, "The research question in social research: what is its role?," *International Journal of Social Research Methodology,* vol. 10, no. 1, pp. 5-20, 2007.

[72] S. B. Hulley, S. R. Cummings, W. S. Browner, D. G. Grady, N. Hearst, and T. Newman, "Conceiving the research question," *Designing clinical research,* p. 335, 2001.

[73] E. E. Lipowski, "Developing great research questions," *American Journal of Health-System Pharmacy,* vol. 65, no. 17, pp. 1667-1670, 2008.

[74] M. Felderer, B. Marculescu, F. G. d. O. Neto, R. Feldt, and R. Torkar, "A Testability Analysis Framework for Non-Functional Properties," *arXiv preprint arXiv:1802.07140,* 2018.

[75] C. Wohlin, P. Runeson, M. Höst, M. C. Ohlsson, B. Regnell, and A. Wesslén, *Experimentation in Software Engineering: An Introduction*. Kluwer Academic Publishers, 2000.

[76] V. Garousi, M. Felderer, M. Kuhrmann, and K. Herkiloğlu, "What industry wants from academia in software testing? Hearing practitioners' opinions," in *International Conference on Evaluation and Assessment in Software Engineering*, Karlskrona, Sweden, 2017, pp. 65-69.

[77] I. Rodriguez, L. Llana, and P. Rabanal, "A General Testability Theory: Classes, properties, complexity, and testing reductions," *IEEE Transactions on Software Engineering,* vol. 40, no. 9, pp. 862-894, 2014.

[78] T. Kanstren, "A study on design for testability in component-based embedded software," in *Software Engineering Research, Management and Applications, 2008. SERA'08. Sixth International Conference on*, 2008, pp. 31-38: IEEE.

[79] A. Lundberg, "Dependency Injection frameworks: an improvement to testability?," ed, 2015.

[80] B. Pettichord, "Design for testability," in *Pacific Northwest Software Quality Conference*, 2002, pp. 1-28.

[81] H. P. Vranken, M. F. Witteman, and R. C. Van Wuijtswinkel, "Design for testability in hardware software systems," *IEEE Design & Test of Computers,* vol. 13, no. 3, pp. 79-86, 1996.

[82] R. V. Binder, "Design for testability in object-oriented systems," *Communications of the ACM,* vol. 37, no. 9, pp. 87-101, 1994.

[83] M. Kim, S. T. Chanson, and S. Yoo, "Design for testability of protocols based on formal specifications," in *Protocol Test Systems VIII*: Springer, 1996, pp. 252-264.

[84] M. Nakazato, S. Ohtake, M. Inoue, and H. Fujiwara, "Design for testability of software-based self-test for processors," in *Test Symposium, 2006. ATS'06. 15th Asian*, 2006, pp. 375-380: IEEE.

[85] H. König, A. Ulrich, and M. Heiner, "Design for testability: a step-wise approach to protocol testing," in *Testing of Communicating systems*: Springer, 1997, pp. 125-140.

[86] Y.-T. Lin, "Design of software components with increased testability," 2004.

[87] K. L. Bellman and C. Landauer, "Designing testable, heterogeneous software environments," *Journal of Systems and Software,* vol. 29, no. 3, pp. 199-217, 1995.

[88] J. E. Payne, R. T. Alexander, and C. D. Hutchinson, "Design-for-testability for object-oriented software," *Object Magazine,* vol. 7, no. 5, pp. 34-43, 1997.

[89] T. M. Khoshgoftaar, R. M. Szabo, and J. M. Voas, "Detecting program modules with low testability," in *Software Maintenance, 1995. Proceedings., International Conference on*, 1995, pp. 242-250: IEEE.

[90] M. Badri and F. Toure, "Empirical Analysis for Investigating the Effect of Control Flow Dependencies on Testability of Classes," in *SEKE*, 2011, pp. 475-480.

[91] P. McMinn, D. Binkley, and M. Harman, "Empirical evaluation of a nesting testability transformation for evolutionary testing," *ACM Transactions on Software Engineering and Methodology (TOSEM),* vol. 18, no. 3, p. 11, 2009.

[92] F. Lammermann, A. Baresel, and J. Wegener, "Evaluating evolutionary testability for structure-oriented testing with software measurements," *Applied Soft Computing,* vol. 8, no. 2, pp. 1018-1028, 2008.

[93] A. Baresel, D. Binkley, M. Harman, and B. Korel, "Evolutionary testing in the presence of loop-assigned flags: A testability transformation approach," in *ACM SIGSOFT Software Engineering Notes*, 2004, vol. 29, no. 4, pp. 108-118: ACM.

[94] L. Badri, M. Badri, and F. Toure, "Exploring empirically the relationship between lack of cohesion and testability in object-oriented systems," in *International Conference on Advanced Software Engineering and Its Applications*, 2010, pp. 78-92: Springer.

[95] J. M. Voas, "Factors that affect software testability," 1991.

[96] M. Khan and R. Srivastava, "FLEXIBILITY: A KEY FACTOR TO TESTABILITY," *International Journal of Software Engineering & Applications,* vol. 6, no. 1, p. 89, 2015.

[97] F. MARTENSSON, H. Grahn, and M. Mattsson, "Forming consensus on testability in software developing organizations," in *Fifth Conference on Software Engineering Research and Practice in Sweden*, 2005, pp. 31-38.

[98] Y. Le Traon and C. Robach, "From hardware to software testability," in *Test Conference, 1995. Proceedings., International*, 1995, pp. 710-719: IEEE.

[99] J. Bach, "Heuristics of software testability," 1999.

[100] R. Coelho, U. Kulesza, and A. Von Staa, "Improving architecture testability with patterns," in *Companion to the 20th annual ACM SIGPLAN conference on Object-oriented programming, systems, languages, and applications*, 2005, pp. 114-115: ACM.

[101] H. Yin and J. M. Bieman, "Improving software testability with assertion insertion," in *Test Conference, 1994. Proceedings., International*, 1994, pp. 831-839: IEEE.

[102] M. C. Benton and N. M. Radziwill, "Improving Testability and Reuse by Transitioning to Functional Programming," *arXiv preprint arXiv:1606.06704,* 2016.

[103] Y. Singh and A. Saha, "Improving the testability of object oriented software through software contracts," *ACM SIGSOFT Software Engineering Notes,* vol. 35, no. 1, pp. 1-4, 2010.

[104] S. Khatri, R. S. Chhillar, and V. Singh, "Improving the testability of object-oriented software during testing and debugging processes," *arXiv preprint arXiv:1308.3320,* 2013.

[105] I. Cabral, M. B. Cohen, and G. Rothermel, "Improving the Testing and Testability of Software Product Lines," in *SPLC*, 2010, vol. 6287, pp. 241-255: Springer.

[106] N. Alshahwan, M. Harman, A. Marchetto, and P. Tonella, "Improving web application testing using testability measures," in *Web Systems Evolution (WSE), 2009 11th IEEE International Symposium on*, 2009, pp. 49-58: IEEE.

[107] D. Deveaux, P. Frison, and J.-M. Jezequel, "Increase software trustability with self-testable classes in java," in *Software Engineering Conference, 2001. Proceedings. 2001 Australian*, 2001, pp. 3-11: IEEE.

[108] S. Kansomkeat, J. Offutt, and W. Rivepiboon, "Increasing Class-Component Testability," in *IASTED Conf. on Software Engineering*, 2005, pp. 156-161.

[109] E. A. Esker, J.-P. Martin, W. R. Simpson, and J. W. Sheppard, "Integrated design for testability and automatic testing approaches," in *AUTOTESTCON'90. IEEE Systems Readiness Technology Conference.'Advancing Mission Accomplishment', Conference Record.*, 1990, pp. 509-514: IEEE.

[110] Z. Al-Khanjari and M. Woodward, "Investigating the partial relationships between testability and the dynamic range-to-domain ratio," *Australasian Journal of Information Systems,* vol. 11, no. 1, 2003.

[111] L. C. Briand, Y. Labiche, and H. Sun, "Investigating the use of analysis contracts to improve the testability of object-oriented code," *Software: Practice and Experience,* vol. 33, no. 7, pp. 637-672, 2003.

[112] R. Bache and M. Müllerburg, "Measures of testability as a basis for quality assurance," *Software Engineering Journal,* vol. 5, no. 2, pp. 86-92, 1990.

[113] B. Baudry, Y. Traon, G. Sunyé, and J.-M. Jézéquel, "Measuring and improving design patterns testability," in *Software Metrics Symposium, 2003. Proceedings. Ninth International*, 2003, pp. 50-59: IEEE.

[114] B. Baudry and Y. Le Traon, "Measuring design testability of a UML class diagram," *Information and software technology,* vol. 47, no. 13, pp. 859-879, 2005.

[115] H.-G. Gross, "Measuring evolutionary testability of real-time software," University of Glamorgan, 2000.

[116] J. H. Hayes, W. Li, T. Yu, X. Han, M. Hays, and C. Woodson, "Measuring Requirement Quality to Predict Testability," in *Artificial Intelligence for Requirements Engineering (AIRE), 2015 IEEE Second International Workshop on*, 2015, pp. 1-8: IEEE.

[117] M. Kumar, P. Sharma, and H. Sadawarti, "Measuring testability of aspect oriented programs," in *Computer Applications and Industrial Electronics (ICCAIE), 2010 International Conference on*, 2010, pp. 345-349: IEEE.

[118] J. FU, M. LU, S. YANG, and Z. LI, "Method to analyzing software testability affecting factors based on testability tree," in *Software Technology And Engineering*: World Scientific, 2009, pp. 206-209.

[119] R. A. Khan and K. Mustafa, "Metric based testability model for object oriented design (MTMOOD)," *ACM SIGSOFT Software Engineering Notes,* vol. 34, no. 2, pp. 1-6, 2009.

[120] M. Bures, "Metrics for automated testability of web applications," in *Proceedings of the 16th International Conference on Computer Systems and Technologies*, 2015, pp. 83-89: ACM.

[121] A. Gonzalez-Sanchez, E. Piel, H.-G. Gross, and A. J. van Gemund, "Minimising the preparation cost of runtime testing based on testability metrics," in *Computer Software and Applications Conference (COMPSAC), 2010 IEEE 34th Annual*, 2010, pp. 419-424: IEEE.

[122] M. Jaring, R. L. Krikhaar, and J. Bosch, "Modeling variability and testability interaction in software product line engineering," in *Composition-Based Software Systems, 2008. ICCBSS 2008. Seventh International Conference on*, 2008, pp. 120-129: IEEE.

[123] R. Srivastava and M. Khan, "Modifiability: A Key Factor To Testability."

[124] A. Jackson and S. Clarke, "MuAspectJ: Mutant Generation to Support Measuring the Testability of AspectJ Programs," Technical report (TCD-CS-2009-38), ACM2009.

[125] B. O. Obele and D. Kim, "On an Embedded Software Design Architecture for Improving the Testability of In-vehicle Multimedia Software," in *Software Testing, Verification and Validation Workshops (ICSTW), 2014 IEEE Seventh International Conference on*, 2014, pp. 349-352: IEEE.

[126] J. Gao, K. Gupta, S. Gupta, and S. Shim, "On building testable software components," in *ICCBSS*, 2002, vol. 2, pp. 108-121: Springer.

[127] Y. Wang, D. Patel, G. King, G. Staples, M. Ross, and M. Fayad, "On built-in test reuse in object-oriented framework design," *ACM Computing Surveys (CSUR),* vol. 32, no. 1es, p. 7, 2000.

[128] Y. Wang, G. King, M. Ross, and G. Staples, "On testable object-oriented programming," *ACM SIGSOFT Software Engineering Notes,* vol. 22, no. 4, pp. 84-90, 1997.

[129] E. J. Weyuker, "On testing non-testable programs," *The Computer Journal,* vol. 25, no. 4, pp. 465-470, 1982.

[130] N. Yevtushenko, A. Petrenko, R. Dssouli, K. Karoui, and S. Prokopenko, "On the design for testability of communication protocols," in *Protocol Test Systems VIII*: Springer, 1996, pp. 265-280.

[131] S. T. Chanson, A. A. F. Loureiro, and S. T. Vuong, "On the design for testability of communication software," in *Test Conference, 1993. Proceedings., International*, 1993, pp. 190-199: IEEE.

[132] L. White and H. K. Leung, "On the edge. Regression testability," *IEEE Micro,* vol. 12, no. 2, pp. 81-84, 1992.

[133] M. Badri, A. Kout, and L. Badri, "On the effect of aspect-oriented refactoring on testability of classes: A case study," in *Computer Systems and Industrial Informatics (ICCSII), 2012 International Conference on*, 2012, pp. 1-7: IEEE.

[134] W. Schutz, "On the testability of distributed real-time systems," in *Reliable Distributed Systems, 1991. Proceedings., Tenth Symposium on*, 1991, pp. 52-61: IEEE.

[135] A. Bertolino and L. Strigini, "On the use of testability measures for dependability assessment," *IEEE Transactions on Software Engineering,* vol. 22, no. 2, pp. 97-108, 1996.

[136] M. Harman, "Open problems in testability transformation," in *Software Testing Verification and Validation Workshop, 2008. ICSTW'08. IEEE International Conference on*, 2008, pp. 196-209: IEEE.

[137] J. M. Voas, K. W. Miller, and J. E. Payne, "PISCES: A tool for predicting software testability," in *Assessment of Quality Software Development Tools, 1992., Proceedings of the Second Symposium on*, 1992, pp. 297-309: IEEE.

[138] V. Chowdhary, "Practicing testability in the real world," in *Software Testing Verification and Validation, 2009. ICST'09. International Conference on*, 2009, pp. 260-268: IEEE.

[139] F. Toure, M. Badri, and L. Lamontagne, "Predicting different levels of the unit testing effort of classes using source code metrics: a multiple case study on open-source software," *Innovations in Systems and Software Engineering,* pp. 1-32, 2017.

[140] T. Yu, W. Wen, X. Han, and J. H. Hayes, "Predicting Testability of Concurrent Programs," in *Software Testing, Verification and Validation (ICST), 2016 IEEE International Conference on*, 2016, pp. 168-179: IEEE.

[141] Y. Singh and A. Saha, "Predicting testability of eclipse: a case study," *Journal of Software Engineering,* vol. 4, no. 2, pp. 122-136, 2010.

[142] T. M. Khoshgoftaar, E. B. Allen, and Z. Xu, "Predicting testability of program modules using a neural network," in *Application-Specific Systems and Software Engineering Technology, 2000. Proceedings. 3rd IEEE Symposium on*, 2000, pp. 57-62: IEEE.

[143] J. Vincent, G. King, P. Lay, and J. Kinghorn, "Principles of built-in-test for run-time-testability in component-based software systems," *Software Quality Journal,* vol. 10, no. 2, pp. 115-133, 2002.

[144] J.-C. Lin, P.-l. Yeh, and S.-C. Yang, "Promoting the software design for testability towards a partial test oracle," in *Software Technology and Engineering Practice, 1997. Proceedings., Eighth IEEE International Workshop on [incorporating Computer Aided Software Engineering]*, 1997, pp. 209-214: IEEE.

[145] J. M. Voas and K. W. Miller, "Putting assertions in their place," in *Software Reliability Engineering, 1994. Proceedings., 5th International Symposium on*, 1994, pp. 152-157: IEEE.

[146] B. Baumgarten and H. Wiland, "Qualitative notions of testability," in *Testing of Communicating Systems*: Springer, 1998, pp. 349-364.

[147] H. Voigt and G. Engels, "Quality plans for measuring testability of models," *Software Engineering*, pp. 1-16, 2008.

[148] M. R. Shaheen and L. du Bousquet, "Quantitative analysis of testability antipatterns on open source java applications," *QAOOSE 2008-Proceedings*, p. 21, 2008.

[149] S. Sohn and P. Seong, "Quantitative evaluation of safety critical software testability based on fault tree analysis and entropy," *Journal of Systems and Software*, vol. 73, no. 2, pp. 351-360, 2004.

[150] H. M. Sneed, "Reengineering for testability," in *Proceedings des Workshops zum Software Engineering, Seite 31ff, Bad Honnef*, 2006.

[151] M. Harman, "Refactoring as testability transformation," in *Software Testing, Verification and Validation Workshops (ICSTW), 2011 IEEE Fourth International Conference on*, 2011, pp. 414-421: IEEE.

[152] E. Martins and V. G. Vieira, "Regression Test Selection for Testable Classes," in *EDCC*, 2005, pp. 453-470: Springer.

[153] V. Izosimov, U. Ingelsson, and A. Wallin, "Requirement decomposition and testability in development of safety-critical automotive components," *Computer Safety, Reliability, and Security*, pp. 74-86, 2012.

[154] M. Chandrasekharan, B. Dasarathy, and Z. Kishimoto, "Requirements-Based Testing of Real-Time Systems. Modeling for Testability," *Computer*, no. 4, pp. 71-80, 1985.

[155] A. Gonzalez-Sanchez, E. Piel, and H.-G. Gross, "RiTMO: A method for runtime testability measurement and optimisation," in *Quality Software, 2009. QSIC'09. 9th International Conference on*, 2009, pp. 377-382: IEEE.

[156] S. Singh, R. Kumar, and R. Singh, "RTM: A Relation based Testability Metric for Object Oriented Systems," *International Journal of Computer Applications*, vol. 67, no. 8, 2013.

[157] A. Gonzalez-Sanchez, E. Piel, H.-G. Gross, and A. J. van Gemund, "Runtime testability in dynamic high-availability component-based systems," in *Advances in System Testing and Validation Lifecycle (VALID), 2010 Second International Conference on*, 2010, pp. 37-42: IEEE.

[158] P. McMinn, "Search-based failure discovery using testability transformations to generate pseudo-oracles," in *Proceedings of the 11th Annual conference on Genetic and evolutionary computation*, 2009, pp. 1689-1696: ACM.

[159] Y. Le Treon, D. Deveaux, and J.-M. Jézéquel, "Self-testable components: from pragmatic tests to design-for-testability methodology," in *Technology of Object-Oriented Languages and Systems, 1999. Proceedings of*, 1999, pp. 96-107: IEEE.

[160] J. M. Voas and K. W. Miller, "Semantic metrics for software testability," *Journal of Systems and Software*, vol. 20, no. 3, pp. 207-216, 1993.

[161] P. B. Wilson, "Sizing Software with Testable Requirements," *Systems Development Management*, pp. 34-10, 2000.

[162] W. Albattah, "Software Maintainability and Testability Predictions Using Package Cohesion," Kent State University, 2014.

[163] J. Voas, J. Payne, R. Mills, and J. McManus, "Software testability: An experiment in measuring simulation reusability," in *ACM SIGSOFT Software Engineering Notes*, 1995, vol. 20, no. SI, pp. 247-255: ACM.

[164] H. H. M. Vo, "Software testability measure for SAE Architecture Analysis and Design Language (AADL)," Clemson University, 2012.

[165] J. Fu, M. Lu, and B. Liu, "Software Testability Measurement Based on Rough Set Theory," in *Computational Intelligence and Software Engineering, 2009. CiSE 2009. International Conference on*, 2009, pp. 1-4: IEEE.

[166] J. M. Voas, "Software Testability Measurement for Assertion Placement and Fault Localization," in *AADEBUG*, 1995, pp. 133-144.

[167] J. Voas, "Software testability measurement for intelligent assertion placement," *Software Quality Journal*, vol. 6, no. 4, pp. 327-336, 1997.

[168] P.-L. Yeh and J.-C. Lin, "Software testability measurements derived from data flow analysis," in *Software Maintenance and Reengineering, 1998. Proceedings of the Second Euromicro Conference on*, 1998, pp. 96-102: IEEE.

[169] K. Karoui, R. Dssouli, and O. Cherkaoui, "Specification transformations and design for testability," in *Global Telecommunications Conference, 1996. GLOBECOM'96.'Communications: The Key to Global Prosperity*, 1996, vol. 1, pp. 680-685: IEEE.

[170] F. Kluźniak and M. Miłkowska, "Spill—a logic language for writing testable requirements specifications," *Science of Computer Programming*, vol. 28, no. 2, pp. 193-223, 1997.

[171] K. R. Pattipati, S. Deb, M. Dontamsetty, and A. Maitra, "Start: System testability analysis and research tool," *IEEE Aerospace and Electronic Systems Magazine*, vol. 6, no. 1, pp. 13-20, 1991.

[172] J. M. Voas and K. W. Miller, "Substituting Voas's testability measure for Musa's fault exposure ratio," in *Communications, 1996. ICC'96, Conference Record, Converging Technologies for Tomorrow's Applications. 1996 IEEE International Conference on*, 1996, vol. 1, pp. 230-234: IEEE.

[173] H. P. Vranken, M. Stevens, M. Segers, and J. Van Rhee, "System-level testability of hardware/software systems," in *Test Conference, 1994. Proceedings., International*, 1994, pp. 134-142: IEEE.

[174] K. Pattipati, V. Raghavan, M. Shakeri, S. Deb, and R. Shrestha, "TEAMS: testability engineering and maintenance system," in *American Control Conference, 1994*, 1994, vol. 2, pp. 1989-1995: IEEE.

[175] J.-P. Corriveau, "Testable requirements for offshore outsourcing," in *SEAFOOD*, 2007, pp. 27-43: Springer.

[176] W. Grieskamp, M. Lepper, W. Schulte, and N. Tillmann, "Testable use cases in the abstract state machine language," in *Quality Software, 2001. Proceedings. Second Asia-Pacific Conference on*, 2001, pp. 167-172: IEEE.

[177] T. B. Nguyen, M. Delaunay, and C. Robach, "Testability analysis applied to embedded data-flow software," in *Quality Software, 2003. Proceedings. Third International Conference on*, 2003, pp. 351-358: IEEE.

[178] T. B. Nguyen, M. Delaunay, and C. Robach, "Testability analysis for software components," in *Software Maintenance, 2002. Proceedings. International Conference on*, 2002, pp. 422-429: IEEE.

[179] K. Drira, P. Azéma, and P. de Saqui Sannes, "Testability analysis in communicating systems," *Computer Networks,* vol. 36, no. 5, pp. 671-693, 2001.

[180] T. B. Nguyen, T. B. Dang, M. Delaunay, and C. Robach, "Testability Analysis Integrated into Scicos Development Environment," in *Computing and Communication Technologies, Research, Innovation, and Vision for the Future (RIVF), 2010 IEEE RIVF International Conference on*, 2010, pp. 1-4: IEEE.

[181] B. Baudry, Y. Le Traon, and G. Sunyé, "Testability analysis of a UML class diagram," in *Software Metrics, 2002. Proceedings. Eighth IEEE Symposium on*, 2002, pp. 54-63: IEEE.

[182] Y. Le Traon and C. Robach, "Testability analysis of co-designed systems," in *Test Symposium, 1995., Proceedings of the Fourth Asian*, 1995, pp. 206-212: IEEE.

[183] T. B. Nguyen, M. Delaunay, and C. Robach, "Testability analysis of data-flow software," *Electronic Notes in Theoretical Computer Science,* vol. 116, pp. 213-225, 2005.

[184] T. B. Nguyen, C. Robach, and M. Delaunay, "Testability analysis of reactive software," in *Testability Assessment, 2004. IWoTA 2004. Proceedings. First International Workshop on*, 2004, pp. 15-25: IEEE.

[185] K. Karoui and R. Dssouli, "Testability analysis of the communication protocols modeled by relations," *Technical Report No. 1050,* vol. 1050, 1996.

[186] G. Cantone, A. Cimitile, and U. de Carlini, "Testability and path testing strategies," *Microprocessing and Microprogramming,* vol. 21, no. 1-5, pp. 371-381, 1987.

[187] P. K. Singh, O. P. Sangwan, A. Pratap, and A. P. Singh, "Testability assessment of aspect oriented software using multicriteria decision making approaches," *World Applied Sciences Journal,* vol. 32, no. 4, pp. 718-730, 2014.

[188] H. Singhani and P. R. Suri, "Testability Assessment of Object Oriented Software Using Internal & External Factor Model and Analytic Hierarchy Process."

[189] M. Nazir and R. A. Khan, "Testability Estimation Model (TEM OOD)," *Advances in Computer Science and Information Technology. Computer Science and Information Technology,* pp. 178-187, 2012.

[190] N. Goel and M. Gupta, "Testability estimation of framework based applications," *Journal of Software Engineering and Applications,* vol. 5, no. 11, p. 841, 2012.

[191] Z. Al-Khanjari and N. Kraiem, "Testability Guidance Using a Process Modeling," *Journal of Software Engineering and Applications,* vol. 6, no. 12, p. 645, 2013.

[192] S. Jungmayr, "Testability measurement and software dependencies," in *Proceedings of the 12th International Workshop on Software Measurement*, 2002, vol. 25, no. 9, pp. 179-202.

[193] D. Abdullah, R. Srivastava, and M. Khan, "Testability Measurement Framework: Design Phase Perspective," *International Journal of Advanced Research in Computer and Communication Engineering,* vol. 3, no. 11, pp. 8573-8576, 2014.

[194] M. Abdullah and R. Srivastava, "Testability Measurement Model for Object Oriented Design (TMMOOD)," *arXiv preprint arXiv:1503.05493,* 2015.

[195] Y. Le Traon and C. Robach, "Testability measurements for data flow designs," in *Software Metrics Symposium, 1997. Proceedings., Fourth International*, 1997, pp. 91-98: IEEE.

[196] B. Lindström, J. Offutt, and S. F. Andler, "Testability of dynamic real-time systems: An empirical study of constrained execution environment implications," in *Software Testing, Verification, and Validation, 2008 1st International Conference on*, 2008, pp. 112-120: IEEE.

[197] M. Bruntink, "Testability of object-oriented systems: a metrics-based approach," *Universiy Van Amsterdam,* vol. 45, 2003.

[198] P. Khanna, "Testability of object-oriented systems: An AHP-based approach for prioritization of metrics," in *Contemporary Computing and Informatics (IC3I), 2014 International Conference on*, 2014, pp. 273-281: IEEE.

[199] R. S. Freedman, "Testability of software components," *IEEE transactions on Software Engineering,* vol. 17, no. 6, pp. 553-564, 1991.

[200] W.-T. Tsai, J. Gao, X. Wei, and Y. Chen, "Testability of software in service-oriented architecture," in *Computer Software and Applications Conference, 2006. COMPSAC'06. 30th Annual International*, 2006, vol. 2, pp. 163-170: IEEE.

[201] M. Harman *et al.*, "Testability transformation," *IEEE Transactions on Software Engineering,* vol. 30, no. 1, pp. 3-16, 2004.

[202] D. Gong and X. Yao, "Testability transformation based on equivalence of target statements," *Neural Computing and Applications,* vol. 21, no. 8, pp. 1871-1882, 2012.

[203] P. McMinn, D. Binkley, and M. Harman, "Testability transformation for efficient automated test data search in the presence of nesting," in *Proceedings of the Third UK Software Testing Workshop*, 2005, pp. 165-182.

[204] M. Harman *et al.*, "Testability transformation–program transformation to improve testability," *Formal methods and testing,* pp. 320-344, 2008.

[205] K. Saleh, "Testability-directed service definitions and their synthesis," in *Computers and Communications, 1992. Conference Proceedings., Eleventh Annual International Phoenix Conference on*, 1992, pp. 674-678: IEEE.

[206] M. R. Woodward and Z. A. Al-Khanjari, "Testability, fault size and the domain-to-range ratio: An eternal triangle," *ACM SIGSOFT Software Engineering Notes,* vol. 25, no. 5, pp. 168-172, 2000.

[207] C. Izurieta and J. M. Bieman, "Testing consequences of grime buildup in object oriented design patterns," in *Software Testing, Verification, and Validation, 2008 1st International Conference on*, 2008, pp. 171-179: IEEE.

[208] S. Do Sohn and P. H. Seong, "Testing digital safety system software with a testability measure based on a software fault tree," *Reliability Engineering & System Safety,* vol. 91, no. 1, pp. 44-52, 2006.

[209] L. Ma, H. Wang, and Y. Lu, "The Design of Dependency Relationships Matrix to improve the testability of Component-based Software," in *Quality Software, 2006. QSIC 2006. Sixth International Conference on*, 2006, pp. 93-98: IEEE.

[210] M. Hays and J. Hayes, "The Effect of Testability on Fault Proneness: A Case Study of the Apache HTTP Server," in *Software Reliability Engineering Workshops (ISSREW), 2012 IEEE 23rd International Symposium on*, 2012, pp. 153-158: IEEE.

[211] B. Baudry, Y. Le Traon, G. Sunyé, and J.-M. Jezequel, "Towards a'safe'use of design patterns to improve OO software testability," in *Software reliability engineering, 2001. ISSRE 2001. Proceedings. 12th international symposium on*, 2001, pp. 324-329: IEEE.

[212] Y. Le Traon and C. Robach, "Towards a unified approach to the testability of co-designed systems," in *Software Reliability Engineering, 1995. Proceedings., Sixth International Symposium on*, 1995, pp. 278-285: IEEE.

[213] S. Ghosh, "Towards measurement of testability of concurrent object-oriented programs using fault insertion: a preliminary investigation," in *Source Code Analysis and Manipulation, 2002. Proceedings. Second IEEE International Workshop on*, 2002, pp. 17-25: IEEE.

[214] R. Dssouli, K. Karoui, A. Petrenko, and O. Rafiq, "Towards testable communication software," in *Protocol Test Systems VIII*: Springer, 1996, pp. 237-251.

[215] A. Rodrigues, P. R. Pinheiro, M. M. Rodrigues, A. B. Albuquerque, and F. M. Gonçalves, "Towards the selection of testable use cases and a real experience," *Best Practices for the Knowledge Society. Knowledge, Learning, Development and Technology for All,* pp. 513-521, 2009.

[216] A. Tahir, S. G. MacDonell, and J. Buchan, "Understanding class-level testability through dynamic analysis," in *Evaluation of Novel Approaches to Software Engineering (ENASE), 2014 International Conference on*, 2014, pp. 1-10: IEEE.

[217] J. Voas and L. Kassab, "Using assertions to make untestable software more testable," *Software Quality Professional,* vol. 1, no. 4, p. 31, 1999.

[218] M. Liangli, W. Houxiang, and L. Yongjie, "Using Component Metadata based on Dependency Relationships Matrix to improve the Testability of Component-based Software," in *Digital Information Management, 2006 1st International Conference on*, 2006, pp. 13-18: IEEE.

[219] J. Voas, L. Morell, and K. Miller, "Using dynamic sensitivity analysis to assess testability," 1990.

[220] S. Almugrin, W. Albattah, and A. Melton, "Using indirect coupling metrics to predict package maintainability and testability," *Journal of Systems and Software,* vol. 121, pp. 298-310, 2016.

[221] A. Bertolino and L. Strigini, "Using testability measures for dependability assessment," in *Software Engineering, 1995. ICSE 1995. 17th International Conference on*, 1995, pp. 61-61: IEEE.

[222] F. Doumbia, O. Laurent, C. Robach, and M. Delaunay, "Using the Testability analysis methodology for the Validation of Airbus Systems," in *Advances in System Testing and Validation Lifecycle, 2009. VALID'09. First International Conference on*, 2009, pp. 86-91: IEEE.

[223] M.-C. Shih, "Verification and measurement of software component testability," 2004.

[224] D. Owen, T. Menzies, and B. Cukic, "What makes finite-state models more (or less) testable?," in *Automated Software Engineering, 2002. Proceedings. ASE 2002. 17th IEEE International Conference on*, 2002, pp. 237-240: IEEE.

[225] D. Deveaux, G. S. Denis, and R. K. Keller, "XML support to design for testability," 2000.